\documentclass[trackchanges,preprint]{aastex701}
\usepackage[utf8]{inputenc}

\usepackage[version=4]{mhchem}
\usepackage{booktabs}
\usepackage{subcaption}
\usepackage{verbatim}
\usepackage{float}
\received{June 17, 2025}
\revised{August 13, 2025}
\accepted{August 27, 2025}

\begin{document}

\title{Modeling Low-Temperature Plasmas Simulating Titan’s Atmosphere}

\author[0000-0003-2769-2089]{David Dubois}
\email{david.f.dubois@nasa.gov}
\affil{NASA Ames Research Center, Moffett Field, CA, USA} 
\affil{Bay Area Environmental Research Institute, Moffett Field, CA, USA}



\author[0000-0002-0596-6336]{Alexander W. Raymond}
\email{david.f.dubois@nasa.gov}
\affil{Jet Propulsion Laboratory, California Institute of Technology, Pasadena, CA, USA} 

\author[0000-0002-1883-552X]{Ella Sciamma-O'Brien}
\email{david.f.dubois@nasa.gov}
\affil{NASA Ames Research Center, Moffett Field, CA, USA} 


\author[0000-0002-6064-4401]{Farid Salama}
\email{david.f.dubois@nasa.gov}
\affil{NASA Ames Research Center, Moffett Field, CA, USA}

\begin{abstract}

In the study presented here, we model the gas phase chemistry induced by plasma discharge at low temperature (150 K) in the NASA Ames COSmIC Simulation Chamber (COSmIC) using a 1-dimensional multi-fluid plasma model named CO-PRISM (COSmIC Plasma Reactivity and Ionization Simulation Model). Our model incorporates an extensive chemical reaction network to simulate the neutral-neutral and ion-neutral reactions occurring in the COSmIC experiments when using \ce{N2-CH4}-based gas mixtures relevant to Titan's atmosphere. Our reaction network now includes crucial reactions involving the first electronically-excited state of atomic nitrogen, recent electron collision cross-sections, and radical chemistry. In particular, we have investigated the influence of \ce{C2H2} on the gas phase polymeric growth and the elemental composition of the chemical products, and we have compared our findings to recently published solid phase analyses. The modeling results are consistent with experimental measurements of \ce{N2-CH4-C2H2} plasmas on COSmIC, showing the production of \ce{C6Hx} intermediates and precursors of larger organics, as well as methanimine in small concentration. Our numerical results point to cationic pathways enabling efficient intermediate-sized and nitrogen-rich \ce{C2H2}-driven chemistry driving tholin production. Comparison of the modeled gas phase elemental composition with elemental composition of the solid phase samples produced in COSmIC reveal similar trends, with C/N increasing when \ce{C2H2} is present in the gas mixture. Finally, our results demonstrate the importance of such synergistic studies using low-temperature plasma chemistry experiments combined with modeling efforts to improve our understanding of cold planetary environments.

\end{abstract}



\section{Introduction} 

Ion-neutral molecular reactions play an important role in shaping the chemistry in solar system and planetary ionospheres \citep[\textit{e.g.,}][]{Huntress1977Ion-moleculeAtmospheres,Harrison2008IonsAtmospheres,Larsson2012,Vuitton2019, Wong2023}. These reactions involve the interaction between ions and neutral molecules, which can result in the formation of various chemical species. For example, in the atmosphere of Mars, ion-neutral reactions are important for the production of nitrogen oxides where charge transfer processes from \ce{N2+}, \ce{Ar+} and \ce{CO+} represent key ionization pathways of \ce{CO2} \citep{Larsson2012}. 
In the outer solar system, where nitrogen and hydrocarbon-based chemistry dominate, cation-neutral reactivity has long been thought to play a role in the transition from gas phase to solid phase. Characterizing these reactions has been crucial for understanding the thermal, chemical, and physical processes that occur in the solar system and for developing accurate models of these environments \citep[\textit{and references therein}]{Larsson2012}. \

Titan, Saturn's largest moon, is the only known moon to have its own dense atmosphere, mostly composed of molecular nitrogen \ce{N2} ($\sim$ 95\%), and methane, \ce{CH4} ($\sim$ 2-5\%). This reducing atmosphere consists of five main atmospheric layers (troposphere, stratosphere, mesosphere, thermosphere and exosphere) hosting planetary-like dynamical, thermal, chemical, and seasonal variations on a global scale. The formation of the haze surrounding Titan is ultimately controlled by the gas phase molecular precursors produced in the upper atmosphere resulting from high-altitude \ce{N2} and \ce{CH4} photolysis and radiolysis. These precursors consist of hydrocarbon radicals (\textit{e.g.}, \ce{CH2}, \ce{CH3}) and more complex hydrocarbons, nitriles and polycyclic aromatic hydrocarbons \citep{Waite2007}. 
Energetic sources triggering high-altitude chemistry include solar UV photons, solar X-rays, Saturn’s magnetospheric energetic electrons \citep{Krasnopolsky2014}. The Cassini spacecraft studied Titan for 13 years and directly measured the molecular mass of ions for the first time in Titan’s upper atmosphere \citep{Waite2007,Crary2009}. These observations uncovered the complexity of Titan’s upper atmospheric chemistry, consisting of radicals, neutrals, positive and negative ions, and the preliminary stages of solid haze particle formation. In some cases, the measured abundances of certain short-lived ions such as \ce{CH4+, HCNH+, C2H5+} exceeded the predicted abundances determined by previous models \citep[\textit{e.g.,}][]{Cui2009}. On the other hand, the abundance of some heavy hydrocarbons and N-bearing neutral molecules measured by INMS were found to be lower than what ion-neutral modeled had estimated \citep[\textit{and references therein}][]{Cui2009b}. These observations highlighted the role of magnetospheric electron precipitation on Titan's nightside and diurnal variations which directly impact the molecular distribution in the upper atmosphere.

Simulating Titan’s ionospheric chemistry in the laboratory constrains the chemical precursors of haze particles formed at high altitude. These particles descend to lower altitudes where they can act as condensation nuclei in the lower stratosphere and where they may settle to the surface. Simulating upper atmospheric chemistry can thus provide insight into the chemical composition in the lower atmosphere.

Chemical models have helped explain the gas phase chemistry which produces the neutral and ion precursors of aerosols found in Titan's atmosphere \citep[\textit{e.g.},][]{Vuitton2019}. Alongside those models, laboratory experiments have helped in vetting the reaction networks through the study of specific channels involving neutral and charged molecular species. Furthermore, photochemical and microphysical models have investigated the dusty nature of Titan’s ionosphere ($>$ 900 km in the atmosphere) and characterized the interaction between the aerosols and charged particles \citep{Lavvas2013}. More specifically, by using a wide array of energy sources and coupling extensive lists of ion-neutral reactions, models have given insights into the first steps linking small hydrocarbons with larger molecules \citep{Dobrijevic2016,Loison2017,Mukundan2018,Vuitton2019}. Ion-molecule reactions are thought to be directly relevant to aerosol growth, and are controlled by the relative abundances of the two initial neutral main constituents, \ce{N2} and \ce{CH4} \citep{Lavvas2013}. The dissociative ionization of \ce{CH4} results in the formation of \ce{CH3+}, an ion predicted \citep{Anicich2003} to participate in the production of the first light hydrocarbons such as \ce{C2H5+}, occurring on both the nightside and dayside of Titan:

\begin{equation}
\label{methane destruction}
\ce{CH_3^+ + CH_4 \longrightarrow C_2H_5^+ + H_2}
\end{equation}

Following Reaction \ref{methane destruction}, a long suite of dissociative recombination reactions have been predicted to occur, forming acetylene (\ce{C2H2}, Reaction \ref{c2h2 prod}), ethylene (\ce{C2H4}, Reaction \ref{c4h2 prod}), and, eventually, benzene (\ce{C6H6}, Reaction \ref{c6H6 prod}), the first and only six-membered ring that was discovered in Titan's lower atmosphere clouds to date \citep{Coustenis1999,Vuitton2008,Vinatier2018}. Other unidentified emission features were identified as potential carriers of polycyclic aromatic hydrocarbons \citep{Dinelli2013,Lopez-Purtas2013}. Benzene is expected to serve as a potential seed for larger macromolecules and aromatics, for example by radical-phenyl pathways \citep{Vuitton2008}.
Larger molecular structures ($>$ C9 hydrocarbons) formed through dissociative recombinations at lower altitudes ($< 1200$ km) have also been proposed \citep[\textit{e.g.,}][]{Crary2009} to be possible precursors to polycyclic aromatic hydrocarbons (PAHs). Both cation and anion pathways have thence been studied to explain the presence of these large precursors observed by Cassini \citep{Zabka2009ReactivityS,Zabka2012}.

\begin{equation}
\label{c2h2 prod}
\ce{C2H5+ + e- \longrightarrow C2H2 +H2 + H}
\end{equation}

\begin{equation}
\label{c4h2 prod}
\ce{C2H5+ + e- \longrightarrow C2H4 + H}
\end{equation}

\begin{equation}
\label{c6H6 prod}
\ce{C6H7+ + e- \longrightarrow C6H6 + H}
\end{equation}

While dissociative recombination constitutes a mechanism for the loss of certain ions, ion-neutral reactions play a major role in the formation of ions. 
For instance, both ion-neutral and dissociative recombination are proposed to explain the production and loss of cyclic \ce{C3H3+}, respectively, which itself depends on the presence of \ce{CH4} and \ce{C2H2} \citep{Vuitton2019}. Reactions involving \ce{C3H3+} can produce benzene. Note that lower in Titan's atmosphere, \ce{C3H3} dimerization can also produce benzene \citep{Vuitton2008,Larsson2012}. 

Along with hydrocarbons and PAHs, nitriles are among the most abundant compounds in Titan's atmosphere. Cassini Ion Neutral Mass Spectrometer (INMS) measurementscoupled with photochemical models yielded peak mole fractions of $2 \times 10^{-4}$ for HCN, the most abundant photochemically-produced nitrile, and $\sim 10^{-6}$ for \ce{HCNH+} at 1100 km \citep{Dobrijevic2016,Vuitton2019}. The nitriles produced in the upper atmosphere have relatively larger proton affinities, thus leading to their more stable closed-shell cation counterparts. \ce{HCN}, for example, is formed through neutral-neutral reactions in the upper atmosphere, while \ce{HCNH+} forms through direct proton transfer to HCN or \ce{N+ + CH4}. Other N-bearing molecules, such as acetonitrile (\ce{CH3CN}) and cyanoacetylene (\ce{HC3N}) are highly reactive and play a significant role in the nitrogen-based ion-neutral chemistry of Titan's atmosphere \citep{Vuitton2007}. In the ionosphere, these neutral and cation nitriles can undergo various reactions with other ions, electrons, and free radicals, leading to the formation of more complex organic molecules in the ionosphere \citep{Berry2018,Vuitton2019,Dubois2020,Carrasco2022,Nixon2024}. 

Some of these compounds are also of astrobiological interest such as methylamine \ce{CH3NH2+}, a precursor to amino acids in the long chain of HCN hydrogenation reactions \citep{Gu2009,Singh2010,Theule2011}. HCN itself is the smallest neutral containing a CN moiety, which is key in the amino acid formation process \citep{Theule2011,Noble2013}. Ion-neutral reactivity therefore plays a key role at every stage of Titan's chemical reactivity: 
(i) the degradation and recombination of small neutrals and free radicals,
(ii) the formation of larger molecules, PAHs, haze particles, and
(iii) the formation of putative molecules of astrobiological interest. However, the processes involved in coupling ion and neutral chemistry and aerosol production are complicated and are an area of ongoing study.

Laboratory experiments have been conducted to simulate Titan's atmospheric chemistry using plasmas created by different energy sources (\textit{e.g.}, UV-induced photochemistry, plasma discharges) to induce chemistry in various precursor gas mixtures representative of Titan’s atmospheric composition \citep[\textit{e.g.,}][]{Scattergood1989,Sagan1992Titan:Chemistry,Imanaka2004,Szopa2006,Pilling2009,Carrasco2012,Couturier-Tamburelli2014a,Sciamma-OBrien2014,Bourgalais2019,Dubois2019a,Dubois2019ApjL,Zabka2012,Cable2012,Raulin2012,Horst2018,Bourgalais2020OnAtmosphere,Bourgalais2021AromaticChemistry}. These experiments along with the development of photochemical and microphysical models have helped better constrain the chemical pathways leading to the formation of larger molecules and solid particles in Titan’s atmosphere, and have substantially advanced our understanding of the chemistry occurring in Titan’s ionosphere \citep{Dubois2025PhotochemicalReview}. However, these numerical simulations by models have faced some obstacles due to a limited number of reactions with known reaction rates at Titan-relevant temperatures.\\

To further advance our understanding of Titan's ionospheric chemistry, experiments were conducted with the COSmIC Simulation Chamber at NASA Ames Research Center to investigate low temperature (150 K) gas phase chemistry in ionized conditions using a plasma discharge in the stream of jet-cooled expansions of \ce{N2-CH4}-based gas mixtures. The COSmIC experiment is designed to simulate low-temperature chemistry in conditions relevant to Titan and planetary atmospheres \citep{Salama2016}. The general goal of our study (\textit{i.e.}, combining chemistry modeling with experimental measurements) is to determine specific relevant pathways to understand the chemistry occurring in the COSmIC experimental setup when simulating Titan’s atmospheric chemistry, and is therefore of interest for the Titan/planetary community.

By comparing the COSmIC experimental data to synthetic mass spectra generated by the CO-PRISM model, we can (1) validate the chemical network considered in CO-PRISM, (2) identify specific pathways or molecules that are missing to fit the experimental data and accordingly implement changes to the model, (3) quantify and identify the species produced in the plasma, (4) assess the impact of specific chemical pathways on the resulting chemical products, and finally (5) guide and optimize future COSmIC experiments relevant to Titan’s atmosphere (\textit{e.g.}, what chemical pathways produce specific molecules). The need for new COSmIC experiments focusing on specific chemical pathways to help in the interpretation of Cassini observations was demonstrated in \cite{Sciamma-OBrien2014}. 

In these experiments, the gas phase was characterized by mass spectrometry to monitor the formation of larger molecular products in the plasma discharge \citep{Sciamma-OBrien2014,Sciamma-OBrien2017}. In parallel, a 1D multi-fluid chemical network model was developed to simulate the chemical reactivity occurring in the plasma discharge \citep{Raymond2018}. This model enables generating simulated mass spectra that can then be compared to experimental mass spectra. In this study, we incorporate updated reaction rates and a total of 45 newly updated reaction pathways into this model nowcalled COSmIC Plasma Reactivity and Ionization Simulation Model (CO-PRISM). We study the impact of changing plasma parameters on the ion chemistry, expanded on the investigation of the plasma parameter space impact on the chemistry from previous studies. A total of 11 N$_2$-CH$_4$-based initial gas mixture conditions are explored. Furthermore, the updated model is used to calculate the C/N elemental composition of the gas-phase products, which we present below with a comparison to recently published solid-phase C/N ratios of Titan aerosol analogs produced in COSmIC. We also discuss the result of a sensitivity analysis conducted with our model considering various energy levels (voltage on the plasma source) and show how the chemistry in COSmIC is affected by electrode potential. In addition, the mole fractions of six  major ions (\ce{CH+, CH3+, C2H3+, C2H5+, C4H3+, C5H5+}) produced in the plasma discharge are computed and we show their evolution in time in the plasma channel and their sensitivity to the initial composition of the gas mixture. Finally, extracted mass spectra are compared to previous laboratory and numerical simulations to demonstrate the importance of combining experiments and modeling to improve our understanding of planetary environments.

\section{Simulation of Titan's Atmospheric Chemistry} \label{sec:methods}\

\subsection{The NASA Ames COSmIC Simulation Chamber (COSmIC)}

\begin{figure}[h]
\centering
\includegraphics[scale=0.4]{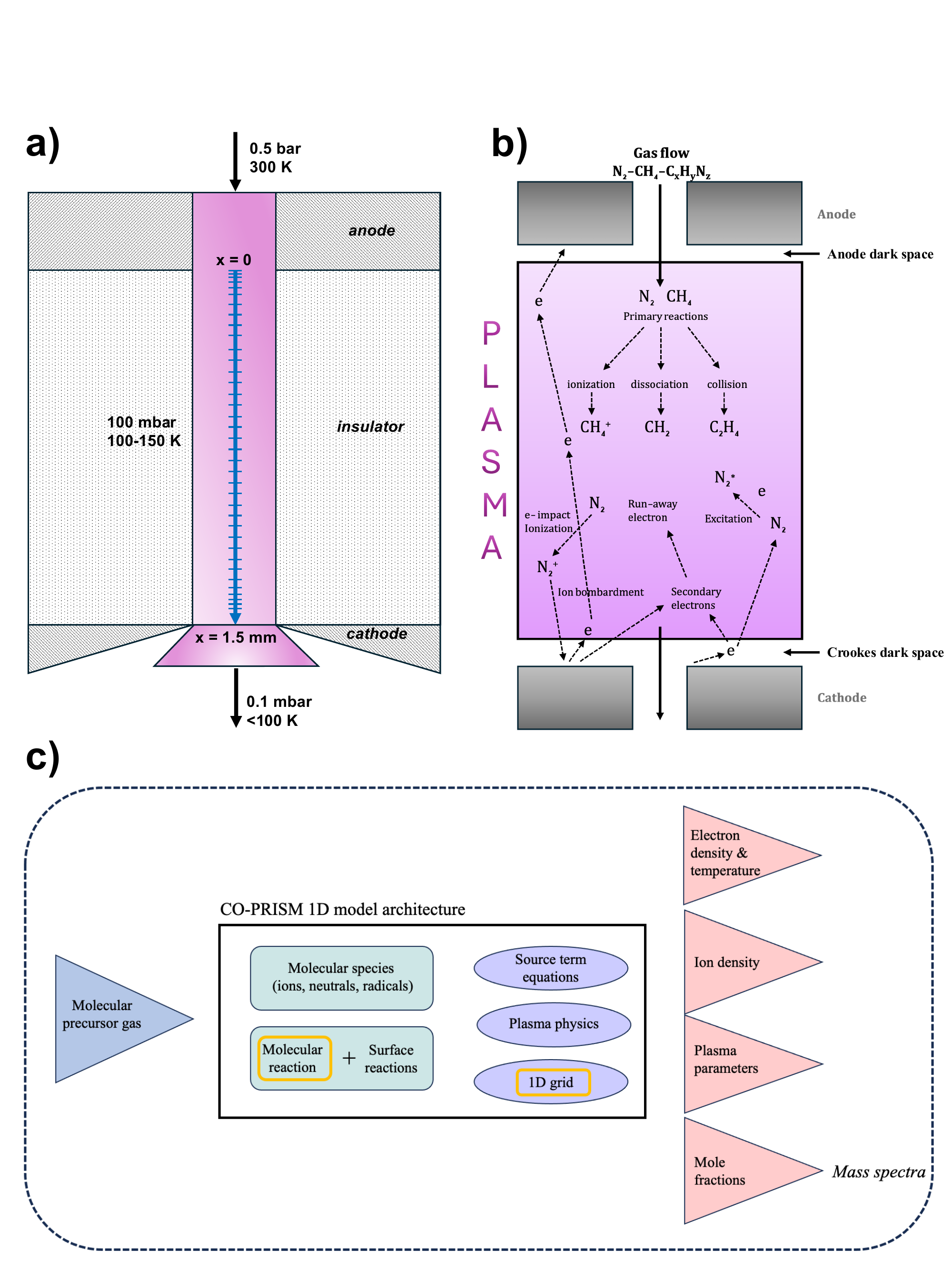}
\caption{a) Cross-section schematic view of the THS plasma channel (not to scale). The pathway along which reactions are calculated by the model in the plasma channel is indicated in blue. The temperature inside the plasma cavity is on the order of T = 150 K \citep{Biennier2006FlowExpansion,Sciamma-OBrien2014}. b) Top-view cross-section schematic view of the plasma cavity (not to scale) indicating the dominant collisional processes occurring in the plasma and controlling the formation of the first building blocks leading to larger products; illustrated in the case of an \ce{N2-CH4}-based atmosphere relevant to Titan. c) Simplified schematic representation of the CO-PRISM model. Each simulation is controlled by the initial precursor gas composition (blue triangle). Green boxes correspond to the chemical input parameters (list of species, molecular reactions, and surface
reactions), purple ellipses correspond to plasma physics and grid geometry parameters, and outputs are represented with red triangles. The model parameters that were updated and modified in the study presented here are circled in orange. Synthetic mass spectra can then be retrieved from the computed mole fractions. \label{fig1}}
\end{figure}

The COSmIC facility was developed at NASA Ames Research Center to simulate astrophysical and planetary environments at low temperature \citep{Salama2016}. Among these environments, Titan's low-temperature (150 K) gas phase chemistry in plasma conditions was of particular interest and led to the development of targeted experiments on COSmIC  \citep[\textit{e.g.}][]{Sciamma-OBrien2014}. In the COSmIC setup, a pulsed discharge nozzle (PDN) is used to (1) cool down a gas mixture by expanding it through a thin 127 $\mu$m x 10 cm slit, and (2) generate a plasma discharge in the stream of the jet-cooled gas expansion by applying a high voltage onto cathodes placed in the stream of the expansion (see Figure \ref{fig1}a,b) \citep{Remy2003, Broks2005a,Broks2005}.

The COSmIC PDN source has been effectively used to simulate analogous conditions to  astrophysical and planetary environments and study both the gas and solid phase. Molecular species present in the plasma expansion have been monitored by Reflectron Time-of-Flight mass spectrometry (ReTOF-MS) \citep{Ricketts2011,Contreras2013,Sciamma-OBrien2014} and cavity ring down spectroscopy \citep{Tan2005,Biennier2006a,Bejaoui2019,Bejaoui2023}. Solid particles, analogs of planetary aerosols \citep{Sciamma-OBrien2017,Sciamma2019,Nuevo2022,Sciamma2023} and cosmic grains \citep{Salama2016,OBrien2020,Marin2020} have been produced in the plasma expansion and analyzed ex situ with various techniques ranging from scanning electron microscopy and infrared spectroscopy, to X-ray absorption near edge structure spectroscopy (XANES). 

In COSmIC, nitrogen-methane-based gas mixtures are used to simulate Titan's atmospheric chemistry. The expansion of the pulsed gas in the cavity between the anode and the cathode results in the cooling of the gas mixture down to low Titan-like temperature relevant to the upper atmosphere ($\sim$ 150 K) before inducing the chemical breakdown within the plasma discharge \citep{Biennier2006a}. The short residence time of the gas in the plasma discharge (less than 4 $\mu$s) results in a truncated chemistry that enables simulating and studying the first steps in the chemical reaction pathways of Titan's atmosphere. 

In the study presented here, we focus on first updating the numerical model that simulates the chemistry in the COSmIC/THS experiments and assess the impact of changing some of the modeling parameters on the numerical results. We then compared our new simulated mass spectra to those obtained experimentally and using the first version of the CO-PRISM model \citep{Sciamma-OBrien2014,Raymond2018}. 
Table \ref{Table1} provides a summary of the parameters used for the experimental and numerical results discussed in this paper.

\begin{table}[h]
\centering
\begin{tabular}{lcccccc}
\toprule
 Gas mixture & Ratio & Gas flow & Voltage & Experiment$^\textit{a}$ & CO-PRISM (2018)$^\textit{b}$ & CO-PRISM (2025) \\ \midrule
 \ce{N2-CH4} & 99-1 & 2000 sccm & 1000 V &  &  & x \\ \midrule
\ce{N2-CH4} & 95-5 & 2000 sccm & 700 V &  & x & x \\ \midrule
\ce{N2-CH4} & 95-5 & 2000 sccm & 800 V &  & x & x \\ \midrule
\ce{N2-CH4} & 95-5 & 2000 sccm & 1000 V & x & x & x \\ \midrule
\ce{N2-CH4} & 90-10 & 2000 sccm & 800/1000 V & x & x & x \\ \midrule
\ce{N2-CH4-C2H2} & 85-10-5 & 2000 sccm & 800/1000 V & x & x & x \\ \midrule
\ce{N2-CH4-C2H4} & 85-10-5 & 2000 sccm & 800/1000 V & x & x & x \\ \midrule
\ce{N2-CH4-C2H6} & 85-10-5 & 2000 sccm & 1000 V &  &  & x \\ \midrule
\ce{N2-CH4-HCN} & 85-10-5 & 2000 sccm & 1000 V &  & & x \\ \midrule
\ce{N2-CH4-CH2NH} & 85-10-5 & 2000 sccm & 1000 V &  & & x \\ \midrule
\ce{N2-CH4-NH3} & 85-10-5 & 2000 sccm & 1000 V &  & & x \\ \bottomrule
\end{tabular}
\caption{List of the parameters used in the experimental and modeling gas phase studies, for various \ce{N2-CH4}-based simulated atmospheres. Gas flows are provided in standard cubic centimeters (sccm). The voltage applied to the cathode to generate the plasma is provided in Volts (V). ($^\textit{a}$) \citet{Sciamma-OBrien2014}, ($^\textit{b}$) \citet{Raymond2018}. Only ultra high purity gases were used for the experimental study and mixing ratios and flows were controlled by MKS 1479 mass flow controllers.\label{Table1}}
\end{table}

\subsection{The COSmIC Plasma Reactivity and Ionization Simulation Model (CO-PRISM)}
\label{section 2.2}

CO-PRISM is the 1D numerical model was first developed by \citet{Raymond2018} to simulate the chemical network in a plasma discharge (Figure \ref{fig1}b) generated with the COSmIC PDN in N$_2$-CH$_4$-based initial gas mixtures. It is a modular model that can be applicable to a wide range of planetary and astrophysical environments simply by adding the reactions relevant for the gas mixture used in the COSmIC experiments. The goal of the \textit{Titan} CO-PRISM model in its present form is to quantify the molecular species (neutrals and positive ions) produced in the COSmIC experiments and generate synthetic mass spectra comparable to experimentally-measured mass spectra acquired with COSmIC's Time-of-Flight (TOF) mass spectrometer \citep{Ricketts2011,Sciamma-OBrien2014}, but also comparabled to mass spectra generated by photochemical models \citep[\textit{e.g.}][]{Vuitton2018}, and \textit{in situ} Cassini mass spectra \citep[\textit{e.g.}][]{Waite2007}. The TOF is equipped with a polarized skimmer that attracts ions into the instrument and detects them directly without fragmentation. As stated above, the gas phase products measured by the TOF can help us better characterize the mixture-dependent composition, thus guiding future experiments needed to help the interpretation of Cassini observations.

In CO-PRISM, plasma physics parameters such as the ionization fractions and the electron temperature are calculated using a fluid mechanical approach \citep{Raymond2018} following previous numerical investigations and characterization of the plasma discharge produced in the COSmIC PDN \citep{Broks2005,Broks2005a, Remy2003} and detailed hereafter. 
CO-PRISM also takes into account the abnormal glow nature of the COSmIC plasma discharge \citep[\textit{e.g.}][]{Remy2003}, \textit{i.e.},  an electron density at the higher end of the range typically deemed normal for glow discharges.

\vspace{0.7 cm}

\textit{2.2.1. The CO-PRISM model}

\vspace{0.7 cm}
Figure \ref{fig1}c shows a simplified flow chart representation of the CO-PRISM 1D model. The main inputs consist of a list of molecular species, the molecular reactions that contribute to their formation or destruction in the plasma (neutral-neutral and neutral-ion
with associated reaction coefficients), and a suite of surface reactions (neutralization reactions and secondary electrons formed near the cathode). 

The CO-PRISM architecture uses a 1D grid with a user-defined grid size. In this study, the grid size was increased to 100 nodes distributed along an arithmetic sequence to help improve the sensitivity of our calculations (see the blue grid in Figure \ref{fig1}a). 

The model outputs include the molar fractions of all the species (cations, neutrals, radicals) formed in the simulated plasma discharge along the plasma channel. Mass spectra can subsequently be computed from the molar fraction obtained at a given position in the plasma channel, and compared against the measured spectra taken by mass spectrometry. Isotopes are not presently included in the CO-PRISM model. This is an element that we are considering for future development.



In CO-PRISM, the electron density and electron energy density outputs are derived following \citet{Broks2005a,Broks2005} from the following equations, respectively:\\
\begin{equation}
 R_e = \frac{\partial n_e}{\partial t} - \nabla \cdot [(\mu_e \cdot E)n_e + \nabla(D_en_e)] 
\label{equ1}
\end{equation}

and

\begin{equation}
 S_{\varepsilon} =  \frac{\partial n_\varepsilon}{\partial t} - \nabla \cdot [(\mu_\varepsilon \cdot E)n_\varepsilon + \nabla(D_{\varepsilon}n_{\varepsilon})] + E \cdot [(\mu_e \cdot E)n_e + \nabla(D_en_e)]
 \label{equ2}
\end{equation}

where the subscripts $e$ and $\varepsilon$ correspond to the electrons and the electron energy density (thermionic energy) in the discharge, respectively. 
$R_e$ is the source term for electrons, 
$S_{\varepsilon}$ is the energy source term accounting for inelastic collisions, $E$ is the electric field, $D$ is the diffusivity, and 
$\mu$ is the mobility\citep{Hagelaar2005}. The concentrations of heavy species (\textit{i.e.}, neutrals, ions, and radicals) 
are calculated using the following formalism:

\begin{equation}
\rho\frac{\partial \omega_k}{\partial t} + \rho(u \cdot \nabla )\omega_k = \nabla \cdot \left[\rho\omega_kD_k \frac{\nabla\omega_k}{\omega_k}-z_k\mu_kE \right] + R_k
\label{equ3}
\end{equation}

where $\omega_k$ represents the 
mass fraction of the $k_{th}$ molecular species and $R_k$ corresponds to the source term
for heavy species. The mixture-average diffusion coefficient $D_k$ is considered negligible ($D_k = 0$) since the supersonic expansion dominates the continuous flow within a plasma pulse \citep{Raymond2018}. For each species, the source term $R_k$ is obtained by computing a network of chemical rate equations. \citet{Raymond2018} simplified  the \citet{Broks2005,Broks2005a} 2D model into the 1D CO-PRISM model in order to allow for a larger chemical network to be simulated. Calculations are conducted along the $x$ axis \citep{Raymond2018}, with the anode occupying the assigned \textit{inlet} position and the cathode the \textit{outlet} position (see Figure \ref{fig1}a). 
For each simulation, the initial gas composition mixing ratio (N$_2$, CH$_4$, with or without the addition of other trace hydrocarbons used as precursors) is defined in the model to be representative of the Titan COSmIC experiments we want to compare the numerical outputs to. In addition, following \cite{Raymond2018} a small number of N$_2$(X, $\nu = 0)^+$ ions are added to the initial gas mixture in the model to insure quasi-neutrality with the seed electrons (Eq. \ref{equ1}). All other species except the initial \ce{N2-CH4}-based carrier gas are initialized with a null abundance.

The model also includes surface reactions to account for ions near the inlet diffusing into the anode as well as heavy species loss, neutralization of ions, and the generation of secondary electrons at the cathode. The rate constants of the surface reactions are calculated following \citep{Motz1960DiffusionBoundary}:
\begin{equation}
r = k_s^f\prod_{k = 1}^{Q}c_k^{\nu}
\label{equ8}
\end{equation}

where the forward rate constant $k_s^f$ is calculated with
\begin{equation}
k_s^f = \left(\frac{\gamma_f}{1-\gamma_f/2}\right)\frac{\prod\sigma_j^\nu j1}{(\Gamma_{tot})^m4}\sqrt{\frac{8RT}{\pi M_k}},
\label{equ5}
\end{equation}

where \textit{T} is the local temperature, $M_k$ the molar mass of the $k_{th}$ ionic species, \textit{R} the ideal gas constant and $\Gamma$ the surface site concentration. $\gamma_f$ is the sticking coefficients and, as explained in \citet{Raymond2018}, is maintained constant at a value of 1 across the plasma channel in the model to focus on the chemistry occurring in the plasma more than on wall effects.
A total of 151 surface reactions are currently present in the model, compared to 64 in \citet{Raymond2018}. The model also parameterizes secondary emission effect (\textit{i.e.}, secondary emission coefficient and mean energy of secondary electrons). Electron impact reactions follow electron collision cross-sections, and include excitation, attachment, ionization and elastic reactions.

Every reaction considered here is irreversible according to the following time-dependent rate parameterization based on the previous Eq. \ref{equ8}:

\begin{equation}
r = k^f\prod_{k = 1}^{Q}c_k^{\nu}
\label{equ6}
\end{equation}

where the rate constants $k^f$ ($m^3/(s\cdot mol))$ determined for each reaction are taken into account:

\begin{equation}
k^f = \gamma \int_{0}^{\infty}\epsilon\sigma_k(\epsilon)f(\epsilon)d\epsilon
\end{equation}

As described in \citet{Raymond2018}, to accommodate for positive ions present near the inlet diffusing into the anode, we use a surface reaction at the inlet boundary that acts both as sink and source term. To account for heavy species loss, ion neutralization (\textit{e.g.} \ce{CH4+ \longrightarrow CH4}) and the generation of secondary electrons at the cathode, we use an open boundary condition and a point surface reaction at the outlet/cathode. Secondary electrons are generated according to a secondary emission coefficient determined to be 0.025 \citep{Phelps1999}. An illustrative view of the dominant chemical processes that occur in the plasma channel is shown in Figure \ref{fig1}. The dark spaces near the anode and cathode help maintain quasineutrality of the plasma, while rapid free electrons are accelerated in the plasma channel inducing electron-impact ionization and dissociative reactions. Secondary electrons are produced at the cathode with a rate fixed by the secondary emission coefficient \citep{Remy2003,Broks2005,Raymond2018}


In this study, calculations were conducted along a one dimensional grid with 100 node points (increased from 40 in \citet{Raymond2018}). Node distribution is user-defined and its sequence follows a symmetric and arithmetic distribution between the anode and cathode (Figure \ref{fig1}a). Our grid contains a maximum element growth rate of 1.3 and a maximum element size of $1.01 \times 10^{-4}$ m. The one-dimensional space chosen by \citet{Raymond2018} accommodates for the larger number of reactions compared to the 2D model used by \citet{Broks2005}, while keeping the computation tractable. The computational grid is based on the COSmIC/THS PDN geometry (Figure \ref{fig1}a), where the gas mixture flows through a 100 mm long and 127 $\mu$m tall inlet at the anode. At the outlet, the cathode is represented by two elkonite conductors. The inner region (\textit{i.e.}, plasma channel or cavity) is confined by two alumina plates (acting as an insulator) and is 1.5 mm wide. The analysis of plasma properties and molecular composition were performed along the width of this plasma channel.

\vspace{0.7 cm}

\textit{2.2.2. Reaction Network Refinements}

\vspace{0.7 cm}

The reaction network in \cite{Raymond2018} tracked 155 molecular species and contained 249 reactions including electron impact, ion-neutral, and neutral-neutral reactions. It also contained 64 surface reactions, neutralizing ions and producing secondary electrons. CO-PRISM now contains 294 electron impact and bimolecular reactions, 67 surface reactions, for a total of 168 species. As described below, the previous CO-PRISM reaction network did not include key electron impact reactions relevant to cold plasma discharges involving \ce{C2H2}, \ce{C2H4}, and \ce{C2H6}.

In the present study, we have updated the reaction rates to refine CO-PRISM and addressed gaps by:

\begin{enumerate}
  \item Updating the reaction rates of 77 reactions in accordance with new values from \citet{Vuitton2019}
  \item Including key electron impact reactions relevant to hydrocarbon-based (\ce{C2H2}, \ce{C4H2}, and \ce{C6H2}) plasma discharges.
  \item Including key reactions forming \ce{NH3} and its derivatives relevant to Titan's upper atmosphere.
  \item Including key reactions involving the first electronically-excited state of atomic nitrogen and methanimine (\ce{CH2NH}), and the very important NH intermediate radical.
\end{enumerate}

\begin{figure}
\centering
\includegraphics[scale=0.4]{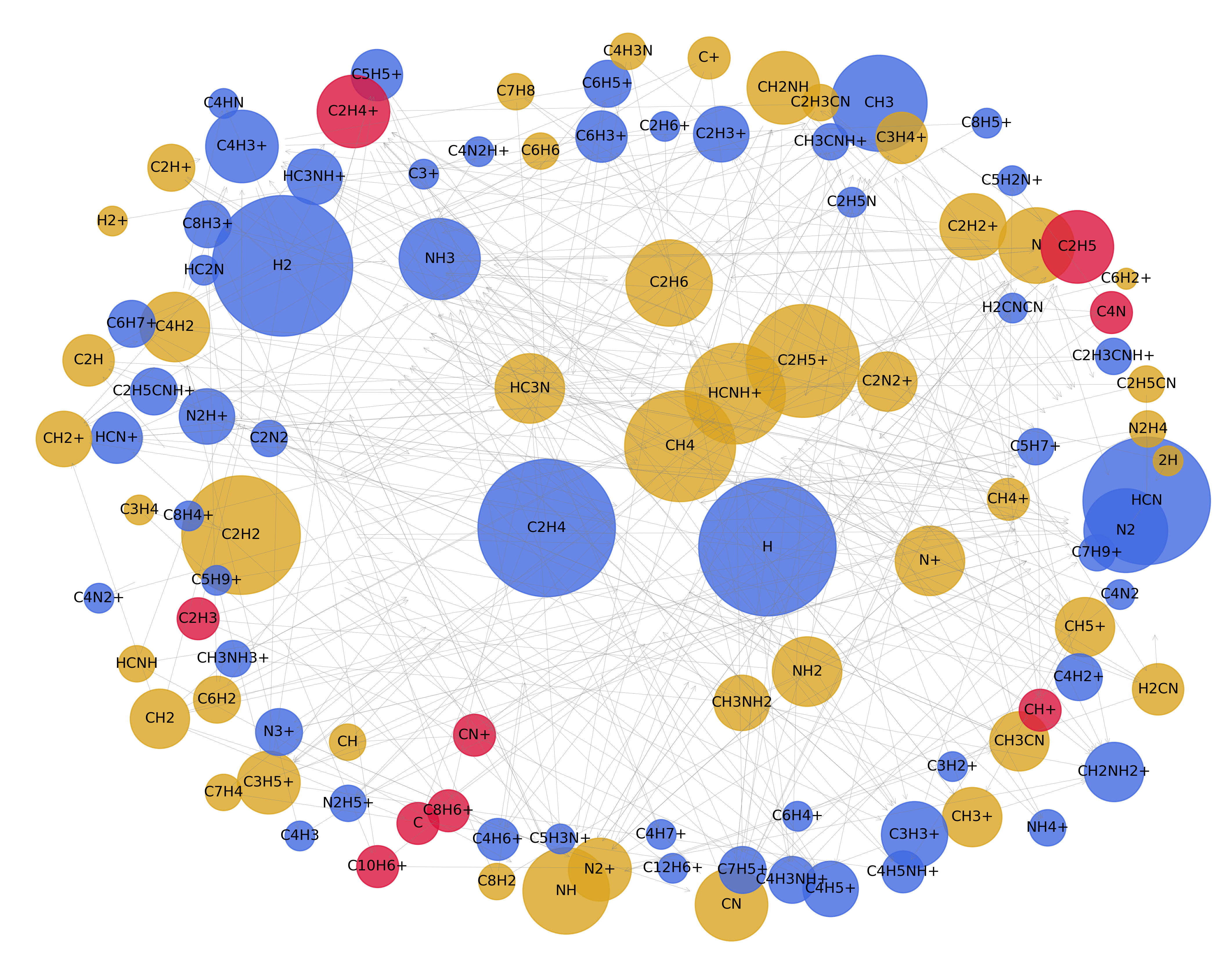}
\caption{Schematic diagram showing all the model pathways leading to the production of both neutrals and cations. Each molecular species indicated here has a node whose size is proportional to the number of times this species is either lost or produced in the overall scheme. The larger the node, the more important that molecule's contribution to the network. In blue are species participating mainly as a formed product. In gold are species participating mainly as a reactant. Species in crimson are in reactant-product equilibrium.
\label{network_global}}
\end{figure}

\begin{figure}
\centering
\includegraphics[scale=0.4]{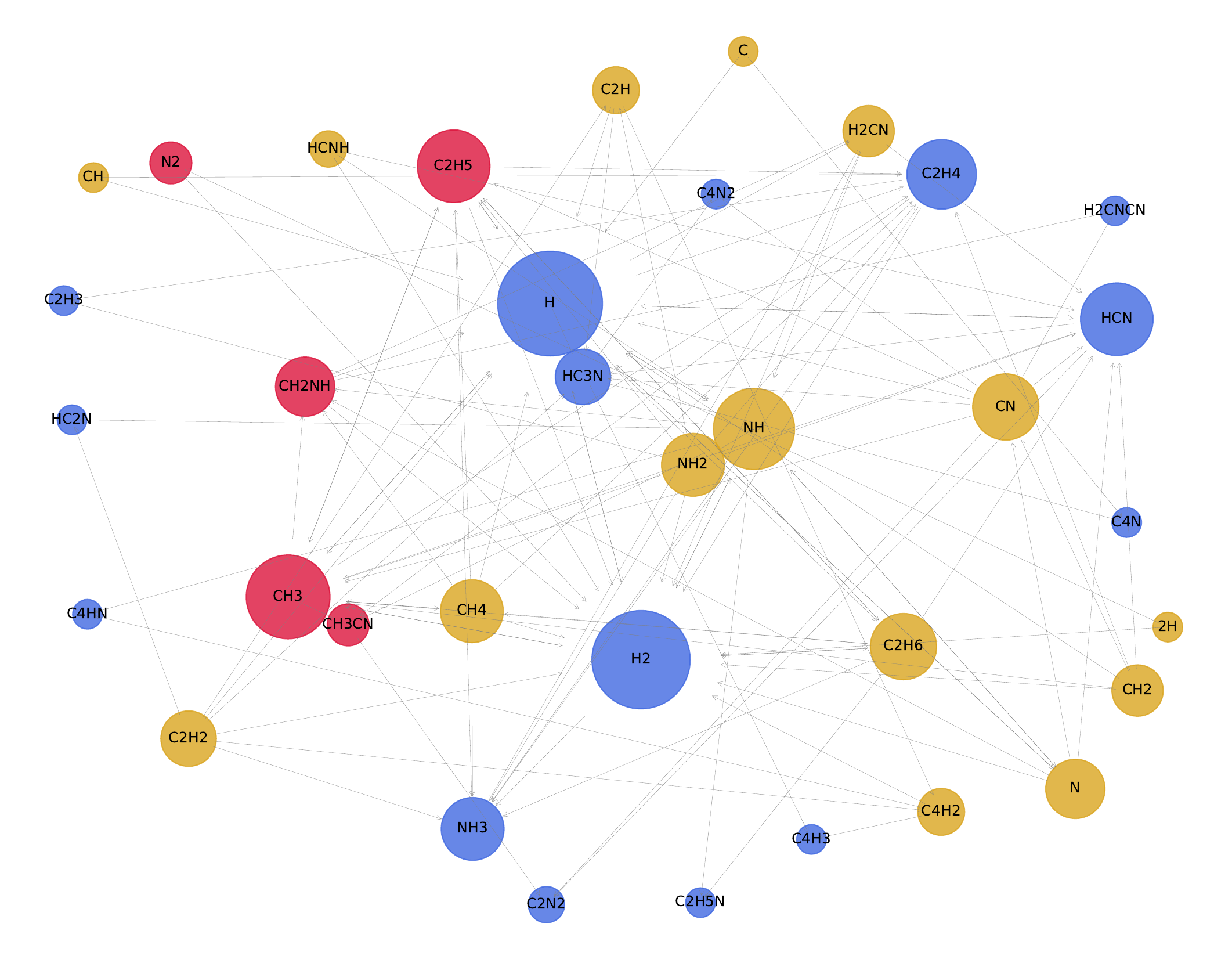}
\caption{Schematic diagram showing the model pathways only involving neutral species in neutral-neutral reactions. The interpretation of the diagram is the same as in Figure \ref{network_global}.
\label{network_neutral-neutral}}
\end{figure}

In order to expand on the scope of the plasma conditions we updated the model with a set of three electron impact dissociative ionization reactions for \ce{C2H2}, \ce{C4H2}, and \ce{C6H2} (see R1, R2, and R3 of Table \ref{Table2}) and a total of 45 new reactions. These, including electron impact excitation reactions in particular, are key for characterizing non-equilibrium gases \citep{Huo2015Electron-impactGas}. Electron collision cross section data for R1, R2, and R3 are obtained from \citet{Tian1998CrossC2H2} and \citet{DeBleecker2006DetailedDischarges}. The global networks including all neutral-ion pathways and all neutral only pathways presently in the model are shown in Figure \ref{network_global} and Figure \ref{network_neutral-neutral}, respectively.

To assess the impact of these changes on the modeled chemistry, we then ran the updated 1D model simulations for gas mixtures listed in Table \ref{Table1} (\ce{N2-CH4}) and computed synthetic mass spectra from the calculated molecular abundances near the outlet before comparing them with the previous synthetic spectra and the experimental spectra presented in \cite{Raymond2017}. We also perform a sensitivity analysis of this new version of CO-PRISM, calculating the molar concentrations of the molecular products (1) across the entire plasma channel (from inlet to outlet) to identify potential wall effects at the electrodes; and (2) when changing some of the experimental parameters (voltage on the electrodes, molecular precursors -- see Table \ref{Table1}) to assess their impact on the chemical pathways and in particular on the nitrogen chemistry in COSmIC and how they relate to the solid phase analyses. The range of initial N$_2$-CH$_4$ mixing ratios considered falls within the typical range of methane concentrations ($\sim$1-10 \% [CH$_4$]$_0$) usually considered in theoretical and experimental atmospheric simulations of Titan (see Introduction).

---

\newpage
\clearpage

\LTcapwidth=\textwidth

\begin{longtable}{cccc}

\toprule
\textbf{No.} & \textbf{Reaction} & \textbf{Rate \textit{k} ($cm^3.s^{-1}$)} & \textbf{Reference} \\




R1 & \ce{C2H2 + e \longrightarrow CH+ + CH + 2e} & collis. cross section & 1 \\
R2 & \ce{C2H4 + e \longrightarrow C4H2+ + 2e} & collis. cross section & 2 \\
R3 & \ce{C2H6 + e \longrightarrow C6H2+ + 2e} & collis. cross section & 2, 17 \\
R4 & \ce{C2H+ + C2H2 \longrightarrow C4H2+ + H} & $1.85 \times 10^{-9}$ & 2, 3, 4, 5 \\
R5 & \ce{C6H2+ + C2H2 \longrightarrow C8H4+} & $4.3 \times 10^{-11}$ & 2, 3, 4, 5 \\
R6 & \ce{C6H5+ + C2H2 \longrightarrow C8H6+ + H} & $7.8 \times 10^{-11}$ & 2, 3, 4, 5 \\
R7 & \ce{C8H6+ + C2H2 \longrightarrow C10H6+ + H2} & $5.0 \times 10^{-11}$ & 2, 3, 4, 5 \\
R8 & \ce{C10H6+ + C2H2 \longrightarrow C12H6+ + H2} & $5.0 \times 10^{-11}$ & 2, 3, 4, 5 \\
R9 & \ce{C2H2 + C4H2+ \longrightarrow C6H4+} & $5.0 \times 10^{-11}$ & 2, 3, 4, 5 \\
R10 & \ce{C2H2 + C4H3+ \longrightarrow C6H5+} & $2.9 \times 10^{-11}$ & 2, 3, 4, 5 \\
R11 & \ce{N + CH3 \longrightarrow HCN + H2} & $1.4 \times 10^{-11}$ & 2, 3, 4, 5 \\
R12 & \ce{N + CH2 \longrightarrow HCN + H} & $5.0 \times 10^{-11}$ & 2, 3, 4, 5 \\
R13 & \ce{N + H2CN \longrightarrow HCN + NH} & $6.7 \times 10^{-11}$ & 2, 3, 4, 5 \\
R14 & \ce{CN + CH4 \longrightarrow HCN + CH3} & $2.9 \times 10^{-11}$ & 2, 3, 4, 5 \\
R15 & \ce{CN + C2H6 \longrightarrow HCN + C2H5} & $1.8 \times 10^{-11}$ & 2, 3, 4, 5 \\
R16 & \ce{H2CN + H \longrightarrow HCN + H2} & $2.9 \times 10^{-11}$ & 2, 3, 4, 5 \\
R17 & \ce{N(^2D) + CH4 \longrightarrow CH2NH + H} & $3.84 \times 10^{-11} e^{-750/T}$ & 6 \\
R18 & \ce{N(^2D) + CH4 \longrightarrow NH + CH3} & $9.6 \times 10^{-12} e^{-750/T}$ & 6 \\
R19 & \ce{NH + CH3 \longrightarrow CH2NH + H} & $3.12 \times 10^{-16} T^{1.55} e^{-103/T}$ & 7, 18 \\
R20 & \ce{CN + CH2NH \longrightarrow H2CNCN + H} & $2.7 \times 10^{-11}$ & 8, 16 \\
R21 & \ce{CH5+ + CH2NH \longrightarrow CH2NH2+ + CH4} & $3.0 \times 10^{-9}$ & 6 \\
R22 & \ce{H + CH2NH \longrightarrow H2CN + H2} & $4.0 \times 10^{-14}$ & 6 \\
R23 & \ce{NH2 + H2CN \longrightarrow NH3 + HCN} & $5.42 \times 10^{-11} (T/300)^{-1.06} e^{-60.8/T}$ & 6 \\
R24 & \ce{CH3+ + NH3 \longrightarrow CH2NH2+ + H2} & $1.0 \times 10^{-9}$ & 9 \\
R25 & \ce{NH + H2 \longrightarrow NH3} & $1.0 \times 10^{-9}$ & 9 \\
R26 & \ce{N(^2D) + H2 \longrightarrow NH + H} & $4.2 \times 10^{-11} e^{-880/T}$ & 6 \\
R27 & \ce{NH2 + H \longrightarrow NH + H2} & $5.25 \times 10^{-12}$ & 10 \\
R28 & \ce{N2+ + CH4 \longrightarrow N2H+ + CH3} & $3.42 \times 10^{-11}$ & 11 \\
R29 & \ce{N2H+ + C2H2 \longrightarrow N2 + C2H3+} & $1.41 \times 10^{-9}$ & 13 \\
R30 & \ce{CH4 + NH2 \longrightarrow CH3 + NH3} & $3.99 \times 10^{-14}$ & 12 \\
R31 & \ce{C2H2 + NH2 \longrightarrow C2H + NH3} & $1.11 \times 10^{-13}$ & 12 \\
R32 & \ce{C2H4 + NH2 \longrightarrow C2H3 + NH3} & $3.42 \times 10^{-14}$ & 12 \\
R33 & \ce{C2H5 + NH2 \longrightarrow C2H4 + NH3} & $4.15 \times 10^{-11}$ & 12 \\
R34 & \ce{C2H6 + NH2 \longrightarrow C2H5 + NH3} & $6.14 \times 10^{-13}$ & 12 \\
R35 & \ce{C2H6 + N+ \longrightarrow NH + C2H5+} & $1.0 \times 10^{-10}$ & 13 \\
R36 & \ce{C2H6 + N+ \longrightarrow NH2 + C2H4+} & $5.5 \times 10^{-10}$ & 13 \\
R37 & \ce{C2H6 + N+ \longrightarrow NH3 + C2H3+} & $2.5 \times 10^{-10}$ & 13 \\
R38 & \ce{C2H6 + N+ \longrightarrow CH4 + HCNH+} & $1.0 \times 10^{-10}$ & 12 \\
R39 & \ce{NH + C2H2 \longrightarrow HC2N + H2} & $2.01 \times 10^{-9} T^{-1.07}$ & 6, 14 \\
R40 & \ce{NH + C2H4 \longrightarrow CH3CN + H2} & $2.3 \times 10^{-12} (T/300)^{-1.09}$ & 6, 14 \\



R41 & \ce{NH + C2H6 \longrightarrow C2H5N + H2} & $6.8 \times 10^{-12}$ & 6, 14 \\
R42 & \ce{NH + C4H2 \longrightarrow C4HN + H2} & $8.24 \times 10^{-9} T^{-1.23}$ & 6, 14 \\
R43 & \ce{CH2NH2+ + e \longrightarrow CH2NH + H} & $1.0 \times 10^{-6} (300/T_e)^{0.7}$ & 6 \\
R44 & \ce{CH2NH2+ + e \longrightarrow CH2 + NH2} & $1.0 \times 10^{-6} (300/T_e)^{0.7}$ & 6 \\
R45 & \ce{N2H+ + e \longrightarrow NH + N} & $5.67 \times 10^{-8} (300/T_e)^{0.51}$ & 9, 15, 16 \\
\caption{List of 45 new reactions added to the new version of the CO-PRISM scheme along with their associated reactions rates \textit{k} (in $cm^3\:s^{-1}$) and their references. $^1$\citet{Tian1998CrossC2H2}; $^2$\citet{DeBleecker2006DetailedDischarges}; $^3$\citet{Angelova2004}; $^4$\citet{Benedikt2010Plasma-chemicalPlasmas}; $^5$\citet{Janalizadeh2003}; $^6$\citet{Yelle2010}; $^7$\citet{Rosi2018}; $^8$\citet{Barone2022}; $^9$\citet{Carrasco2012}; $^{10}$\citet{Harada2010}; $^{11}$\citet{Plessis2010}; $^{12}$\citet{Hebrard2009}; $^{13}$\citet{Anicich2003}; $^{14}$\citet{Mullen2005}; $^{15}$\citet{Geppert2004ApJ}; $^{16}$\citet{Vazart2015}; $^{17}$\citet{MaoJPhys2008};$^{18}$\citet{Bourgalais2019}}\\

\label{Table2}
\end{longtable}

\section{Results and Discussion}

\subsection{Comparison of mass spectra}\ \label{sec:3.1}

To first assess the impact of the changes applied to the reaction scheme on the calculated synthetic mass spectra, we ran the model using the same gas mixture used with the original version of the CO-PRISM model \citep{Raymond2018}, \textit{i.e.}, \ce{N2-CH4} (90:10), \ce{N2-CH4-C2H2} (85:10:5), and \ce{N2-CH4-C2H4} (85:10:5) gas mixtures. We considered a gas flow of 2000 sccm, a gas temperature of 150 K, and a high voltage of -800 V in the model, to match experimental parameters. A seed electron density of $n_e =$ 10$^{7}$ cm$^{-3}$ was considered. Positive ion synthetic mass spectra were computed from the output molar fraction of all the species calculated near the cathode, \textit{i.e.}, close to their collection point as probed by the mass spectrometry in the experiments \citep{Sciamma-OBrien2014}. 

\begin{figure}
\centering
\begin{subfigure}[b]{0.35\textwidth}
   \includegraphics[scale=0.4]{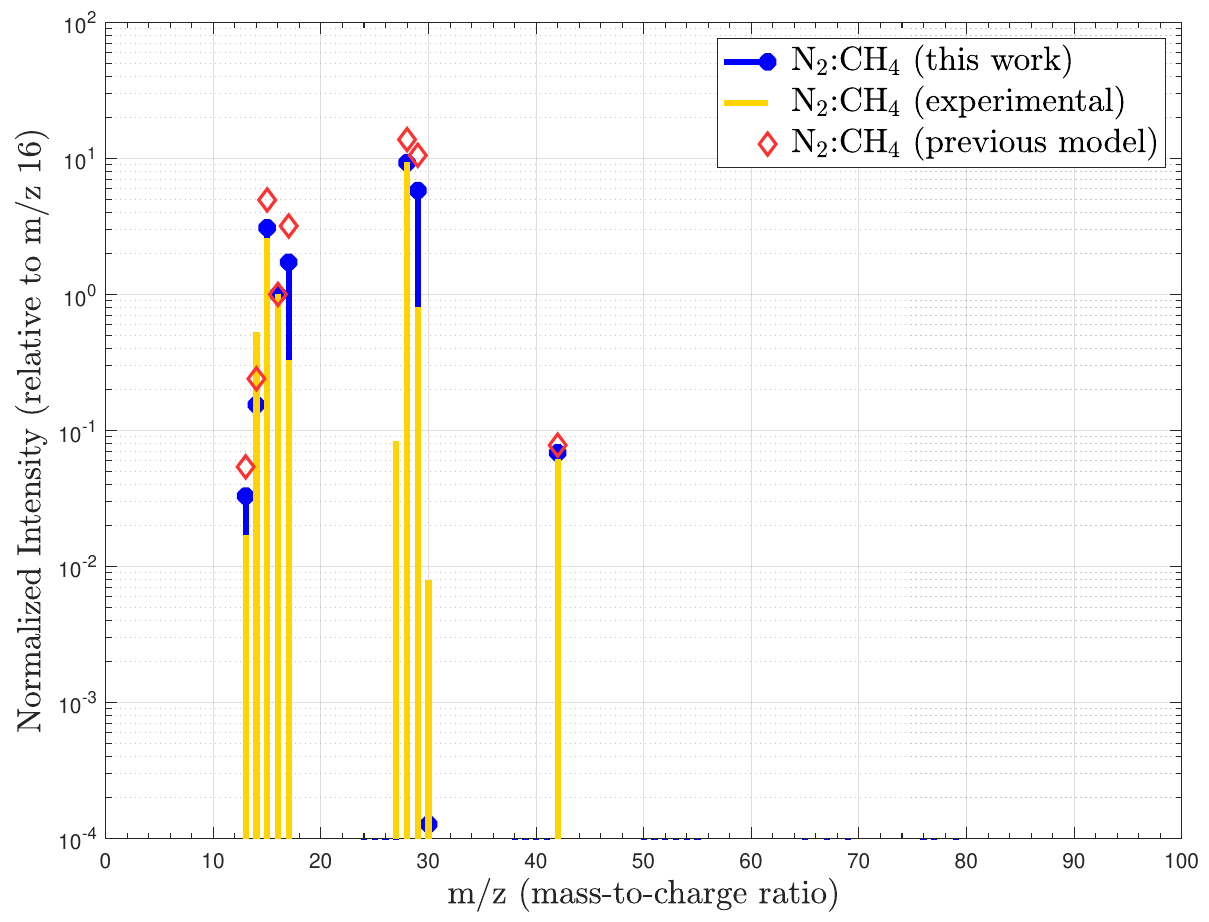}
   \label{fig:Ng1} 
\end{subfigure}

\begin{subfigure}[b]{0.35\textwidth}
   \includegraphics[scale=0.4]{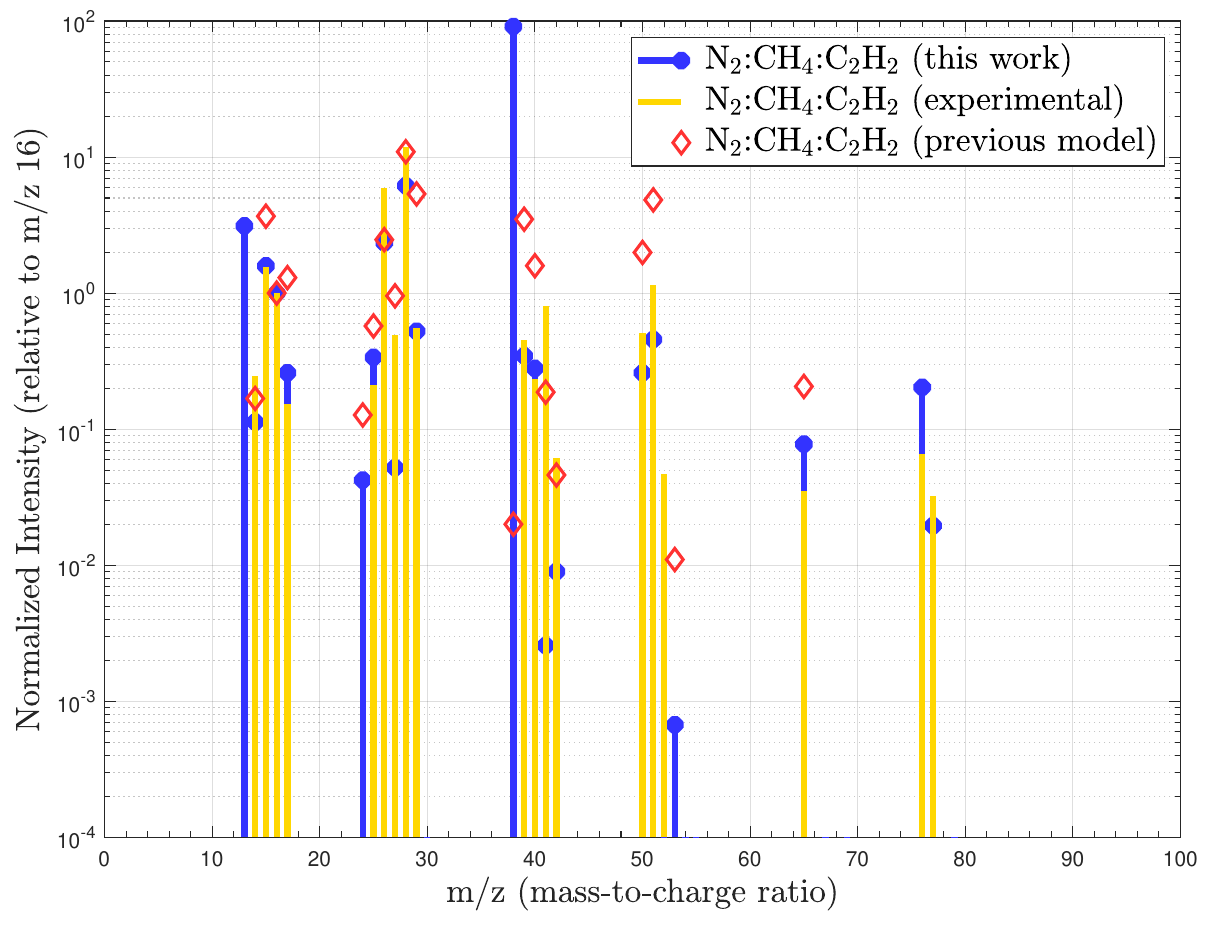}
   \label{fig:Ng2}
\end{subfigure}

\begin{subfigure}[b]{0.35\textwidth}
   \includegraphics[scale=0.4]{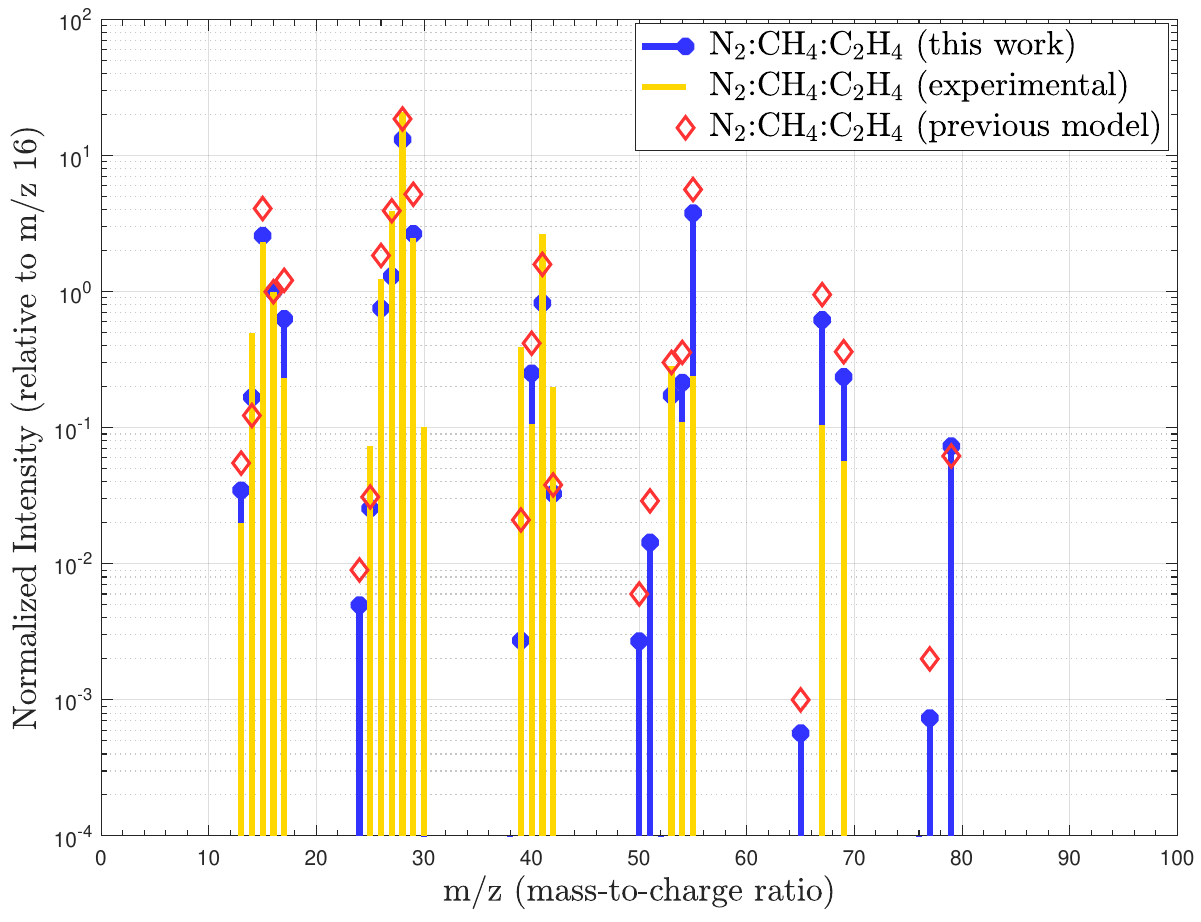}
   \label{fig:Ng2}
\end{subfigure}
\caption{Comparison of the original calculated and experimental mass spectra (in red diamonds from \citet{Raymond2018} and yellow bars from \citet{Sciamma-OBrien2014}, respectively) with newly calculated mass spectra (in blue circles, this study) in \ce{N2-CH4} (90-10\%), \ce{N2-CH4-C2H2} (85-10-5\%), and \ce{N2-CH4-C2H4} (85-10-5\%) gas mixtures. Intensities are normalized over the intensity at \textit{m/z} 16, and all measurements and calculations were performed with a -800 V cathode potential. \label{figure6MS_calc_expt_comparisons}}
\end{figure}

Figure \ref{figure6MS_calc_expt_comparisons} shows the comparison of the model spectra obtained with the original version of CO-PRISM (red diamonds) and experimental mass spectra (yellow bars), with newly calculated mass spectra (blue circles, this study) showing species ranging from \textit{m/z} 13 up to \textit{m/z} 79. All spectra hereafter are normalized to the intensity found at \textit{m/z} 16 to stay consistent and comparable with the normalizations in \cite{Sciamma-OBrien2014} and \cite{Raymond2018}.

The updated reaction rates in the model affect the molecular abundance of the product species, and result in the production of new species compared to the previous version of the model. Almost all major ions observed experimentally are reproduced by the model. The new reaction network generally exhibits lower ion abundance compared to the previous model (red diamonds to blue circles), bring the intensities generally closer to the experimental value (\textit{e.g.}, \textit{m/z} 15, 28, C5, and C6 species). Although there are still some discrepancies, previous intensities at these masses were higher than the experimental ones.

In the simple \ce{N2-CH4} (90-10\%) mixing ratio, no hydrocarbons are formed above \textit{m/z} 30; a single ion at \textit{m/z} 42 is observed (corresponding to \ce{N3+}). With the new version of CO-PRISM, in the \ce{N2-CH4} (90-10\%) gas mixture, we observe a small contribution of \textit{m/z} 30 which we attribute to protonated methanimine (\ce{H2C=NH2+}) produced by the newly added reactions R21 and R24 (see Table \ref{Table2}). Protonated methanimine (and its parent molecule methanimine) production relies on the presence of \ce{CH3+}, \ce{CH5+}, and neutral methanimine itself. Although the intensity is low, this represents the first measurable amount of protonated methanimine calculated in the COSmIC plasma. It was not produced in the previous version of the model. Including reaction pathways for the formation of methanimine and protonated methanimine allows for the modeling of a small contribution from these molecules to the peaks observed experimentally at \textit{m/z} 29 and \textit{m/z} 30, respectively. The main contributor to the 29 m/z peak remains \ce{C2H5+}, with an abundance of $\sim10^{-7}$ in the model (Figure 10). We interpret the discrepancy observed in the \textit{m/z} 30 peak as missing chemical pathways in the CO-PRISM chemical network that would lead to the formation of protonated methanimine and ionized ethane (\ce{C2H6+}). 
Methanimine being a transient species, it is also possible that it is rapidly lost through barrierless reactions with other ions through unknown pathways not incorporated into our reaction scheme. 
At present, only three loss mechanisms are included in CO-PRISM for the loss of methanimine (Table \ref{Table2}), R20, R21, and R22. Other destruction pathways contributing to the loss of \ce{CH2NH} and the production of \ce{CH2NH2+} through bimolecular reactions with \ce{C2H5+}, \ce{C3H5+}, and \ce{C4H3+} also exist but are not included in our scheme \cite{Plessis2010}.

To improve agreement between modeled spectra and experiments involving \ce{C2H6} and \ce{N2H+} ion-neutral reactions, it may be useful to incorporate the reactions mechanisms obtained from \ce{H2}-based cold plasma discharges in \cite{Tanarro2007IonN2} in future work. It is also possible that methanimine is more favorably produced under direct UV/EUV irradiation conditions \citep{Bourgalais2019}, stemming from a favorable photolysis of \ce{N2} by UV photons \cite{Yelle2010}.
Future measurements out of the scope of this study utilizing isotopically labeled \ce{CH4} would confirm whether methanimine is efficiently produced in the COSmIC plasma. Such measurements would also guide future modeling analyses including additional reaction pathways leading to methanimine production.
On the other hand, \textit{m/z} 27 is observed in \ce{N2-CH4} experiment but is not predicted by the model. In the more complicated mixtures containing \ce{C2H2} and \ce{C2H4} however, products are observed at \textit{m/z} 27 both experimentally and via modeling and predicted to be \ce{C2H3+} and \ce{HCN+}. Their production is facilitated by reactions R29, and R32 in Table \ref{Table2}. The species \ce{NH2} and \ce{N2H+} appear to be a missing link in the production of \ce{HCN+} and \ce{C2H3+} in the \ce{N2-CH4} modeling. Two main pathways currently include the production of \ce{C2H3+} (see Reactions 21 and 137 in the Appendix), but future work should also incorporate other bimolecular reactions such as \ce{CH+ + CH4 -> C2H3+ + H2} \citep{McEwan2007}.

Adding \ce{C2H2} to the precursor mixtures allows the chemistry in the plasma to further evolve and results in the production of more complicated C4, C5, and C6 hydrocarbons. The revised reaction network leads to improved agreement with the experimental spectrum for the peaks at \textit{m/z} 17, 25, 29, 39, 40, and 65. These peaks correspond mainly to \ce{CH5+}, \ce{C2H+}, \ce{C2H5+}, \ce{C3H3+}, \ce{C3H4+}, and \ce{C5H5+}, respectively. Reactions R1, R2, R3, and R35 all facilitate the production of these species, and ions resulting from the collisional dissociation of \ce{C2H2} and the addition of the new collisional cross sections. We note however that the absence of calculated products at \textit{m/z} 52 (contributed by \ce{C4H4+} and \ce{HC3NH+}) is due to limited knowledge of \ce{C4H4+} reaction rates, and minimal production of \ce{HC3NH+}. Future inclusion of \ce{C4H4+} production via the bimolecular reaction between \ce{N2+} and \ce{C6H6} may explain the lack of peak at \textit{m/z} 52 \citep{Arnold1999FlowK}, although formation of \ce{C4H4+} under low-pressure plasma conditions is expected to be marginal \citep{Bera2015HydrocarbonFragments}. Products at \textit{m/z} 38 can either correspond to \ce{CNC+} or \ce{C3H2+}, the latter being more abundant. The inclusion of new \ce{C2H2} pathways to the updated reaction scheme results in substantial formation of \ce{C3H2+} at \textit{m/z} 38, through \ce{C2H2 + CH+ \longrightarrow H + C3H2+}. The discrepancy observed for this peak between experimental and predicted peak is likely due to missing loss channels or inaccurate rates. The \ce{C3H2+ + CH4 -> CH3 + C3H3+} channel may in the future fill this gap, leading to a loss of \ce{C3H2+} (\textit{m/z} 38) and the formation of \ce{C3H3+} (\textit{m/z} 39, \cite{McEwan2007}). An improvement now observed in the model outputs, compared to the original model, is the predicted production of molecules at masses \textit{m/z} 76 and 77 in the \ce{N2-CH4-C2H2} plasma, as observed experimentally. These peaks correspond to the benzene fragments \ce{C6H4+} and \ce{C6H5+}, both observed in \citet{Sciamma-OBrien2014} (and also found to be the two most important C6 ions in a different experimental setup \citep{Dubois2020}). It is likely that these products could then be the starting points for larger products to be formed \citep{Sciamma-OBrien2014}. Future work might incorporate pathways including even larger products and thus examine \ce{C2H2} contribution further.

The mixture involving \ce{C2H4} is the most complex and begins to show the limitations of the modeling to describe the experiment.  Overall, C1 through C3 peaks are still in good agreement; however, in blocks C4 through C5, there are bins for which the model predicts the production of species that are not observed in the experiment (\textit{m/z} 50, 51, 65, 77, 79).  \textit{m/z} 50 corresponds to \ce{C4H2+}, \textit{m/z} 51 corresponds to \ce{C4H3+}, \textit{m/z} 65 corresponds to \ce{C5H5+}, \textit{m/z} 77 corresponds to \ce{C6H5+}, and \textit{m/z} 79 corresponds to \ce{C6H7+}. Many of those predicted abundances are at or below the limit of lowest-intensity experimental detections observed across any of the mixtures ($\sim10^{-2}$). Those species may be present in the experiment but in too low a quantity to be detected. It is also possible that some reactions might be missing from our reaction scheme that would account for the loss of these species. Another possibility is that at least some of those species may contribute to the formation of solids (\textit{tholin}). The loss of gas product due to solid formation is not accounted for in the model. 

Adding \ce{C2H4} increases the population of the C5 block, dominated by \ce{C5H7+} and \ce{C5H9+} due to the direct loss of neutral \ce{C2H4} and ion \ce{C2H4+} \citep{Raymond2018}. From the updated reaction scheme, it appears that both \ce{C2H4+} and \ce{HCNH+} should be highly sensitive to the \ce{N2}/\ce{CH4} ratio (R1, R38, and \ce{C2H4 + e \longrightarrow C2H4+ + 2e}), while neutral \ce{C2H4} loss can readily occur through the following reaction \ce{N + C2H4 \longrightarrow C2H3N + H}. In addition, a shift in the C6 block is observed when adding \ce{C2H4} to the \ce{N2-CH4} plasma, with in particular a large increase in intensity of the \textit{m/z} 79 peak, resulting from the protonation of benzene due to proton transfer between benzene and \ce{C2H5+}/\ce{HCNH+} \citep{Vuitton2019}. The lack of any experimental detection in that region in an \ce{N2-CH4-C2H4} mixture could be due to instrument limitation, \ce{C6H6} limitation, loss of gas molecular products due to \textit{tholin} formation, or a combination of these three possible factors. In any case, these calculations demonstrate that the presence of \ce{C2H2} largely influence the production of \ce{C6H6} and its (de)protonated derivatives.

As seen in Figure \ref{figure6MS_calc_expt_comparisons}, for all three gas mixtures, the largest product species observed experimentally in the plasma have \textit{m/z} of 77 and 79. The largest calculated species participating in the reaction network, however, is \ce{C12H6+} with \textit{m/z} of 150. The non-detection in the experimental spectra of any species larger than \textit{m/z} 79 when adding \ce{C2H4} might mean that (1) we are at the detection limit of the instrument with insufficient quantities being produced to be detected with our instrument and/or (2) these species are solid phase precursors and are therefore lost to the solid phase, with the remaining gas phase abundance too low for detection. Also, production of \ce{C12H6+} (thanks to R8) is limited by the availability of \ce{C8H6+} and \ce{C10H6+} (R6 \& R7). \citet{Sciamma-OBrien2014} demonstrated larger compounds could be detected experimentally when benzene was present in the initial gas mixture, which would be interesting to investigate in future modeling studies.

\subsection{Calculated mass spectra in other nitrogen- and hydrocarbon-based conditions}

In addition to conducting the comparative study with CO-PRISM models run with plasma parameters (voltage, gas mixtures) similar to those used in the \citet{Raymond2018} study, we also modeled plasmas operated with an applied voltage of -1000V for various gas mixtures. \ce{N2-CH4} (90-10) and \ce{N2-CH4-C2H2} (85-10-5) gas mixtures were first chosen to be consistent with two recent solid phase studies conducted with COSmIC \citep{Nuevo2022,Sciamma-OBrien2023FirstAerosols} where the elemental composition and optical properties of the solids and the impact of the plasma parameters were investigated. We then expanded the range of plasma parameters by modeling additional gas mixtures in a -1000V plasma such as: \ce{N2-CH4} (99-1), \ce{N2-CH4-C2H4} (85-10-5), \ce{N2-CH4-C2H6} (85-10-5), \ce{N2-CH4-NH3} (85-10-5), \ce{N2-CH4-HCN} (85-10-5) and \ce{N2-CH4-CH2NH} (85-10-5), to see the effect of the molecular precursors on the chemistry.

\begin{figure}
\centering
\includegraphics[scale=0.9]{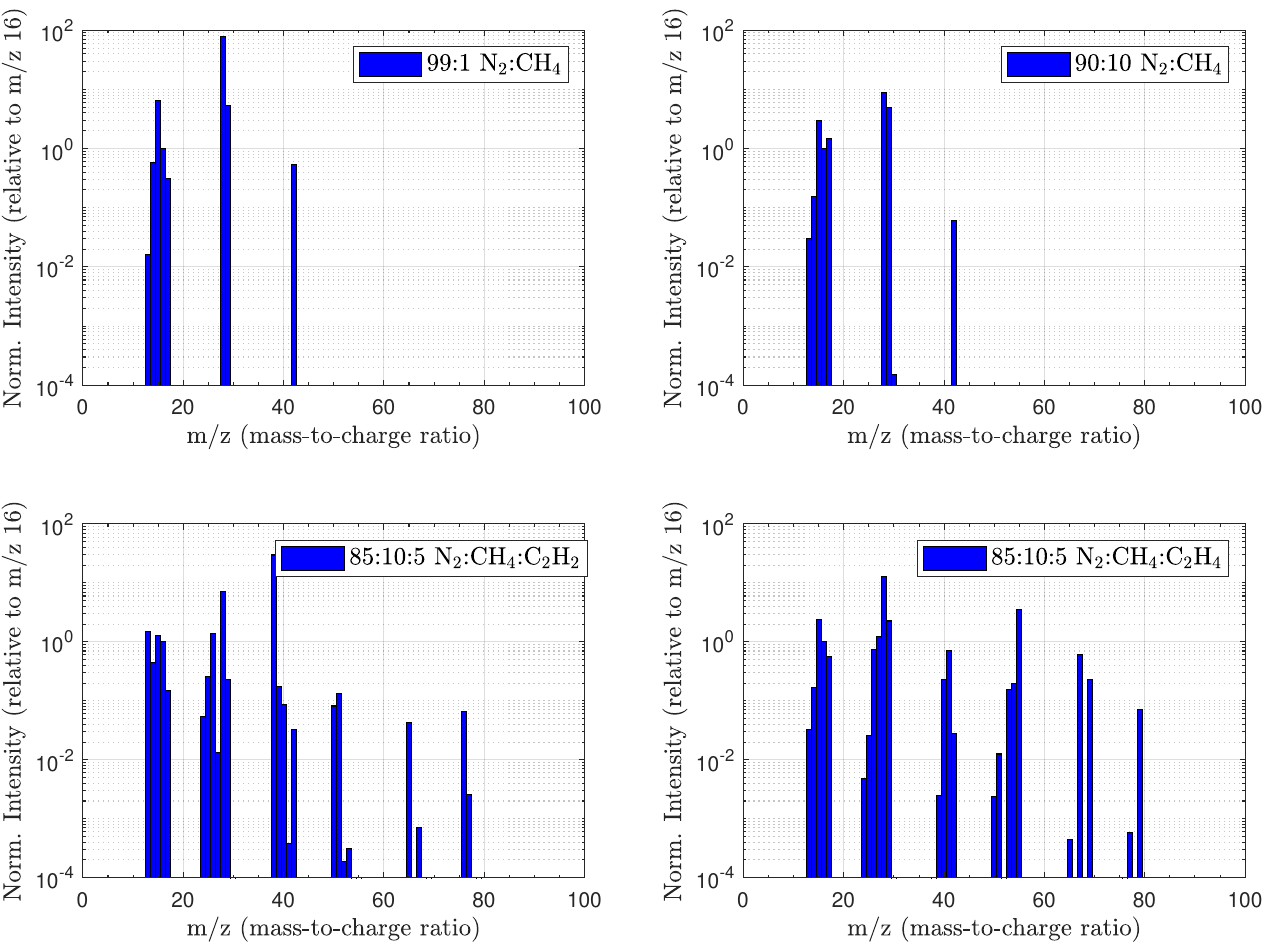}
\caption{Mass spectra of various \ce{N2}:\ce{CH4}-based plasmas computed with 1\% and 10\% \ce{CH4}, and 5\% of \ce{C2H2} and \ce{C2H4}. Intensities are normalized to the peak at m/z = 16. 
\label{fig7a}}
\end{figure}

Figure \ref{fig7a} shows the calculated mass spectra obtained with the model for the two \ce{N2-CH4} mixtures, as well as the \ce{N2-CH4-C2H2} and \ce{N2-CH4-C2H4} mixtures. The calculated mass spectra for the other mixtures are provided in the Appendix, in Figure A\ref{fig7b}). When comparing the \ce{N2-CH4} (99-1\%) plasma model to the \ce{N2-CH4} (90-10\%) plasma model , we note an increase in \ce{CH3+} at \textit{m/z} 15 and \ce{CH5+} at \textit{m/z} 17 in \ce{N2-CH4} (99-1\%), as well as an absence of any protonated methanimine \ce{CH3N} at \textit{m/z} 30. In spite of a lower amount of \ce{CH4} available in the 99-1\% mixture, \ce{CH3+} production remains important through Reaction \ref{CH4 + N2(X)} \citep{Pintassilgo1999MethanePressures}.

\begin{equation}
\label{CH4 + N2(X)}
\ce{CH4 + N2(X)+ \longrightarrow N2(X) + H + CH3+}
\end{equation}

Other peaks show very small variations ($< 5\%$) between the two mixtures. The overall aspects of the calculated mass spectra for the \ce{N2-CH4-C2H2} and \ce{N2-CH4-C2H4} modeled plasma at -1000 V are similar to those obtained at -800 V with no significant changes observed by increasing the voltage. For the other four \ce{N2-CH4}-based mixtures containing \ce{C2H6}, \ce{NH3}, \ce{HCN}, and \ce{CH2NH}, the calculated mass spectra appear identical to the \ce{N2-CH4} (90-10\%) (Figure A\ref{fig7b} in the Appendix).

\subsection{Sensitivity analysis along the plasma cavity} \label{sec:3.2}\

In addition to conducting a comparative analysis between the original and updated version of CO-PRISM for \ce{N2-CH4}-based plasmas produced in COSmIC, we also conducted a parameter space study to expand on our previous work \citep{Raymond2018} and assess the sensitivity of the model on the resulting plasma chemistry when looking at different locations in the plasma channel and when changing the plasma conditions (\textit{e.g.}, different potentials) and the precursor gas composition (\textit{e.g.}, different concentrations of \ce{CH4} in \ce{N2} and addition of other minor species). Understanding the impact of plasma parameters allows us to optimize the experiment for Titan simulations.

\subsubsection{Plasma chemistry along the plasma channel}

\begin{figure}
\centering
\includegraphics[scale=0.5]{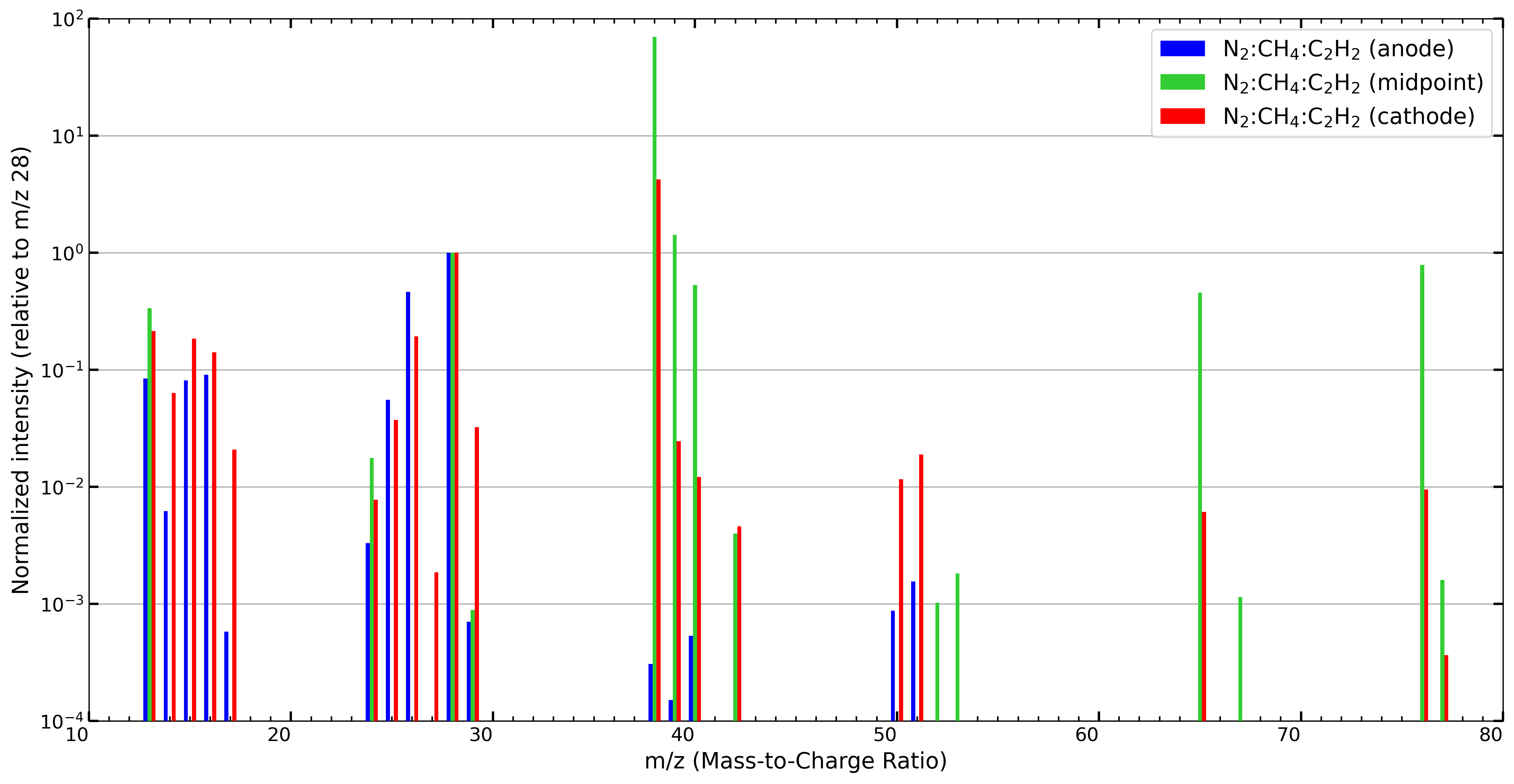}
\caption{Mass spectra of a modeled plasma generated in a \ce{N2-CH4-C2H2} (85-10-5\%) gas mixture with the updated version of CO-PRISM at three different locations in the plasma channel: x = 0.004 mm (blue, anode), x = 0.075 mm (green, midpoint), and x = 1.496 mm (red, cathode). Peaks are normalized by the intensity at \textit{m/z} 28. \label{fig9}}
\end{figure}

We recalculated the mass spectra of a simulated \ce{N2}:\ce{CH4} (90:10) plasma discharge in COSmIC at the plasma channel edge (\textit{x} = 1.496 mm) and also at upstream locations: \textit{x} = 0.750 mm (midpoint) and \textit{x} = 0.004 mm (near anode). The spectra for an \ce{N2-CH4-C2H2} (85-10-5) plasma at -1000 V (Figure \ref{fig9}) show primary methane ions near the anode, with \ce{C2+}, \ce{C2H+}, and \ce{C2H2+}. Small amounts ($< 10^{-3}$) of \ce{C2H5+/N2H+/N3+} appear. Midpoint products like \ce{C3H2+} at \textit{m/z} 38 reach high intensities, indicating significant methane consumption essential for chemical growth. Benzene products at \textit{m/z} 76 and 77 emerge as \ce{C6H4+} becomes abundant. Near the cathode, C1 and C2 species rebound, while C3 decreases notably. C4 species appear, and C5 and C6 species drop significantly, highlighting their role as intermediates in forming larger organics and aerosols, evidenced by changes observed between midpoint and cathode regions.

\subsubsection{Electron density as a function of voltage, time, and position in the plasma channel}

\begin{figure}
\centering
\begin{subfigure}[b]{0.7\textwidth}
   \includegraphics[scale=0.5]{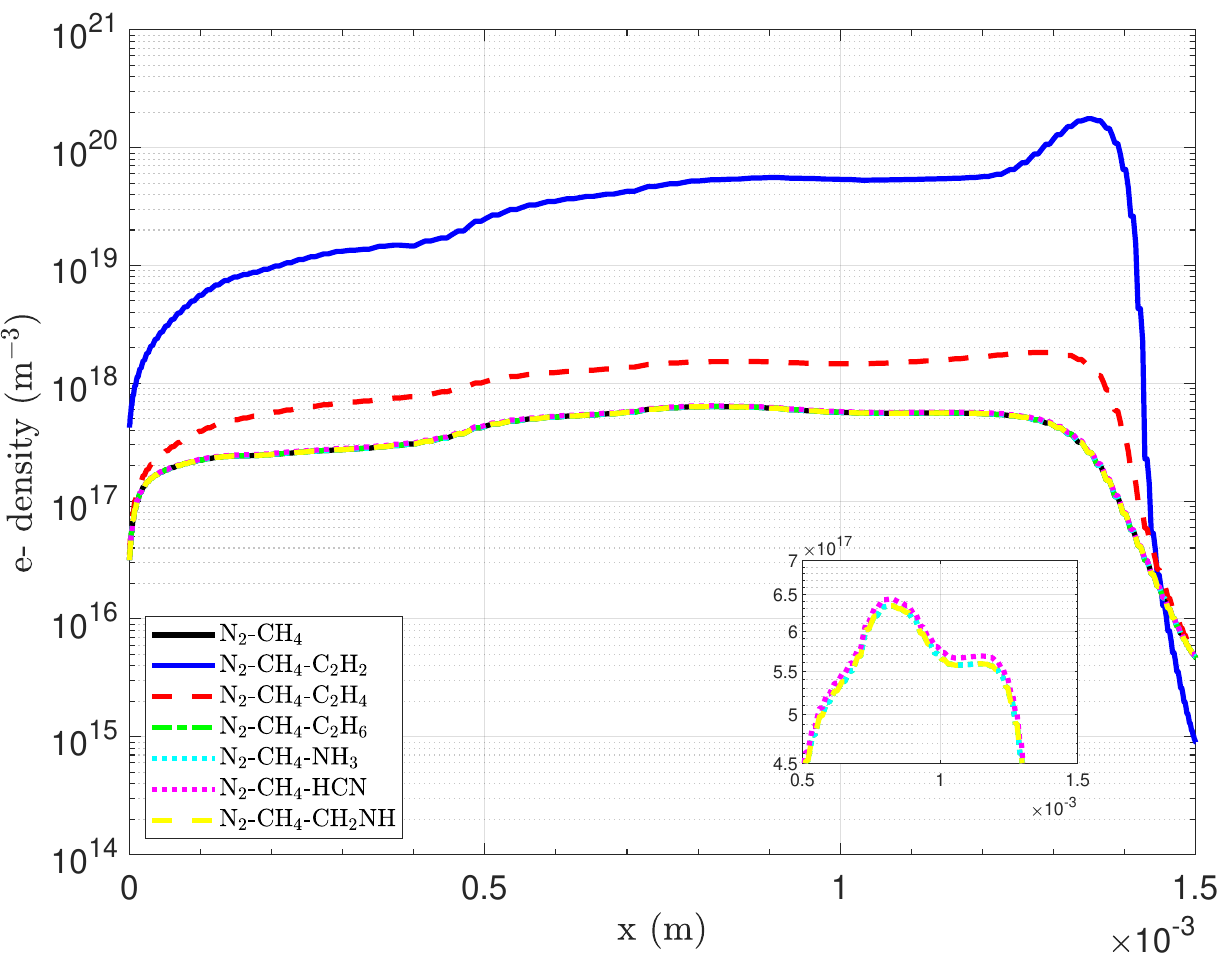}
   \caption{}
   \label{fig:10a} 
\end{subfigure}

\begin{subfigure}[b]{0.7\textwidth}
   \includegraphics[scale=0.5]{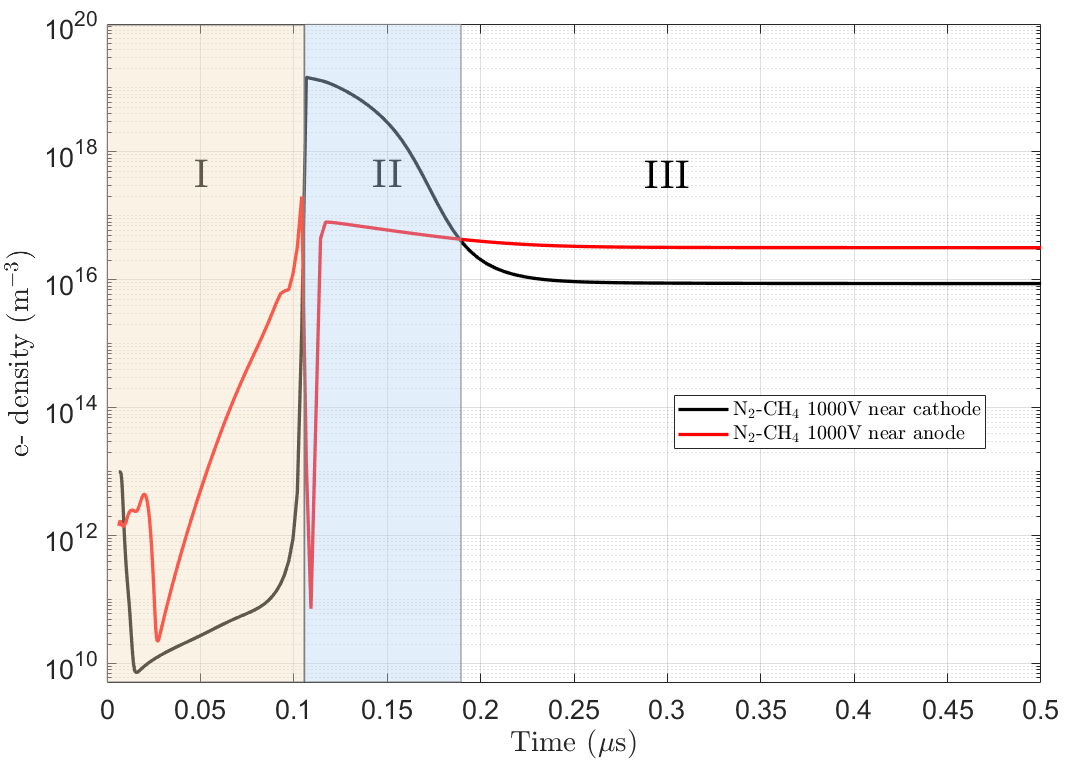}
   \caption{}
   \label{fig:10b}

\end{subfigure}

\caption[Two numerical solutions]{(a) Electron density profiles calculated along the plasma channel in 7 different mixing ratio conditions. Origin coordinate, $x=0$ mm, starts at the anode (see Figure \ref{fig1}). Inset shows a zoom on the lower 5 similar curves. (b) Electron density profiles near the anode and near the cathode calculated as a function of time during one plasma pulse, showing the electron density steady-state conditions are reached around 0.25 $\mu$s. Phase I near breakdown shows the migration of electrons towards the anode in the first 0.1 $\mu$s. Phase II starts once a maximum density of $\sim$ $1.5 \times 10^{19}$ is reached with a migration of electrons to the cathode, until steady-state conditions are reached around 0.25 $\mu$s (Phase III).
\label{figure6combined}}
\end{figure}

Electron density is a crucial plasma parameter, influencing reaction schemes, electron energy distribution, collision frequency, and degree of ionization \citep{Stoykov2001,Bleecker2006b,DeBleecker2006DetailedDischarges,MaoJPhys2008,Raymond2018}. We calculated electron density profiles along the plasma channel from anode to cathode for seven gas mixtures (Table \ref{Table1}) at a cathode potential of -1000 V. The profiles, shown in Figure \ref{fig:10a}, are categorized into three classes: low density (peak \(< 10^{18} \, \text{m}^{-3}\)), intermediate density (peak \(\sim 2 \times 10^{18} \, \text{m}^{-3}\)), and high density (peak \(> 10^{20} \, \text{m}^{-3}\)). Low electron density distributions are observed in \(\text{N}_2\text{-CH}_4\), \(\text{N}_2\text{-CH}_4\text{-C}_2\text{H}_6\), \(\text{N}_2\text{-CH}_4\text{-HCN}\), and \(\text{N}_2\text{-CH}_4\text{-CH}_2\text{N}\) plasmas, with similar overall profiles. Intermediate electron density is seen in \(\text{N}_2\text{-CH}_4\text{-C}_2\text{H}_4\) plasmas, while the highest density is found in \(\text{N}_2\text{-CH}_4\text{-C}_2\text{H}_2\) plasmas, attributed to the release of twice as many electrons upon \(\text{C}_2\text{H}_2\) dissociation (R1 and Appendix Table). Near the cathode, a drop in electron density indicates a larger Debye length, consistent with reduced shielding \citep{Remy2003,Broks2005,Broks2005a}. Drops in electron density near the anode and cathode highlight sheath regions where electrons are accelerated, with the anode's dark space being thinner due to pressure differences \citep{Remy2003}. Near the anode, the micrometer-range mean free path results in a nearly collisionless sheath. Within the cathode's Crookes dark space, electrons are repelled by the negative potential, causing a significant drop in electric potential and electron density \citep{Broks2005}. 

We also calculated the electron density profiles at two given positions (near the anode and near the cathode) as a function of time to observe the different phases leading to the establishment of the steady state. Figure \ref{fig:10b} shows the different phases of the electron density profile from the beginning of the plasma pulse to steady state conditions for an \ce{N2}-\ce{CH4} (90-10) plasma discharge at -1000V. As described in \citet{Raymond2018}, at the start of the discharge (phase I, highlighted in yellow), the negative cathode voltage (black curve) increases, which results in an electron migration towards the anode (red curve) near breakdown, up to $\sim$ 0.1 $\mu$s after the start of the pulse. During this period, the electron density increases by more than six orders of magnitude near the anode and two orders of magnitude near the cathode. 0.1 $\mu$s after the start of the pulse, the electron density near the cathode reaches a peak value of $1.5 \times 10^{19}$ m$^{-3}$. An electron density inversion and the second electron migration phase begin near 0.1 $\mu$s, this time toward the cathode (Phase II, highlighted in blue), before reaching equilibrium (steady-state conditions) by 0.25 $\mu$s where the near-anode electron density dominates again (Phase III) during steady-state.

The dynamics and inversions observed here are linked to electron mobility in the plasma region and the relationship between the glow region and sheaths \citep{Remy2003}.
The mean free path of neutrals and electrons is much larger near the cathode (mm) than near the anode ($\mu$m). This implies that electrons can enter the plasma region with high relative velocities. In the plasma they lose their kinetic energy through collision, forming 
the first positive ions in the glow region. 
Note that given the low degree of ionization and electron density in the COSmIC plasma discharge, due to its abnormal glow nature \citep[see][]{Remy2003}, electron-electron collisions are outnumbered by inelastic collisions between the carried neutral gas and electrons \citep{Broks2005}.\\

\subsubsection{Electron energy and inelastic collision frequency}

\begin{figure}
\centering
\includegraphics[scale=0.5]{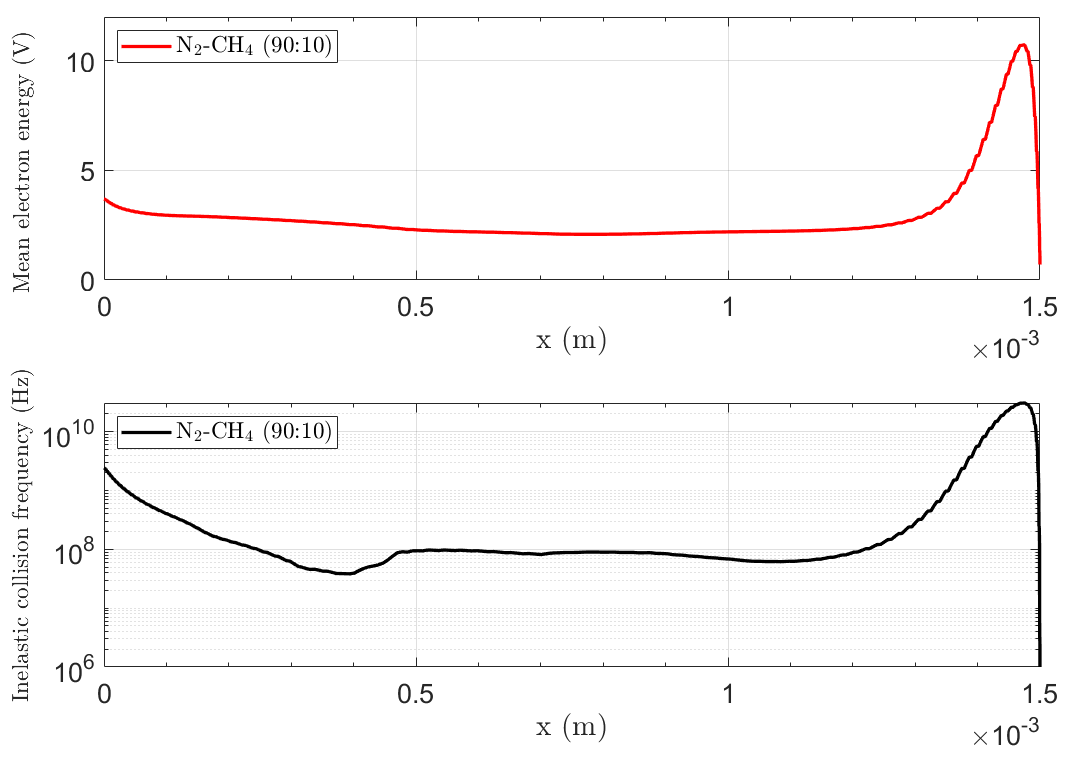}
\caption{Top: Mean electron energies (or temperature) in eV calculated across the plasma channel for a \ce{N2}:\ce{CH4} (90:10) plasma discharge (red). Origin coordinate starts at $x=0$ mm (see Figure \ref{fig1}). Bottom: Inelastic electron collision frequency across the plasma channel calculated in steady-state conditions with a source voltage of -1000 V (black). Inelastic collisions (source term $S_e$ see Equation \ref{equ2}) lead to direct ionization and excitation of neutrals.\label{fig8sub}}
\end{figure}

As described in Equation \ref{equ2}, \( S_e \) is the energy source term for inelastic collisions, accounting for ionization and excitation between the electrodes. In the plasma region between the anode and cathode, trapped, runaway, and secondary electrons ionize \(\ce{N2}\) and \(\ce{CH4}\) neutrals. Secondary electrons are produced at a rate characterized by the secondary emission coefficient (0.025), and electron-electron collisions are not considered. Inelastic collisions, which depend on gas composition and pressure, are responsible for direct ionization and excitation of neutrals. The bottom panel of Figure \ref{fig8sub} shows the steady-state inelastic collision frequency in an \(\ce{N2}-\ce{CH4}\) (90-10) gas mixture with a source voltage of -1000 V. The inelastic collision frequency is relatively steady from the anode to the plasma center (\(10^8-10^9\) Hz) but increases sharply near the cathode by over two orders of magnitude to \(10^{10}\) Hz due to high-velocity electrons causing more collisions, ions, and secondary electron generation. The top panel of Figure \ref{fig8sub} shows the mean electron energy, which peaks at 10-11 eV near the cathode, consistent with previous Ar plasma modeling. This energy is sufficient for \(\ce{N2}\) direct dissociation (requiring $>$9.76 eV) and electron-impact ionization. The ionization fraction in a \(\ce{N2}-\ce{CH4}\) plasma discharge on COSmIC (\(10^{-8}\) to \(10^{-5}\)) is comparable to Titan's upper atmosphere (\(10^{-7}\) to \(10^{-5}\)) \citep{Remy2003, Broks2005, Dutuit2013, Raymond2018}.

\newpage

\subsubsection{Positive ion density}

\begin{figure}
\centering
\includegraphics[scale=0.5]{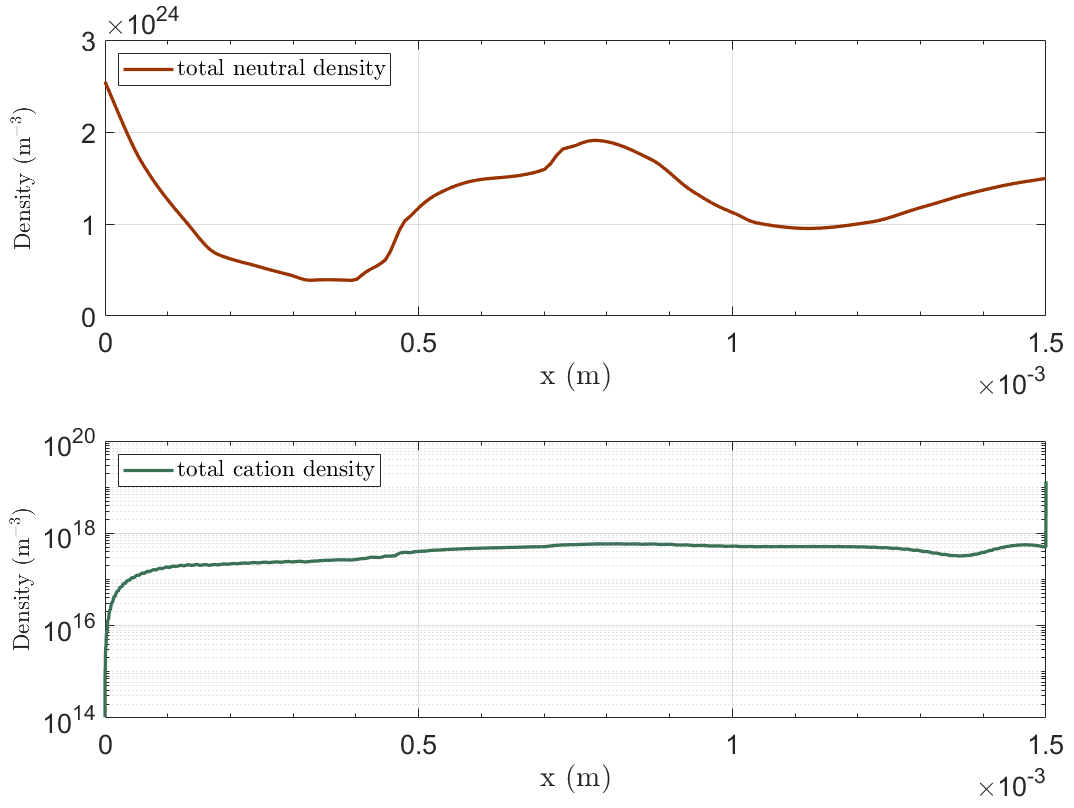}
\caption{Total neutral and positive ion density profiles calculated in an \ce{N2-CH4} (90-10) gas mixture with a -1000 V source voltage. \label{fig12}}
\end{figure}

Figure \ref{fig12} shows the calculated total neutral and positive ion density profiles calculated at steady state in an \ce{N2-CH4} (90-10) gas mixture with a source voltage of -1000 V. The total neutral density is relatively constant on the order of $10^{24}$ m$^{-3}$ across the plasma channel. The total positive ion density, about six orders of magnitude lower, reaches $\sim 6 \times 10^{17}$ $m^{-3}$ in the plasma cavity near the cathode (x = 1.496), which is at least one order of magnitude higher than the electron density at that location, and on the same order of magnitude as the electron density calculated at midpoint in the plasma cavity. Given the electron and positive ion densities near the cathode, it is expected that electron impact ionization rates are on the same order of magnitude as ion neutralization in this region \citep{Raymond2018}. Closer to the anode, however, the positive ion density is calculated to be several orders of magnitude lower comparatively, as the positive ions do not migrate as much upstream towards the anode. Quasi-neutrality is relatively constant throughout the plasma channel. As positive ions are formed in the plasma region, the products are then carried into the ion-neutral network, where they will be incorporated with breakdown-induced ions and neutrals. It is thus important to characterize the molecular species and plasma parameters since these will be dependent upon the initial Titan-relevant gas mixture to be studied. \\

\newpage

\subsection{Impact of initial gas mixture composition on the gas phase products} \label{sec:3.4}

\subsubsection{Light-mass hydrocarbon molar fractions}

\begin{figure}[h]
    \centering
    \begin{subfigure}{0.48\textwidth}
        \centering
        \includegraphics[scale=0.3]{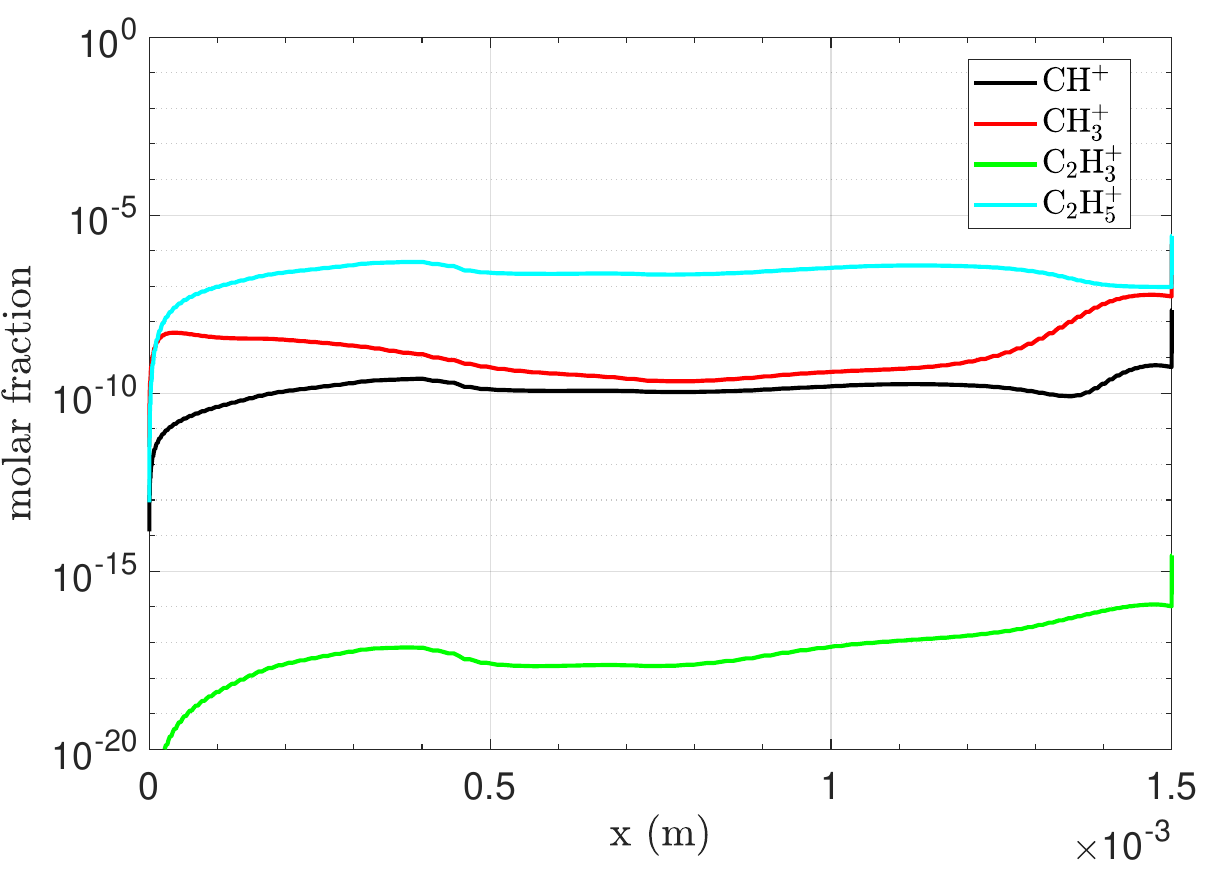}
        \caption{}
        \label{figure13a}
    \end{subfigure}
    \begin{subfigure}{0.48\textwidth}
        \centering
        \includegraphics[scale=0.2]{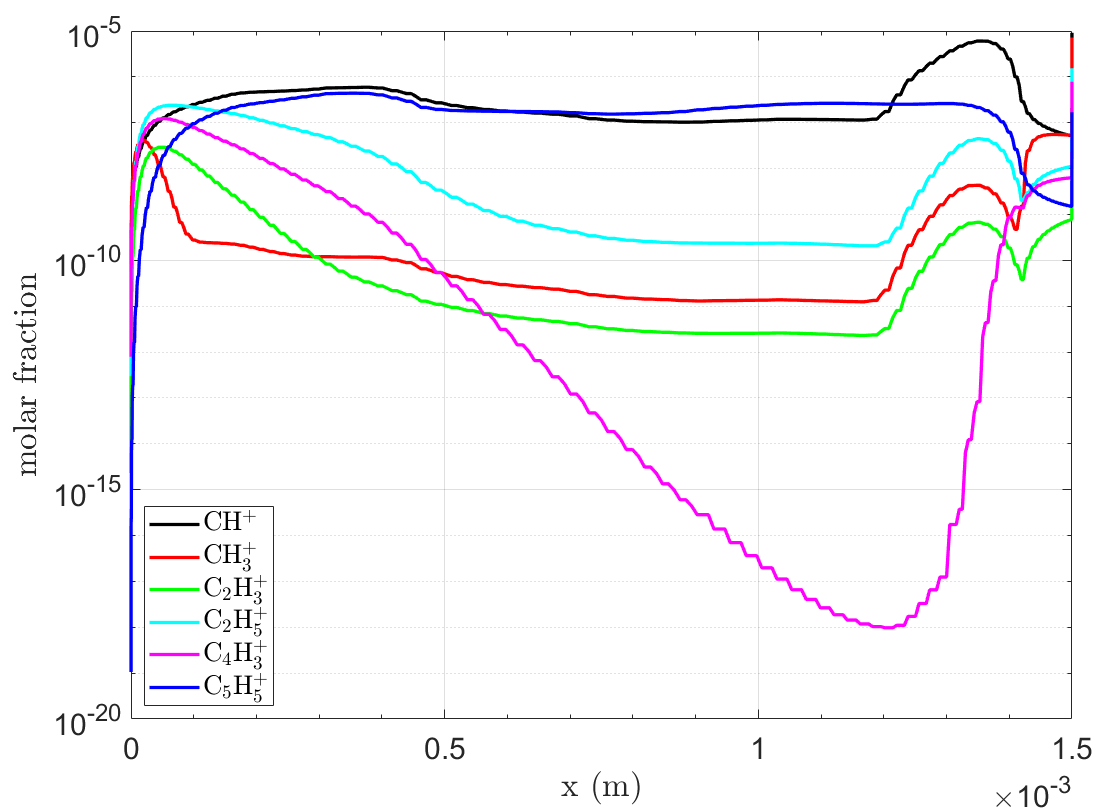}
        \caption{}
        \label{figure13b}
    \end{subfigure}
    \caption{(a) Steady state molar fraction profiles, along the plasma channel, of \ce{CH+}, \ce{CH3+}, \ce{C2H3+}, and \ce{C2H5+} calculated in an \ce{N2-CH4} (90-10) plasma with a -1000 V voltage. The molar fractions of \ce{C4H3+} and \ce{C5H5+} were also calculated but their values were below $10^{-20}$, which is considered to be the physical threshold. (b) Molar fraction profiles of the same molecules, including \ce{C4H3+} and \ce{C5H5+} which have values above the physical threshold this time, calculated in an \ce{N2-CH4-C2H2} (85-10-5) plasma with -1000V voltage. Values approaching a mole fraction of $10^{-20}$ represent the lower limit of what is physically possible inside the volume of the plasma cavity. Lower values close to $10^{-20}$ approach what is physically possible in terms of mole fraction; they are shown here for comparison purposes between the multiple species.}
    \label{figure13}
\end{figure}

As shown in Figure \ref{fig7a}, adding heavier hydrocarbon precursors such as \ce{C2H2} or \ce{C2H4} has an impact on the chemistry and results in changes in the production yield of ion species in the modeled spectra. The addition of \ce{C2H2} has also been proven to have a large impact on the elemental composition of the Titan aerosol analogues (or \textit{tholins}) formed in the COSmIC plasma experiments \citep{Nuevo2022}.
In the study presented here, the impact of \ce{C2H2} on the plasma chemistry leading to the production of \textit{tholins} was assessed by comparing the calculated molar fractions of several light-mass hydrocarbons at steady-state for modeled plasmas generated in initial \ce{N2-CH4} gas mixtures with and without \ce{C2H2}.
These include the following cations \ce{CH+}, \ce{CH3+}, \ce{C2H3+}, \ce{C2H5+}, \ce{C4H3+}, and \ce{C5H5+}. In the modeled \ce{N2-CH4} plasma (Figure \ref{figure13a}), \ce{C4H3+} and \ce{C5H5+} show the lowest predicted abundances ($\sim 10^{-28}$) while \ce{C2H5+} reaches the highest concentration ($\sim 10^{-7}$ or on the order of 0.1 ppm), which is consistent with the relative intensities seen in the experimental mass spectrum (Figure \ref{figure6MS_calc_expt_comparisons}). The overall profile of these species remains relatively constant throughout the plasma channel, indicating little perturbance by chemistry other than the limited migration of ions near the anode, as discussed above. This also suggests that in \ce{N2-CH4} plasmas, the main engine to hydrocarbon growth is driven by a handful of cations such as \ce{C2H3+}, and \ce{C2H5+}. \ce{C2H5+} in particular appears to be in excess in the model compared to experimental measurements (see Figure \ref{figure6MS_calc_expt_comparisons}), which likely suggests that there are still missing reactions to account for the observed discrepancies. Moreover, due to the truncated nature of the pulsed plasma discharge and our reaction scheme, \ce{C2H5+} production is facilitated over that of HCN, with which it reacts to produce \ce{HCNH+} \citep{Raymond2018}. As a result, radical-induced growth of C2 and C3 species may be accentuated \citep{Benedikt2010Plasma-chemicalPlasmas, Raymond2018}.

First we observe that \ce{C5H5+} (formed primarily through the channel \ce{C2H2 + C3H5+ -> H2 + C5H5+}) becomes the most abundant species reaching  $\sim10^{-7}$ (or on the order of 0.1 ppm) in the middle of the plasma cavity, at x = 0.75 mm. \ce{CH+} is also heavily produced with a molar fraction also around 0.1 ppm at x = 0.75 mm and reaching almost 10 ppm closer to the cathode and back down to 0.1 ppm at the outlet of the plasma cavity. This value is higher than the \ce{CH+} concentration of $\sim$0.01 ppm at the outlet of the plasma cavity obtained by \citep{Raymond2018}. This difference is explained by the fact that \citet{Raymond2018} modeled a \ce{N2-CH4-C2H2} plasma with a -800 V voltage while in our calculation, we used a higher voltage of -1000 V. Close to the cathode, at x = 1.3-1.4 mm, we observe a sudden sharp increase in molar fractions for several species: \ce{CH+} and \ce{CH3+}, \ce{C2H3+}, and \ce{C2H5+}. This can be explained by an increase in inelastic collision occurring in that region, specifically two channels producing \ce{C2H2+} and free electrons: 

\begin{equation}
\label{free elec prod 1}
\ce{e + C2H2 \longrightarrow C2H2+ + 2e}
\end{equation}

and

\begin{equation}
\label{free elec prod 2}
\ce{e + C2H4 \longrightarrow H2 + C2H2+ + 2e}
\end{equation}\\

An increase in the electron density is also observed in this region (Figure \ref{fig:10a}). The higher concentration of reactive electrons, radicals, and ionic species in this region could result in higher \textit{tholin} growth rates following the formation of larger molecular species, particularly through a cationic pathway (Reaction \ref{cluster growth}, after multiple acetylene or ethylene additions) recently proposed by \citet{Janalizadeh2003}:

\begin{equation}
\label{cluster growth}
\ce{C_{2n}H_{2m}+ + C2H2 \longrightarrow C_{2n+2}H_{2m+2}+ + xH2}
\end{equation}\\

where $n = 1-5, x = 0$ or $1$. A drastic consumption of \ce{C4H3+} (magenta, Figure \ref{figure13b}) is also observed which begins early on in the plasma cavity until it increases again near the cathode, at $x < 1.3$ mm. As soon as \ce{C4H3+} is formed near the cathode, it is quickly consumed through reactions with the much more abundant \ce{C2H2} and \ce{C2H4} molecules to form the larger cations \ce{C6H3+} and \ce{C6H5+} (Reactions \ref{C4H3+ R1} and \ref{C4H3+ R2}). This step is expected to be a key transition towards the formation of the first C6 compounds \citep{Peverati2016INSIGHTSPATHWAYS,Raymond2018, Vuitton2019}.

\begin{equation}
\label{C4H3+ R1}
\ce{C4H2 + C4H3+ -> C2H2 + C6H3+}
\end{equation}

\begin{equation}
\label{C4H3+ R2}
\ce{C2H2 + C4H3+ -> C6H5+}
\end{equation}\\

\begin{figure}[htbp]
    \centering
    \begin{minipage}{0.45\textwidth}
        \centering
        \begin{subfigure}[b]{\textwidth}
            \centering
            \includegraphics[width=0.8\textwidth]{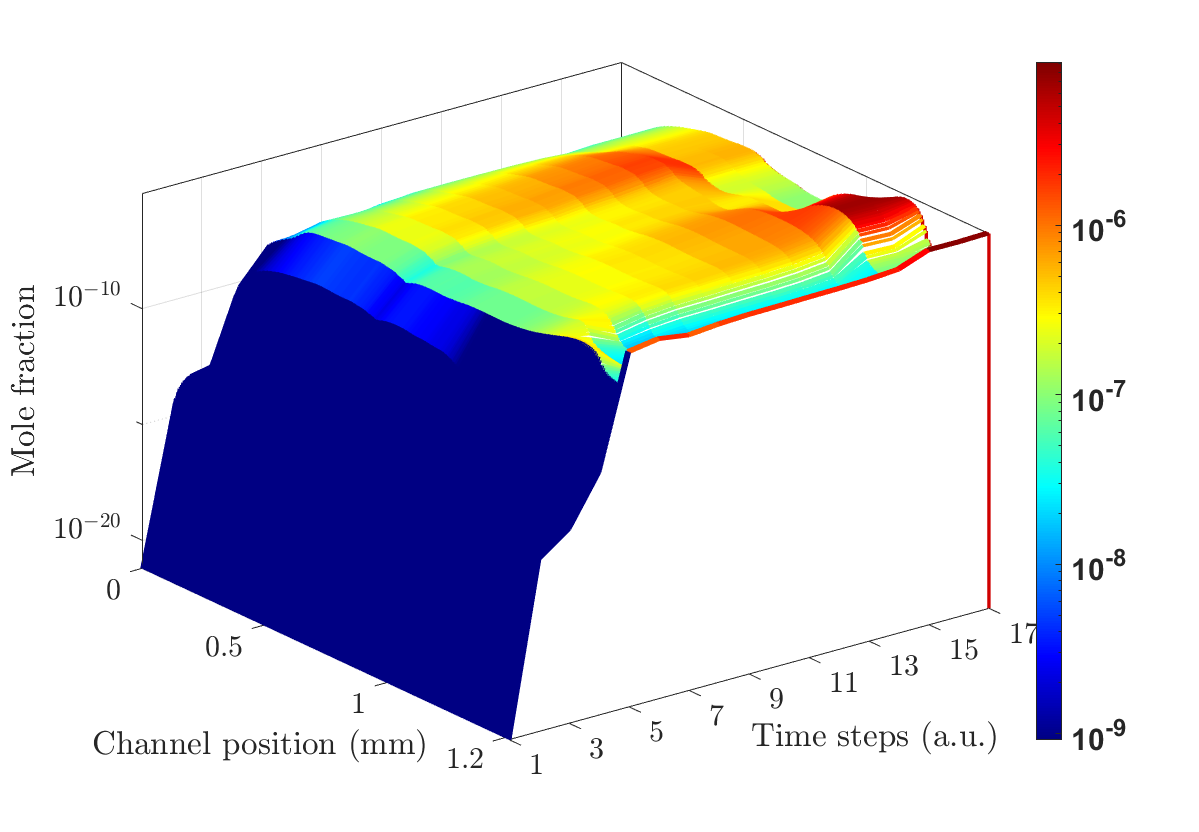}
            \caption{\ce{CH+}}
        \end{subfigure}

        \begin{subfigure}[b]{\textwidth}
            \centering
            \includegraphics[width=0.8\textwidth]{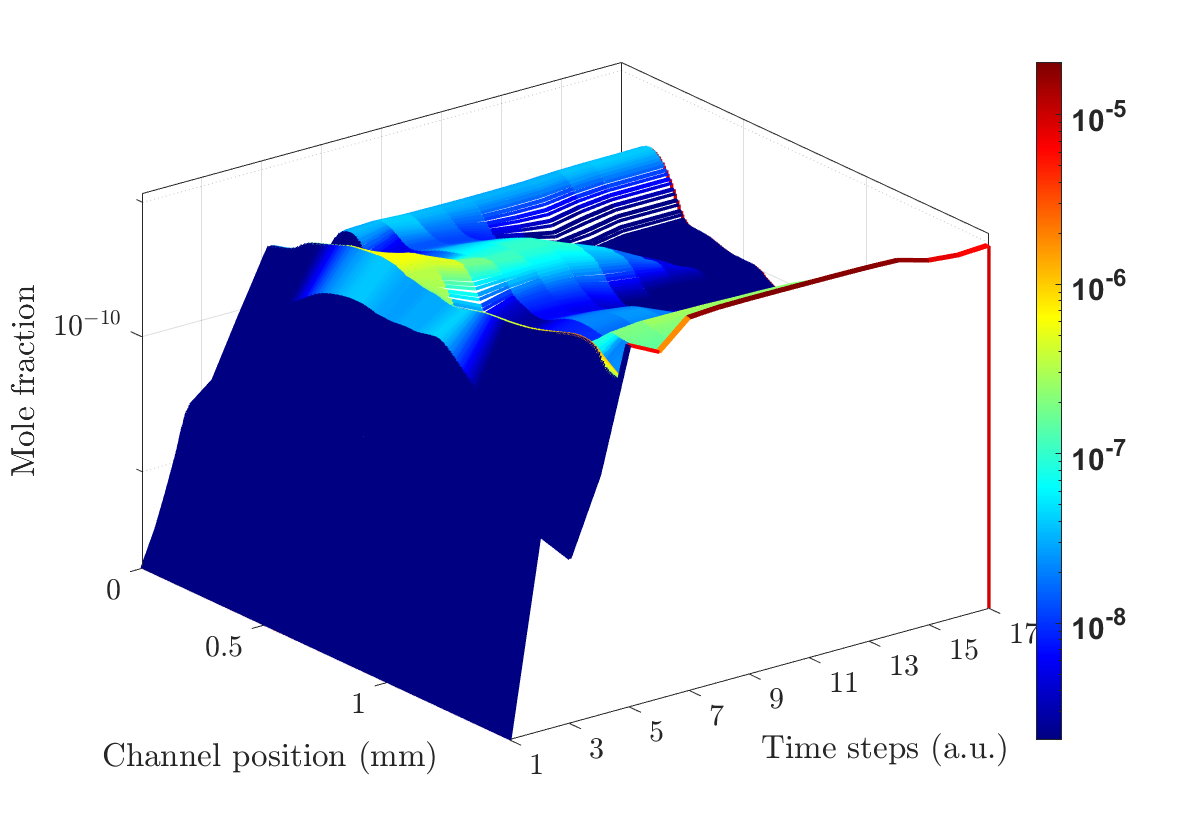}
            \caption{\ce{CH3+}}
        \end{subfigure}

        \begin{subfigure}[b]{\textwidth}
            \centering
            \includegraphics[width=0.8\textwidth]{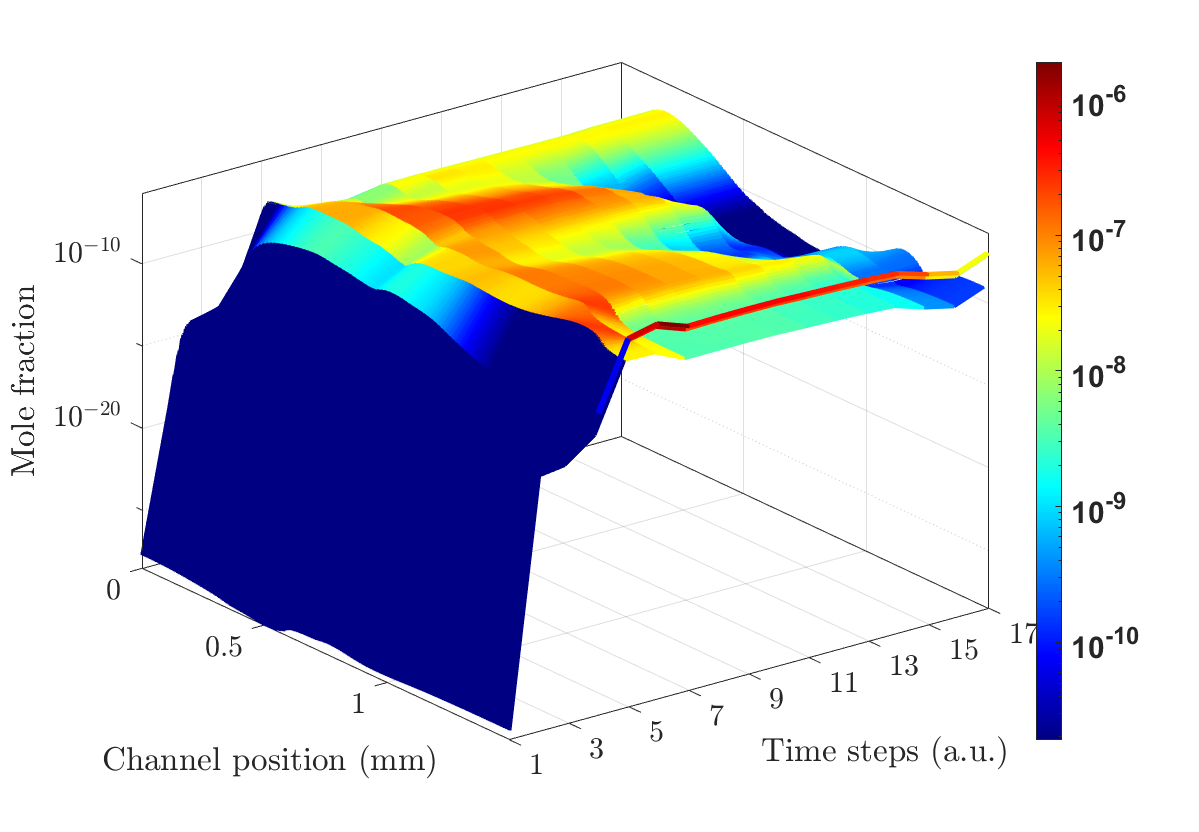}
            \caption{\ce{C2H3+}}
        \end{subfigure}
    \end{minipage}
    \hfill
    \begin{minipage}{0.45\textwidth}
        \centering
        \begin{subfigure}[b]{\textwidth}
            \centering
            \includegraphics[width=0.8\textwidth]{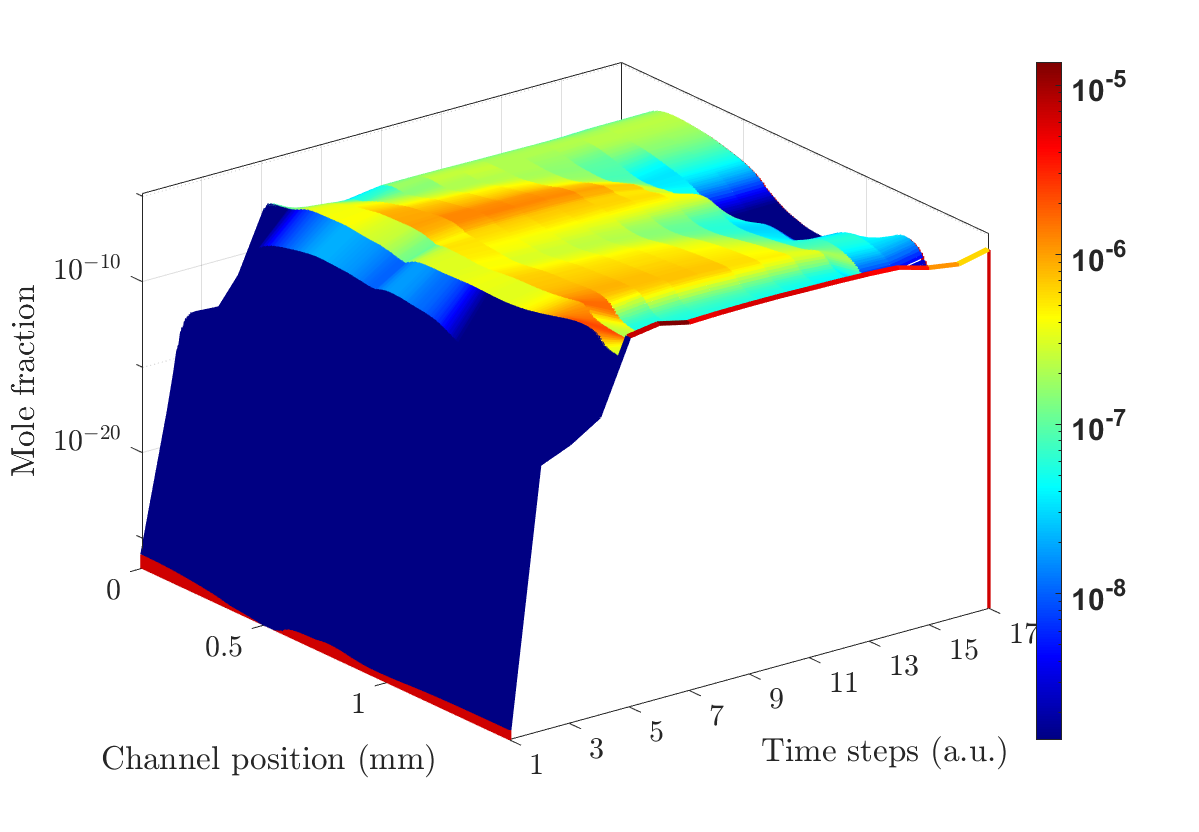}
            \caption{\ce{C2H5+}}
        \end{subfigure}

        \begin{subfigure}[b]{\textwidth}
            \centering
            \includegraphics[width=0.8\textwidth]{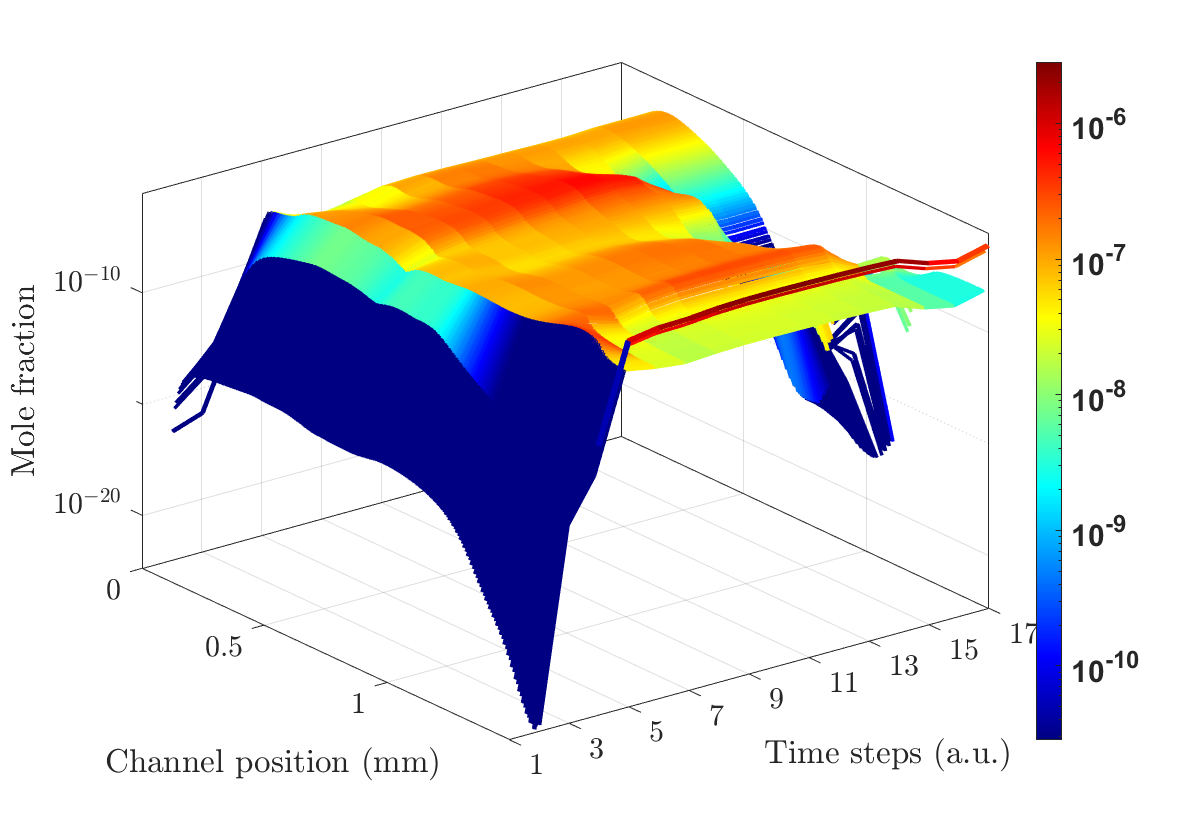}
            \caption{\ce{C4H3+}}
        \end{subfigure}

        \begin{subfigure}[b]{\textwidth}
            \centering
            \includegraphics[width=0.8\textwidth]{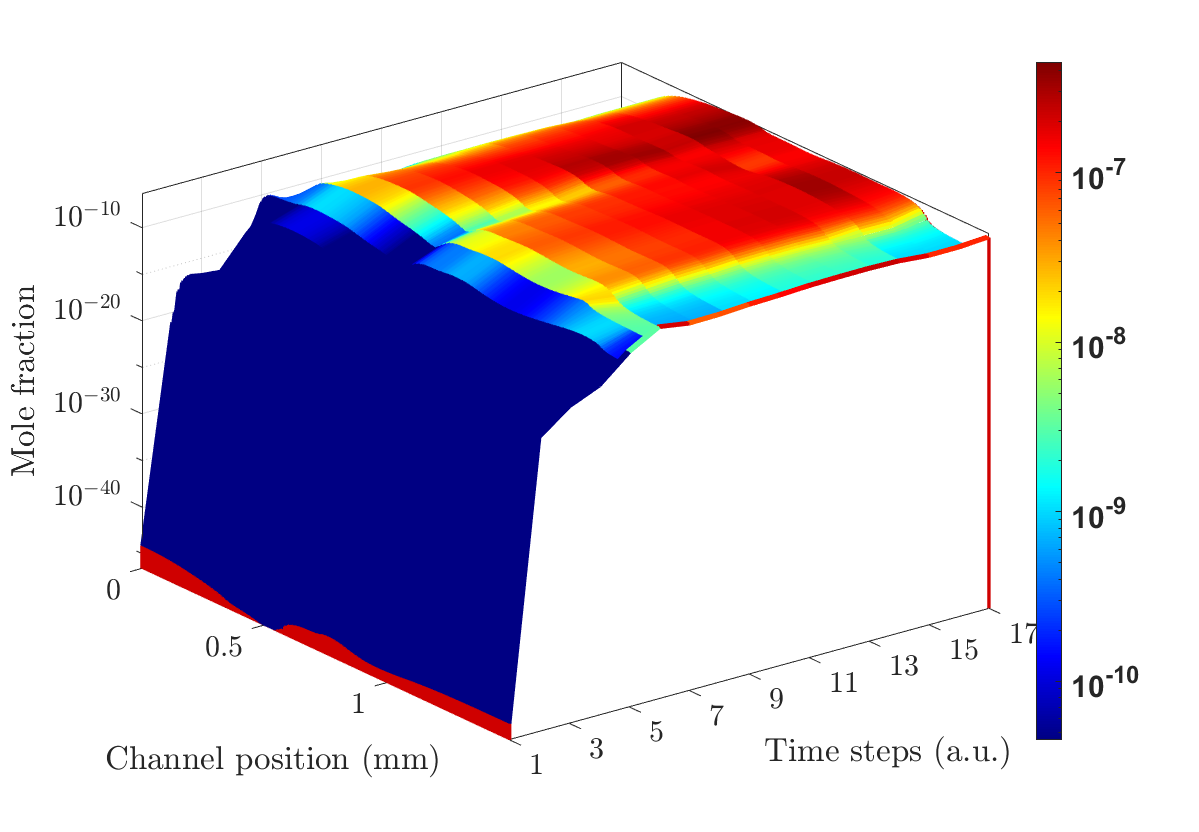}
            \caption{\ce{C5H5+}}
        \end{subfigure}
    \end{minipage}
    
    \caption{3D representations of the molar fractions of each cation and its time-dependency calculated during one plasma pulse. The x axis represents arbitrary time steps chosen from the beginning of each simulation (t = 1) until steady-state conditions are reached by t = 17. The y axis is the channel distance along the plasma channel (in mm, where x = 0 mm is the anode), and the z axis is the color coded molar fractions.}
    \label{figure14_3D}
\end{figure}

The observed sensitivity of the plasma products, and thus the entire network, on the presence of \ce{C2H2}, show intrinsic evolutions that depend on the precursor gas density profile, electron mobility, and collisional efficiency. These trends can also be visualized as a function of time, as shown in Figure \ref{figure14_3D} where the molar fractions are represented in 3D as a function of time and position in the plasma cavity. \ce{CH+} and \ce{C5H5+} abundances do not vary as much once the electron steady state plateau is reached, whereas \ce{CH3+}, \ce{C2H3+}, \ce{C4H3+}, and to a lesser extent \ce{C2H5+} go through a much more dynamical range of concentrations. \ce{C2H3+} and \ce{C4H3+} undergo an important production phase early in the plasma pulse (red increase) before being consumed (blue dips). On the other hand, \ce{C5H5+} remains relatively stable across the pulse duration. Note that these trends do not include loss effects due to the formation of much larger molecular products such as polycyclic aromatic compounds and solid particles (\textit{tholins}) as they are not included in the CO-PRISM model.

\vspace{0.3 cm}
\subsubsection{
\ce{C2H2} and \ce{HC3N} ionization and mechanisms towards the formation of benzonitrile in a \ce{N2}-\ce{CH4}-\ce{C2H2} plasma}
\vspace{0.3 cm}

\ce{C2H2} and acetonitrile (\ce{HC3N}) are both important molecules produced and detected in a variety of planetary atmospheres, molecular cloud, and star-forming regions \citep[\textit{e.g.}][]{Freeman1978,Snell1981,Teanby2006,Rimmer2021}. \ce{HC3N} is a simple linear molecule which possesses two triple-bonded carbons, one of which with a terminal nitrile function. It has been found to play an important role in the formation of prebiotic molecules on Earth and in the ISM \citep{Orgel2002,Ritson2022}, along with other nitriles such as HCN \citep{Ferus2017,Stein2020} and azole compounds as an intermediate for nucleotide and amino acid synthesis \citep{Shapiro1999,Ritson2022}. More recently, \ce{HC3N} (and its association with \ce{C2H2}) was found to be an intermediate candidate in the preliminary stages of aromatic molecules formed in the ISM \citep{Jose2021}, particularly in the formation of benzonitrile \ce{C7H5N}. Computationally, \cite{Jose2021} showed that ionized Van der Waals clusters containing \ce{C2H2} and \ce{HC3N} could lead to the formation of ISM-relevant molecules, and in particular, benzonitrile. In the ionization process, an important ion, \ce{C5H3N+}, is formed through \ce{C2H2} and \ce{HC3N} bonding. This alyphatic ion can then serve as a backbone of larger ($>$C5) linear or cyclic molecules, and 
was listed as one of the most abundant C5 candidate species in the ion-neutral reaction network by \cite{Carrasco2008}. This highlights the potential role of \ce{C2H2} and \ce{HC3N} (and the intermediate ions) as crucial precursors to Titan's complex organic products. In our model, Reaction \ref{C5H3N+ production} follows a modified Arrhenius expression with the reaction rate $k = 2 \times 10^{-12}$ cm$^3$ s$^{-1}$ \citep{Raymond2018}. 

\begin{equation}
\label{C5H3N+ production}
\ce{HC3N + C2H2+ \longrightarrow C5H3N+}
\end{equation}\\

In steady-state regime with an \ce{N2-CH4-C2H2} mixture, we find a maximum \ce{C5H3N+} mole fraction of $10^{-21}$ near the cathode, as opposed to $\sim 10^{-46}$ at the same location in an \ce{N2-CH4} without acetylene. This 25 order of magnitude difference is particularly appreciable since the initial concentration of \ce{C2H2} is of only 5\%. It shows that \ce{C5H3N+} is sensitive to the initial concentration of \ce{C2H2} which also controls the production of cyanoacetylene via 

\begin{equation}
\label{HC3N}
\ce{C2H2 + CN \longrightarrow H + HC3N}
\end{equation}\\

with a rate $k = 2.7 \times 10^{-10}$ cm$^3$ s$^{-1}$ \citep{Wakelam2012}. Thus, in addition to \ce{C2H2} acting as an important precursor to double- and triple-bonded (\ce{C=C}, \ce{C=N}, and \ce{C#N}) species, 
it also favors the gas phase formation of \textit{tholin} precursors with nitrile groups attached to aromatic species. The increase in ions such as \ce{C5H3N+} may thus support their role as an important gas phase precursor of aromatic and unsaturated characters. Future work may be able to assess whether species like benzonitrile may be favorably formed in \ce{N2-CH4}-based plasmas or not \citep{Esposito2025AIntermediates}. \\

\vspace{0.3 cm}
\subsubsection{Nitrogenated products and incorporation of N-bearing ions: formation of the stable HCN dimer \ce{H2CNCN} and role of \ce{CH2NH}}
\vspace{0.3 cm}

If we further examine the implications for N-rich chemistry, \ce{H2CNCN}, known as cyanomethanimine, is an inorganic compound of relevant interest. It consists of a carbon-nitrogen framework where the central carbon is bonded to both an amine group (\ce{NH2}) and a nitrile group (CN). This structure gives cyanomethanimine both nucleophilic and electrophilic properties, allowing it to react in a variety of ways. It is commonly used in the synthesis of various organic compounds, including imidazoles and other nitrogen-containing heterocycles. Cyanomethanimine can also undergo hydrolysis to form ammonia and formic acid under acidic conditions. \ce{H2CNCN} has recently gained astrochemical interest since its detection in the Galactic Center \citep{SanAndres2024} and spectroscopic properties and formation pathways \citep{Fortenberry2024}. While condensed phase pathways are being investigated \citep{Fortenberry2024}, gas phase pathways (Table \ref{Table2}, R20) have so far \citep{Vazart2015} been constrained to the following pathway:

\begin{equation}
\label{H2CNCN}
\ce{CN + CH2NH \longrightarrow H2CNCN + H}
\end{equation}

\begin{figure}
\centering
\includegraphics[scale=0.5]{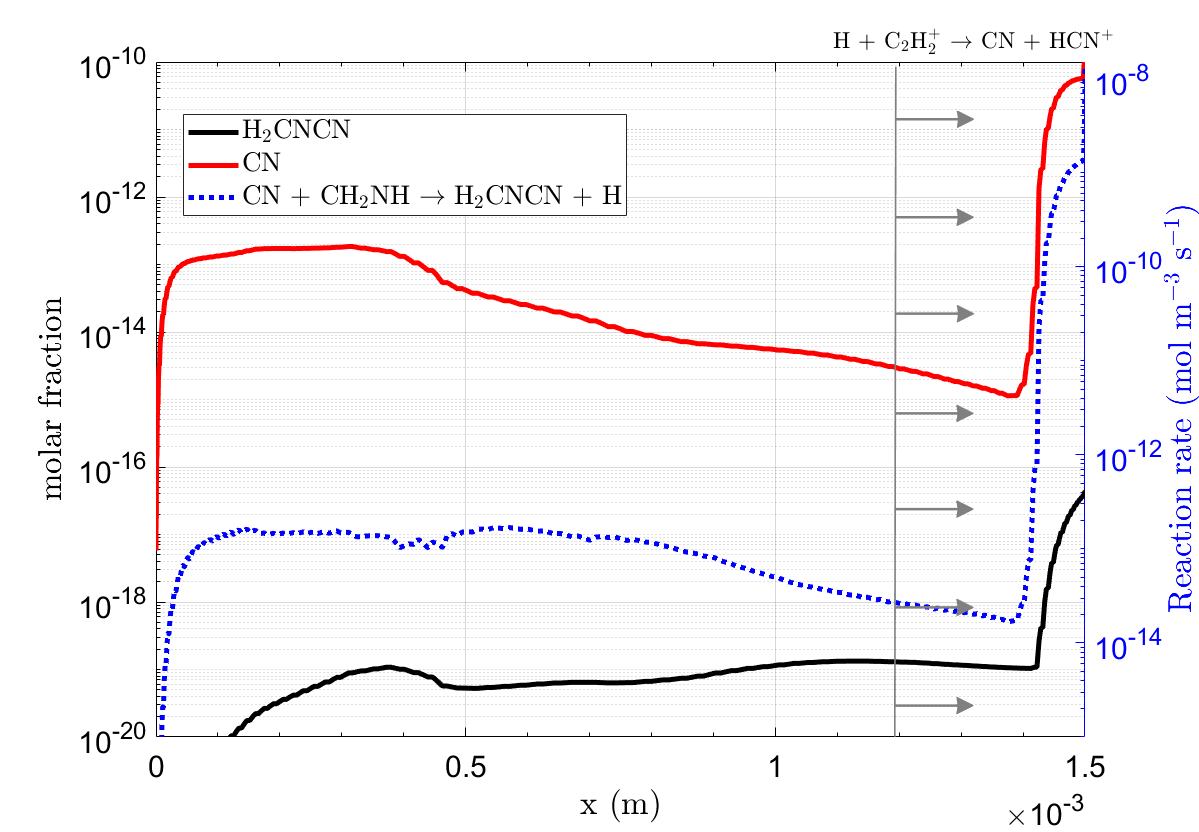}
\caption{Calculated molar fractions for \ce{H2CNCN} (black) and \ce{CN} (red) in a \ce{N2-CH4-C2H2} (85-5-5) plasma with -1000 V voltage. The reaction rate of the main reaction incorporated into our model producing \ce{H2CNCN} is also shown (dotted blue line). \label{figure15}}
\end{figure}

Although this neutral pathway remains elusive in that it cannot reproduce the recent observations of \ce{H2CNCN} \citep{SanAndres2024}, a putative approach was considered here to assess whether this HCN dimer may be formed under a \ce{C2H2}/\ce{CN}-enriched, Titan-like gas mixture.
Figure \ref{figure15} shows the \ce{H2CNCN} abundance calculated using Eq. \ref{H2CNCN} and a rate $k = 2.7 \times 10^{-11}$ cm$^3$ s$^{-1}$ with CO-PRISM for an \ce{N2-CH4-C2H2} modeled plasma with -1000 V voltage.
\ce{H2CNCN} (black line) reaches a molar fraction ranging from $\sim 10^{-19}$ to $\sim 10^{-16}$ (or $10^{-10}$ to $10^{-7}$ ppb). While these values are extremely small, they highlight however the important role \ce{C2H2} (and \ce{C2H2+}) can play in the catalysis of nitrogen-rich molecules, according to any of the following pathways: 

\begin{equation}
\label{C2H2+ no1}
\ce{C2H2 + CN \longrightarrow HC3N + H}
\end{equation}

\begin{equation}
\label{C2H2+ no2}
\ce{HC3N + C2H2+ \longrightarrow C5H3N+}
\end{equation}

\begin{equation}
\label{C2H2+ no3}
\ce{N(S) + C2H2+ \longrightarrow HCN + CH+}
\end{equation}\\

where $k_{21} = (2.71 \times 10^{-9} \text{ cm}^3\text{s}^{-1}) \cdot (T/1[K])^{-0.52}~ \text{exp}[-20[K]/T]$, $k_{22} = (2.0 \times 10^{-12} \text{ cm}^3\text{s}^{-1}) \cdot (T/300[K])^{-0.25}$, and $k_{23} = 2.5 \times 10^{-11}$, respectively \citep{Wakelam2012}.
Along with acetylene, the radical \ce{CN} (molar fraction shown in red) also contributes to this scheme by reacting with methanimine \ce{CH2NH}, reaching non-negligeable amounts of $\sim$0.1 ppm. \ce{C2H2}, CN, and \ce{CH2NH} (formed from R17, Table \ref{Table2}) all contribute to the formation of \ce{H2CNCN}. In Figure \ref{figure15}, the temperature-dependent reaction rate for Equ. \ref{H2CNCN} is shown in blue. The vertical bar and arrows indicate the starting point region of a substantial uptake ($\sim$ 40 orders of magnitude) in the production of \ce{C2H2+} via two main channels: \ce{e + C2H2 \longrightarrow C2H2+ + 2e} and \ce{e + C2H4 \rightarrow H2 + C2H2+ + 2e} \citep{Endstrasser2009}. This uptake in \ce{C2H2+} thus creates near the cathode a region highly conducive to \ce{C2H2+}-based reactions. One of these reactions, \ce{H + C2H2+ \rightarrow CN + HCN+} ($k = 4.96 \times 10^{-10}$ cm$^3$ s$^{-1}$) is shown in the top-right corner of the figure. The increase in reaction rate observed closer to the cathode is consistent with a high \ce{C2H2+} production in that region (symbolized by the vertical bar and arrows pointing towards the cathode). The important rise in acetylene ionization occurring in this region of the plasma enables an efficient formation of CN radicals, thence available to react with \ce{CH2NH}. This uptake explains the important increase in CN and R20's reaction rate (as well as other species, as seen above) spanning several orders of magnitude.\\

Recently, work by \citet{Bourgalais2019} interestingly found that \ce{CH2NH} was the only stable nitrogen-bearing molecule formed by low-pressure EUV photochemistry experiments in \ce{N2-CH4} (90-10) mixtures, which was then confirmed by their 0D photochemical model. With the addition in CO-PRISM of reactions R17 and R19 (Table \ref{Table2}) involving the first electronically excited state of atomic nitrogen \ce{N(^2D)} and ammonia, respectively, we are able to corroborate the results found in \citet{Bourgalais2019}, as well as support the formation of the protonated methanimine \ce{CH2NH2+}. Methanimine being a transient species, this highlights the role of \ce{N(^2D)} as a good contributor to nitrogen-bearing molecules. 
Despite the detection of methanimine in laboratory experiments \citep{Carrasco2012,Horst2018,Bourgalais2019}, it has been difficult to determine its exact abundance in Titan's atmosphere \citep{Bourgalais2019}. Future work combining modeling, data analysis, and experiments might elucidate the abundances inferred from Cassini \citep{Vuitton2007}.

\vspace{0.3 cm}
\subsubsection{Gas-tholin elemental compositions and astrochemical implications}
\vspace{0.3 cm}

The unsaturated hydrocarbons shown in Figure \ref{figure13} could play a crucial role in the formation of solid particles in Titan's atmosphere and are expected to contribute to the molecular diversity of \textit{tholins} formed in COSmIC. 
A recent XANES analysis of tholins produced in COSmIC in \ce{N2-CH4} (95-5) and \ce{N2-CH4-C2H2} (94.5-5-0.5) plasmas showed the impact of \ce{C2H2} on the incorporation of nitrogen in the solid phase \citep{Nuevo2022}. We have used CO-PRISM to determine the overall elemental composition, and in particular the C/N ratio, of the gas phase products in plasmas modeled with the same initial gas mixtures as the XANES study.
As shown in Table \ref{table CN ratios}, we observe a similar trend in C/N in the gas phase to that found in tholins, when adding acetylene, \textit{i.e.}, less nitrogen is being incorporated in the molecular products both in the gas and solid phase when \ce{C2H2} is present in the initial gas mixture, and the way nitrogen is incorporated in the chemical network changes. Nitrile bands are more prevalent with the addition of \ce{C2H2}, controlled \textbf{by} \ce{C4H2} and \ce{HCN}. As a consequence, this may result in a larger proportion of nitrile groups attached to aromatic structure, as opposed to aliphatic ones \citep{Nuevo2022}. These general similar trends give robustness to CO-PRISM.

Furthermore, \ce{C2H2} reactions with \ce{C2H} and \ce{C4H} also drive the formation of the first C4 and C6 molecules, diacetylene \ce{C4H2} and triacetylene \ce{C6H2}, respecitvely \citep{Peverati2016INSIGHTSPATHWAYS,Pentsak2024} as follows:

\begin{equation}
\label{C2H2+C2H}
\ce{C2H2 + C2H \longrightarrow C4H2 + H}
\end{equation}

\begin{equation}
\label{C2H2+C4H}
\ce{C2H2 + C4H \longrightarrow C6H2 + H}
\end{equation}

The \ce{C4H2+} and \ce{C4H3+} cations are important Titan tholin precusors, as was seen previously in \citet{Sciamma-OBrien2014, Raymond2018} and in Figure \ref{figure6MS_calc_expt_comparisons} at \textit{m/z} 50 and 51, as well as in other Titan experimental studies \citep{Carrasco2012,Berry2018,Dubois2019a,Dubois2020}. These C4 molecules play an intermediate role towards the formation of larger linear and even cyclic \textit{tholin} precursors \citep{Bera2015HydrocarbonFragments,Pentsak2024}. The formation of \ce{C4H2+} occurs primarily through the ion-molecule reactions between \ce{C2H2} and \ce{C2H2+} (R141 in the Appendix) and between \ce{C2H2} and \ce{C2H+} (R253). These pathways are important since the calculated abundances depend on the initial \ce{C2H2} concentration, and some C6 cyclic isomers may be formed according to Reaction R141 \citep{Bera2015HydrocarbonFragments}. Meanwhile, diacetylene \ce{C4H2} may further react with \ce{C2H5+} to produce \ce{C2H4} and \ce{C4H3+} (R88 Appendix), thus highlighting the importance of constraining these pathways leading to the formation of key C4 Titan \textit{tholin} precursors. Similarly, at \textit{m/z} 51 and an even higher abundance, \ce{C4H3+} is formed through Reaction R141 with a branching ratio of 0.49 \citep{Bera2015HydrocarbonFragments}. This closed shell radical is produced at a fast rate of $\sim10^{-9}$~cm$^3$~s$^{-1}$, and the relatively good agreement we obtain with the experimental measurements indicates that the reaction pathways aforementioned play a central role in the precursor chemistry synthesizing Titan \textit{tholins}. Triacetylene \ce{C6H2}, although found in the interstellar medium but undetected on Titan, future work regarding its formation and subsequent loss through ion chemistry would help understand the role of C6 radical chemistry toward polyyne formation, relevant to acetylene-rich \textit{tholin} formation \citep{Cernicharo2001,Contreras2013,Bera2015HydrocarbonFragments, Nixon2024}.

The rich chemistry induced by \ce{C2H2} may also lead to the formation of polyynes and cyanopolyynes, first forming propiolonitrile \ce{HC3N}, to eventually \ce{C3N} or \ce{C5N} via neutral-neutral or ionic channels \citep{Dubois2019ApjL, Pentsak2024}, which could help to better characterize the gas phase precursors before their incorporation into the tholins. Future incorporation of barrierless reactions other than the favorable \ce{CN + C2H2} or even the formation of triacetylene \ce{C6H2} by the reaction between diacetylene and the ethynyl radical \citep{Pentsak2024} could assist mass spectrometry or spectroscopic measurements in identifying more precursors.

Furthermore, we may also be able to constrain our chemical network and the underlying mechanisms for ring formation (so far not accounted for in CO-PRISM) with the help of flame-burning models, which would have application not only for Titan's atmosphere but also for other astrophysical environments, and in particular circumstellar envelopes. Several models have provided guidance to cyclization pathways via barrierless recombination reactions relevant to circumstellar envelopes \citep{Pentsak2024}. The presence of quantifiable C6 species at \textit{m/z} 76 and 77 (Figure \ref{figure6MS_calc_expt_comparisons} and section \ref{sec:3.1}) warrants a numerical focus on low-temperature barrierless pathways for which products may be within the sensitivity range of mass spectrometry and other techniques. In any case, this gas phase investigation supports the experimental findings between the correlation of acetylene and the higher formation yield towards aromatic compounds conducted experimentally \citep{Contreras2013,Sciamma-OBrien2014,Nuevo2022}.

\begin{table}[]
    \centering
    \begin{tabular}{cccc}
    \toprule
    C/N & Gas phase (this work) & Tholin (Nuevo et al. 2022) & Main functional groups\\ \midrule
    \toprule
    \ce{N2}-\ce{CH4} (90:10\%)   & 5.3 & 1.3 & C=H\\ \midrule
    \ce{N2}-\ce{CH4}-\ce{C2H2} (95-5-0.5\%) & 8.6 & 2.4 & \ce{C#N}\\ \bottomrule
    \end{tabular}
    \caption{Comparison of the modeled gas phase C/N ratio obtained by CO-PRISM and the experimental solid phase C/N ratio obtained with XANES for \ce{N2-CH4} (95-5) and \ce{N2-CH4-C2H2} (94.5-5-0.5) plasmas \citep{Nuevo2022}. The main functional groups determined in the gas phase and solid phase are also given.}
    \label{table CN ratios}
\end{table}

\subsubsection{Perspectives for future model expansions}

Through this work we have shown an improvement in the level of agreement between the CO-PRISM modeling and experimental TOF measurements. In particular, computed mass spectra better reflect experimental intensities of C2 and C3 ions, and notably now, C6 species associated with the production of benzene at \textit{m/z} 76 and 77 when adding \ce{C2H2} to the gas mixture. Nonetheless some limitations to the model are present and can be refined in future studies in order to further improve agreements with experimental measurements and guide future experiments. First, more realistic sticking coefficient values, instead of the $\gamma_k=1$ value currently considered in CO-PRISM, could better account for surface chemistry at the walls, and the potential importance of \ce{NH3} surface chemistry. 
Second, including particle formation mechanisms that would deplete the gas medium of certain species, currently not included in CO-PRISM, would be a more realistic representation of what is observed experimentally, as studies have shown that \textit{tholins} are formed in the COSmIC plasma discharge \citep[\textit{e.g.,}][]{Sciamma-OBrien2017}. The mechanisms leading to the formation of solid particles from gas precursors are however not yet well understood. Finally, including isotopologues and negative ions in the chemical network would lead to a more comprehensive picture of the plasma chemistry occurring in COSmIC.

\newpage
\section{Conclusions}\ \label{sec:Conclusions}

Using a newly updated chemical network model, we calculated ion-neutral reactions to simulate the NASA Ames Research Center COsmic Simulation Chamber-Titan Haze Simulation setup, a low-pressure and low-temperature supersonic plasma discharge. We conducted an analysis of \ce{N2-CH4}-based gas phase reactivity relevant to Titan's upper atmosphere conditions and compared model outputs to experimental measurements using time-of-flight mass spectrometry. This analysis relied on updating and expanding the reaction rates part of our reaction scheme with the latest available data. Mass spectra were computed for plasmas containing varying concentrations of \ce{CH4} and other precursors (\textit{e.g.}, \ce{C2H2}, \ce{C2H4}). The simulated plasma composition (gas phase products, electron density and temperature, and molecular densities) was characterized in these conditions. The main results are as follows:

\begin{itemize}

    \item We carried out a substantial expansion on the already existing model by adding 45 new reactions along with updating 77 other reactions according to the most recent data available in the literature. This included key electron impact ionization cross-sections for hydrocarbon-based plasmas. These changes resulted in the modification of the mass distribution in the computed mass spectra, particularly (i) a better agreement with the experimental spectra for some C3, C4, and C5 species, (ii) the partial contribution of the transient and important methanimine and its protonated form \ce{CH2NH2+} due to \ce{C2H2}-driven chemistry, and (iii) the reproduction of experimentally-detected benzene C6 fragments. These observations have helped constrain updated reactions and/or reaction rates and their respective sensitivity on the polymeric gas phase growth.
    \item The polymeric growth progressing through the plasma cavity was assessed by generating spectra along the supersonic flow, giving insights on the role small hydrocarbons have on larger hydrocarbons.
    \item The role of ionization efficiency in the plasma cavity was characterized, which highlighted the influence primary ions and electrons have on the chemical inventory.
    \item While updates to the reaction network have improved the level of agreement between model and experiment, we show that the chosen reactions do not yet perfectly capture the complicated evolution of the plasma. There continues to be experimental detections not observed in the model prediction and \textit{vice-versa}. The high degree of sensitivity to the precursor species (like \ce{C2H2}) has helped to assess the impact of these specific precursors, which as a result can help define which molecules or specific pathways might still be missing from our reaction scheme.
    \item An intense ionization regime induced by high \ce{C2H2} consumption has been associated with ionic pathways leading to the formation larger compounds (C5 and C6) and was compared with a recent study on carbonaceous nanoparticle growth by hydrocarbon addition. We obtain good agreement for abundances of \ce{C6H4+} and \ce{C6H5+} between experiments and computations, resulting from the reaction between \ce{C2H2} with \ce{C4H2+} and \ce{C2H2} with \ce{C4H3+}, respectively. Thus, these reactions are found to be key to explain the presence of C6 precursor molecules of Titan \textit{tholins}.
    \item These changes to the CO-PRISM model enabled a theoretical investigation of the production of the HCN dimer, its relationship to methanimine, and nitrogen incorporation into larger intermediates such as benzonitrile precursors. These advancements were made possible by the incorporation of key reactions involving the first electronically-excited state of atomic nitrogen and NH radicals, highlighting the influence of acetylene on cold, astrophysically-relevant environments. While solid phase (\textit{tholin}) chemistry is not taken into consideration in our calculations, only gas phase chemistry, current developments to also include negative ion chemistry and other prebiotically-relevant species are underway.
    \item The COSmIC experiment simulates low-temperature chemistry relevant to Titan and other planetary atmospheres. Our study aims to integrate chemical modeling with experimental measurements to identify key pathways in Titan's atmospheric chemistry within the COSmIC setup. By comparing COSmIC data with synthetic mass spectra from the CO-PRISM model, we can: (1) validate CO-PRISM's chemical network, (2) identify and add missing pathways or molecules, (3) quantify plasma-produced species, (4) evaluate the impact of specific pathways on chemical products, and (5) guide future COSmIC experiments to better understand Titan's atmosphere. The necessity for these targeted experiments to interpret Cassini observations was highlighted in previous research.
\end{itemize}

\section{Acknowledgements}

We thank two anonymous reviewers for their insightful comments which contributed to improving the quality of the manuscript. Funding was provided through appointment with the NASA Postdoctoral Program and through the NASA SMD CDAP R\&A Program. A portion of this work was carried out by the Jet Propulsion Laboratory, California Institute of Technology, under a contract with NASA.

\appendix

\begin{figure}[H]
\centering
\includegraphics[scale=0.5]{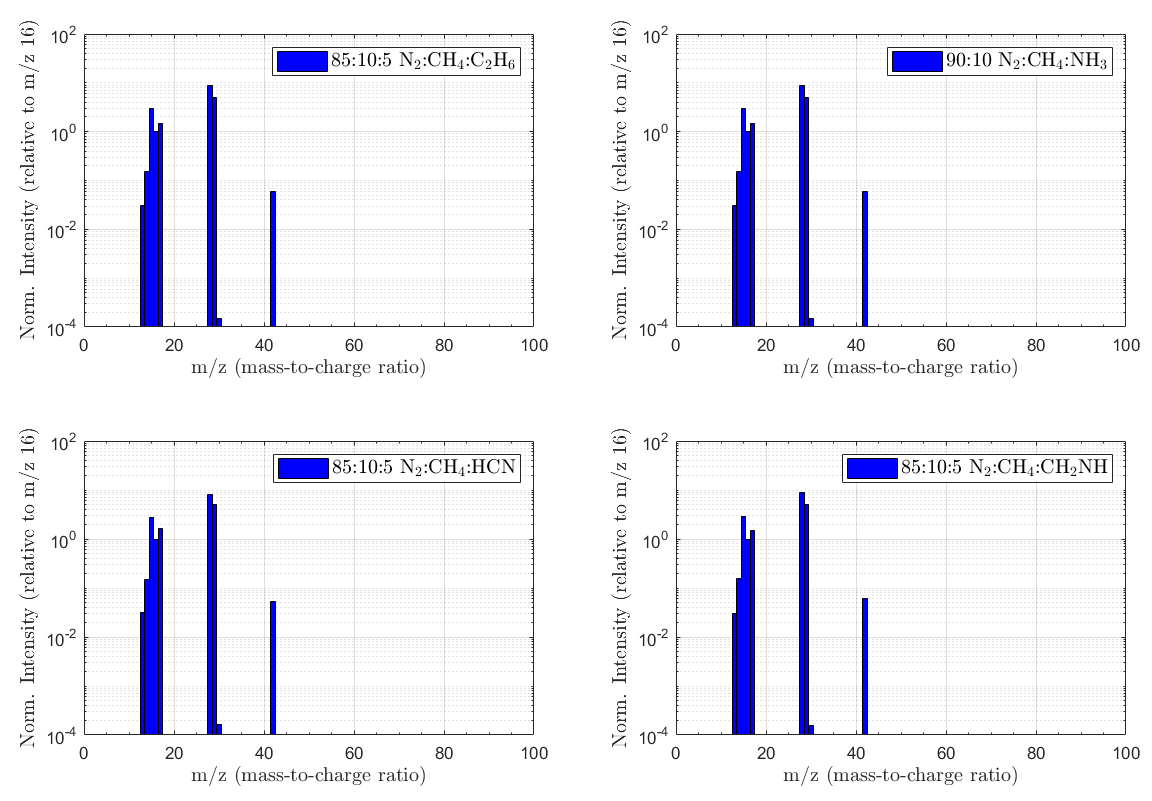}
\caption{Mass spectra computed in a \ce{N2}:\ce{CH4}:\ce{C2H6}, \ce{N2}:\ce{CH4}:\ce{NH3}, \ce{N2}:\ce{CH4}:\ce{HCN}, and \ce{N2}:\ce{CH4}:\ce{CH2NH} (85:10:5) plasma conditions. Intensities are normalized to the peak at m/z 16.
\label{fig7b}}
\end{figure}

Figure \ref{fig7b} shows computed mass spectra in four additional conditions containing \ce{C2H6}, \ce{NH3}, \ce{HCN}, and \ce{CH2NH} in minor amounts. No notable changes $> 1\%$ in these mass spectra are observed. More extensive networks including these species and their parent molecules are likely missing from our reaction scheme to explain the substantial production of new species. The absence of meaningful changes in the calculated mass spectra for these four plasma conditions could be due to (i) an incomplete list of reactions using any of these molecules as reactant in the model, or (ii) any minor species production remaining overwhelmed by \ce{CH4}-induced chemistry. In other words, any pathways related to the added minor species are eventually superseded by \ce{CH4}-induced chemistry. This could create a condition where \ce{CH4}-induced chemistry with a \ce{CH4} concentration of 10 \% reaches a stage where readily available low-mass ions such as \ce{H+}, \ce{CH+} and \ce{CH2} preferentially participate in the production of species that would otherwise just as well be formed in a simple \ce{N2-CH4} model, without much influence of other minor species. In reality, it is likely that a combination of both hypotheses may explain the lack of other observed products. To remediate this issue, future work should explore the addition of new reactions involving these precursors in conjunction with experimental measurements. This is however outside the scope of this study. In addition, most pathways towards \ce{C2H6} and \ce{NH3} production involve very slow bimolecular reaction \citep{Raymond2018} and compared to \ce{C2H2}-based chemistry as seen in this study, are almost completely negligible.

\newpage

\LTcapwidth=\textwidth
{\scriptsize
\begin{longtable}{lll}
\caption{Full Chemical Scheme used in CO-PRISM} \\

\toprule
\textbf{No.} & \textbf{Reaction} & \textbf{Rate \textit{k} ($cm^3.s^{-1}$)} \\
\toprule
\endfirsthead

\multicolumn{3}{c}{(Continued)}\\
\toprule
\textbf{No.} & \textbf{Reaction} & \textbf{Rate \textit{k} ($cm^3.s^{-1}$)}  \\
\toprule
\endhead

\toprule
\endfoot

\endlastfoot

1 & \ce{e + CH4 -> 2e + CH+ + 3H} & $f(\bar{\epsilon})$  \\
2 & \ce{e + CH4 -> 2e + CH2 + 2H} & $f(\bar{\epsilon})$  \\
3 & \ce{e + CH4 -> 2e + CH3 + H} & $f(\bar{\epsilon})$  \\
4 & \ce{e + CH4 -> 2e + CH4+} & $f(\bar{\epsilon})$ \\
5 & \ce{e + CH4 -> e + CH + 3H} & $f(\bar{\epsilon})$ \\
6 & \ce{e + CH4 -> e + CH2 + H2} & $f(\bar{\epsilon})$  \\
7 & \ce{e + CH4 -> e + CH3 + H} & $f(\bar{\epsilon})$ \\
8 & \ce{e + C2H2 -> 2e + C + CH2+} & $f(\bar{\epsilon})$  \\
9 & \ce{e + C2H2 -> 2e + CH2 + C+} & $f(\bar{\epsilon})$  \\
10 & \ce{e + C2H2 -> 2e + C2H2+} & $f(\bar{\epsilon})$  \\
11 & \ce{e + C2H2 -> 2e + H + C2H+} & $f(\bar{\epsilon})$  \\
12 & \ce{e + C2H2 -> 2e + H2 + C2+} & $f(\bar{\epsilon})$  \\
13 & \ce{e + C2H4 -> 2e + 2H2 + C2+} & $f(\bar{\epsilon})$ \\
14 & \ce{e + C2H4 -> 2e + CH + CH3+} & $f(\bar{\epsilon})$  \\
15 & \ce{e + C2H4 -> 2e + CH2 + CH2+} & $f(\bar{\epsilon})$  \\
16 & \ce{e + C2H4 -> 2e + CH2 + H2 + C+} & $f(\bar{\epsilon})$ \\
17 & \ce{e + C2H4 -> 2e + CH3 + CH+} & $f(\bar{\epsilon})$  \\
18 & \ce{e + C2H4 -> 2e + C2H2 + H2+} & $f(\bar{\epsilon})$  \\
19 & \ce{e + C2H4 -> 2e + C2H3 + H+} & $f(\bar{\epsilon})$  \\
20 & \ce{e + C2H4 -> 2e + C2H4+} & $f(\bar{\epsilon})$  \\
21 & \ce{e + C2H4 -> 2e + H + C2H3+} & $f(\bar{\epsilon})$  \\
22 & \ce{e + C2H4 -> 2e + H + H2 + C2H+} & $f(\bar{\epsilon})$  \\
23 & \ce{e + C2H4 -> 2e + H2 + C2H2+} & $f(\bar{\epsilon})$ \\
24 & \ce{e + H2 -> e + 2H} & $f(\bar{\epsilon})$  \\
25 & \ce{e + N2(A) -> 2e + N2(A)+} & $f(\bar{\epsilon})$ \\
26 & \ce{e + N2(A) -> 2e + N2(B)+} & $f(\bar{\epsilon})$ \\
27 & \ce{e + N2(A) -> 2e + N2(X)+} & $f(\bar{\epsilon})$ \\
28 & \ce{e + N2(A) -> e + 2N(S)} & $f(\bar{\epsilon})$  \\
29 & \ce{e + N2(A) -> e + N2(B)} & $f(\bar{\epsilon})$  \\
30 & \ce{e + N2(A) -> e + N2(X)} & $f(\bar{\epsilon})$  \\
31 & \ce{e + N2(A) -> e + N2(a)} & $f(\bar{\epsilon})$ \\
32 & \ce{e + N2}(A) $\longrightarrow$ e + N2(a') & $f(\bar{\epsilon})$  \\
33 & \ce{e + N2(A)+ -> 2N(S)} & $f(\bar{\epsilon})$  \\
34 & \ce{e + N2(A)+ -> e + N2(B)+} & $f(\bar{\epsilon})$ \\
35 & \ce{e + N2(A)+ -> e + N2(X)+} & $f(\bar{\epsilon})$ \\
36 & \ce{e + N2(B) -> 2e + N2(A)+} & $f(\bar{\epsilon})$ \\
37 & \ce{e + N2(B) -> 2e + N2(B)+} & $f(\bar{\epsilon})$ \\
38 & \ce{e + N2(B) -> 2e + N2(X)+} & $f(\bar{\epsilon})$ \\
39 & \ce{e + N2(B) -> e + 2N(S)} & $f(\bar{\epsilon})$  \\
40 & \ce{e + N2(B) -> e + N2(A)} & $f(\bar{\epsilon})$  \\
41 & \ce{e + N2(B) -> e + N2(X)} & $f(\bar{\epsilon})$  \\
42 & \ce{e + N2(B) -> e + N2(a)} & $f(\bar{\epsilon})$  \\
43 & \ce{e + N2}(B) $\longrightarrow$ e + N2(a') & $f(\bar{\epsilon})$ \\
44 & \ce{e + N2(B)+ -> 2N(S)} & $f(\bar{\epsilon})$  \\
45 & \ce{e + N2(B)+ -> e + N2(A)+} & $f(\bar{\epsilon})$ \\
46 & \ce{e + N2(B)+ -> e + N2(X)+} & $f(\bar{\epsilon})$ \\
47 & \ce{e + N2}(B') $\longrightarrow$ e + N2(X) & $f(\bar{\epsilon})$  \\
48 & \ce{e + N2(C) -> e + N2(X)} & $f(\bar{\epsilon})$  \\
49 & \ce{e + N2(E) -> e + N2(X)} & $f(\bar{\epsilon})$  \\
50 & \ce{e + N2(W) -> e + N2(X)} & $f(\bar{\epsilon})$  \\
51 & \ce{e + N2(X) -> 2e + N(S)+N+} & $f(\bar{\epsilon})$ \\
52 & \ce{e + N2(X) -> 2e + N2(A)+} & $f(\bar{\epsilon})$ \\
53 & \ce{e + N2(X) -> 2e + N2(B)+} & $f(\bar{\epsilon})$  \\
54 & \ce{e + N2(X) -> 2e + N2(X)+} & $f(\bar{\epsilon})$  \\
55 & \ce{e + N2(X) -> e + 2N(S)} & $f(\bar{\epsilon})$  \\
56 & \ce{e + N2(X) -> e + N2(A)} & $f(\bar{\epsilon})$ \\
57 & \ce{e + N2(X) -> e + N2(B)} & $f(\bar{\epsilon})$  \\
58 & \ce{e + N2}(X) $\longrightarrow$ e + N2(B') & $f(\bar{\epsilon})$  \\
59 & \ce{e + N2(X) -> e + N2(C)} & $f(\bar{\epsilon})$  \\
60 & \ce{e + N2(X) -> e + N2(E)} & $f(\bar{\epsilon})$  \\
61 & \ce{e + N2(X) -> e + N2(W)} & $f(\bar{\epsilon})$ \\
62 & \ce{e + N2(X) -> e + N2(X)} & $f(\bar{\epsilon})$  \\
63 & \ce{e + N2(X) -> e + N2(a)} & $f(\bar{\epsilon})$  \\
64 & \ce{e + N2(X)} $\longrightarrow$ e + N2(a') & $f(\bar{\epsilon})$  \\
65 & \ce{e + N2(X)} $\longrightarrow$ e + N2(a'') & $f(\bar{\epsilon})$  \\
66 & \ce{e + N2(X) -> e + N2(w1)} & $f(\bar{\epsilon})$  \\
67 & \ce{e + N2(X) -> 2N(S)} & $f(\bar{\epsilon})$  \\
68 & \ce{e + N2(X)+ -> e + N2(A)+} & $f(\bar{\epsilon})$ \\
69 & \ce{e + N2(X)+ -> e + N2(B)+} & $f(\bar{\epsilon})$  \\
70 & \ce{e + N2(a) -> 2e + N2(A)+} & $f(\bar{\epsilon})$ \\
71 & \ce{e + N2(a) -> 2e + N2(B)+} & $f(\bar{\epsilon})$ \\
72 & \ce{e + N2(a) -> 2e + N2(X)+} & $f(\bar{\epsilon})$ \\
73 & \ce{e + N2(a) -> e + 2N(S)} & $f(\bar{\epsilon})$  \\
74 & \ce{e + N2(a) -> e + N2(A)} & $f(\bar{\epsilon})$ \\
75 & \ce{e + N2(a) -> e + N2(B)} & $f(\bar{\epsilon})$  \\
76 & \ce{e + N2(a) -> e + N2(X)} & $f(\bar{\epsilon})$  \\
77 & \ce{e + N2(a) -> 2e + N2(A)+} & $f(\bar{\epsilon})$ \\
78 & \ce{e + N2(a) -> 2e + N2(B)+} & $f(\bar{\epsilon})$ \\
79 & \ce{e + N2(a) -> 2e + N2(X)+} & $f(\bar{\epsilon})$  \\
80 & \ce{e + N2}(a') $\longrightarrow$ \ce{e + 2N(S)} & $f(\bar{\epsilon})$  \\
81 & \ce{e + N2}(a') $\longrightarrow$ e + N2(A) & $f(\bar{\epsilon})$  \\
82 & \ce{e + N2}(a') $\longrightarrow$ e + N2(B) & $f(\bar{\epsilon})$  \\
83 & \ce{e + N2}(a') $\longrightarrow$ e + N2(X) & $f(\bar{\epsilon})$  \\
84 & \ce{e + N2}(a'') $\longrightarrow$ \ce{e + N2(X)} & $f(\bar{\epsilon})$  \\
85 & \ce{e + N2(w1) -> e + N2(X)} & $f(\bar{\epsilon})$  \\
86 & \ce{2N2(X) + N+ -> N2(X) + N3+} & $1.8 \times 10^{-29} \text{ cm}^6\text{s}^{-1}$  \\
87 & \ce{C + HC3N -> H + C4N} & $1 \times 10^{-10} \text{ cm}^3\text{s}^{-1}$  \\
88 & \ce{C4H2 + C2H5+ -> C2H4 + C4H3+} & $3.0 \times 10^{-9} \text{ cm}^3\text{s}^{-1}$  \\
89 & \ce{C4H2 + C4H3+ -> C2H2 + C6H3+} & $7.4 \times 10^{-10} \text{ cm}^3\text{s}^{-1}$  \\
90 & \ce{C4H2 + C5H5+ -> C2H2 + C7H5+} & $2.2 \times 10^{-10} \text{ cm}^3\text{s}^{-1}$  \\
91 & \ce{C4H2 + HCNH+ -> HCN + C4H3+} & $1.8 \times 10^{-9} \text{ cm}^3\text{s}^{-1}$  \\
92 & \ce{C6H2 + C4H3+ -> C2H4 + C6H3+} & $3.0 \times 10^{-9} \text{ cm}^3\text{s}^{-1}$  \\
93 & \ce{C6H2 + C4H3+ -> C2H2 + C8H3+} & $7.4 \times 10^{-10} \text{ cm}^3\text{s}^{-1}$  \\
94 & \ce{C6H2 + HCNH+ -> HCN + C6H3+} & $3.0 \times 10^{-9} \text{ cm}^3\text{s}^{-1}$  \\
95 & \ce{C8H2 + C2H5+ -> C2H4 + C8H3+} & $3.0 \times 10^{-9} \text{ cm}^3\text{s}^{-1}$  \\
96 & \ce{C8H2 + HCNH+ -> HCN + C8H3+} & $3.0 \times 10^{-9} \text{ cm}^3\text{s}^{-1}$  \\
97 & \ce{CH + CH4 -> C2H4 + H} & $(1.06 \times 10^{-10} \text{ cm}^3\text{s}^{-1}) \cdot (T/1[K])^{-1.04} \text{exp}[-36.1[K]/T]$  \\
98 & \ce{CH + N+ -> H + CN+} & $(6.6 \times 10^{-10} \text{ cm}^3\text{s}^{-1}) \cdot [0.62 + 1.59(300[K]/T)^{0.5}]$  \\
99 & \ce{CH2 + CH2 -> C2H4} & $1.7 \times 10^{-13} \text{ cm}^3\text{s}^{-1}$  \\
100 & \ce{CH2 + CH4 -> CH3 + CH3} & $(2.14 \times 10^{-11} \text{ cm}^3\text{s}^{-1}) \cdot (T/1[K])^{0.5}$  \\
101 & \ce{CH2 + C+ -> C + CH2+} & $(6.7 \times 10^{-10} \text{ cm}^3\text{s}^{-1}) \cdot [0.62 + 0.67(300[K]/T)^{0.5}]$  \\
102 & \ce{CH2 + H2 -> CH3 + H} & $(3.34 \times 10^{-11} \text{ cm}^3\text{s}^{-1}) \cdot (T/1[K])^{0.5}$  \\
103 & \ce{CH2NH + C2H5+ -> C2H4 + CH2NH2+} & $3.0 \times 10^{-9} \text{ cm}^3\text{s}^{-1}$  \\
104 & \ce{CH2NH + HCNH+ -> HCN + CH2NH2+} & $3.0 \times 10^{-9} \text{ cm}^3\text{s}^{-1}$  \\
105 & \ce{CH3 + CH2 -> C2H4 + H} & $7 \times 10^{-11} \text{ cm}^3\text{s}^{-1}$  \\
106 & \ce{CH3 + CH3 -> C2H5 + H} & $1.3 \times 10^{-11} \text{ cm}^3\text{s}^{-1}~\cdot exp[-13275[K]/T]$  \\
107 & \ce{CH3 + CH3 -> C2H6} & $(4 \times 10^{-10} \text{ cm}^3\text{s}^{-1}) \cdot (T/1[K])^{-0.4}$  \\
108 & \ce{CH3CN + CH3+ -> C2H5CNH+ } & $9.0 \times 10^{-11} \text{ cm}^3\text{s}^{-1}$  \\
109 & \ce{CH3CN + C2H5+ -> C2H4 + CH3CNH+} & $3.8 \times 10^{-9} \text{ cm}^3\text{s}^{-1}$  \\
110 & \ce{CH3CN + HCNH+ -> HCN + CH3CNH+} & $3.8 \times 10^{-9} \text{ cm}^3\text{s}^{-1}$  \\
111 & \ce{CH3NH2 + CH3+ -> CH4 + CH2NH2+} & $1.4 \times 10^{-9} \text{ cm}^3\text{s}^{-1}$  \\
112 & \ce{CH3NH2 + C2H5+ -> C2H4 + CH3NH3+} & $1.9 \times 10^{-9} \text{ cm}^3\text{s}^{-1}$  \\
113 & \ce{CH3NH2 + HCNH+ -> HCN + CH3NH3+} & $3.0 \times 10^{-9} \text{ cm}^3\text{s}^{-1}$  \\
114 & \ce{CH3NH2 + N2(X)+ -> N2(X) + H + CH2NH2+} & $8.8 \times 10^{-10} \text{ cm}^3\text{s}^{-1}$  \\
115 & \ce{CH4 + CH2+ -> H + C2H5+} & $3.9 \times 10^{-10} \text{ cm}^3\text{s}^{-1}$  \\
116 & \ce{CH4 + CH2+ -> H2 + C2H4+} & $9.1 \times 10^{-10} \text{ cm}^3\text{s}^{-1}$  \\
117 & \ce{CH4 + CH3 -> C2H5 + H2} & $(1.7 \times 10^{-11} \text{ cm}^3\text{s}^{-1}) \cdot \text{exp}[-11500[K]/T]$  \\
118 & \ce{CH4 + CH3+ -> H2 + C2H5+} & $1.1 \times 10^{-9} \text{ cm}^3\text{s}^{-1}$  \\
119 & \ce{CH4 + CH4+ -> CH3 + CH5+} & $1.1 \times 10^{-9} \text{ cm}^3\text{s}^{-1}$  \\
120 & \ce{CH4 + C2H2+ -> H2 + C3H4+} & $1.87 \times 10^{-10} \text{ cm}^3\text{s}^{-1}$  \\
121 & \ce{CH4 + C2H3+ -> H2 + C3H5+} & $1.9 \times 10^{-10} \text{ cm}^3\text{s}^{-1}$  \\
122 & \ce{CH4 + C2H+ -> H2 + clC3H3+} & $1.87 \times 10^{-10} \text{ cm}^3\text{s}^{-1}$  \\
123 & \ce{CH4 + C2H+ -> H2 + lC3H3+} & $1.87 \times 10^{-10} \text{ cm}^3\text{s}^{-1}$  \\
124 & \ce{CH4 + H -> H2 + CH3} & $(2.2 \times 10^{-20} \text{ cm}^3\text{s}^{-1}) \cdot \text{exp}[-4045[K]/T]$  \\
125 & \ce{CH4 + HCN+ -> CH3 + HCNH+} & $1.1 \times 10^{-9} \text{ cm}^3\text{s}^{-1} $  \\
126 & \ce{CH4 + N2(X)+ -> N2(X) + H + CH3+} & $1.3 \times 10^{-9} \text{ cm}^3\text{s}^{-1}$  \\
127 & \ce{CH4 + N2H+ -> N2(X) + CH5+} & $8.9 \times 10^{-10} \text{ cm}^3\text{s}^{-1}$  \\
128 & \ce{CH4 + N+ -> H2 + HCNH+} & $4.1 \times 10^{-10} \text{ cm}^3\text{s}^{-1}$  \\
129 & \ce{CN + CH3CN -> C2N2 + H} & $(6.46 \times 10^{-11} \text{ cm}^3\text{s}^{-1}) \cdot \text{exp}[-1190[K]/T]$  \\
130 & \ce{CN + C2H2 -> HC3N + H} & $2.71 \times 10^{-9} \text{ cm}^3\text{s}^{-1} \cdot (T/1[K])^{-0.52}~\cdot~ \text{exp}[-20[K]/T]$  \\
131 & \ce{CN + HCN -> C2N2 + H} & $4.3 \times 10^{-13} \text{ cm}^3\text{s}^ \cdot (T/1[K])^{1.71}~\cdot~ \text{exp}[-770[K]/T]$  \\
132 & \ce{CN + HC3N -> C4N2 + H} & $1.7 \times 10^{-11} \text{ cm}^3\text{s}^{-1}~\cdot~(T/300[K])^{-0.67}$  \\
133 & \ce{C2H + C+ -> H + C3+} & $1.7 \times 10^{-11} \text{ cm}^3\text{s}^{-1}~\cdot [0.62 + 0.64(300[K]/T)^{0.5}]$  \\
134 & \ce{C2H2 + C4H2+ -> C6H4+} & $2.7 \times 10^{-10} \text{ cm}^3\text{s}^{-1}$  \\
135 & \ce{C2H2 + CH3+ -> H2 + clC3H3+ } & $2.9 \times 10^{-10} \text{ cm}^3\text{s}^{-1}$  \\
136 & \ce{C2H2 + CH3+ -> H2 + lC3H3+} & $2.9 \times 10^{-9} \text{ cm}^3\text{s}^{-1}$  \\
137 & \ce{C2H2 + CH5+ -> CH4 + C2H3+} & $1.5 \times 10^{-9} \text{ cm}^3\text{s}^{-1}$  \\
138 & \ce{C2H2 + CH+ -> H + C3H2+} & $1.2 \times 10^{-9} \text{ cm}^3\text{s}^{-1}$  \\
139 & \ce{C2H2 + CN -> H + HC3N} & ($2.7 \times 10^{-10}) \text{ cm}^3\text{s}^{-1} ~\cdot~(T/300[K])^{-0.52}~\textbf{exp}[-20[K]/T]$  \\
140 & \ce{C2H2 + C2H2+ -> H + C4H3+} & $9.52 \times 10^{-10} \text{ cm}^3\text{s}^{-1}$  \\
141 & \ce{C2H2 + C2H2+ -> H2 + C4H2+} & $4.5 \times 10^{-10} \text{ cm}^3\text{s}^{-1}$  \\
142 & \ce{C2H2 + C2H4+ -> CH3 + clC3H3+} & $6.5 \times 10^{-10} \text{ cm}^3\text{s}^{-1}$  \\
143 & \ce{C2H2 + C2H4+ -> CH3 + lC3H3+} & $6.5 \times 10^{-10} \text{ cm}^3\text{s}^{-1}$  \\
144 & \ce{C2H2 + C2H4+ -> H + C4H5+} & $(1.93 \times 10^{-10} \text{ cm}^3\text{s}^{-1}) \cdot \text{exp}[-11500[K]/T]$  \\
145 & \ce{C2H2 + C2H5+ -> CH4 + clC3H3+} & $6.8 \times 10^{-11} \text{ cm}^3\text{s}^{-1}$  \\
146 & \ce{C2H2 + C2H5+ -> CH4 + lC3H3+} & $6.8 \times 10^{-11} \text{ cm}^3\text{s}^{-1}$  \\
147 & \ce{C2H2 + C2N2+ -> H + C4N2+} & $7.0 \times 10^{-11} \text{ cm}^3\text{s}^{-1}$  \\
148 & \ce{C2H2 + C4N2+ -> H2 + C4N2+} & $3.0 \times 10^{-11} \text{ cm}^3\text{s}^{-1}$  \\
149 & \ce{C2H2 + C3H5+ -> H2 + C5H5+} & $3.8 \times 10^{-10} \text{ cm}^3\text{s}^{-1}$  \\
150 & \ce{C2H2 + C4H3+ -> C6H5+} & $2.2 \times 10^{-10} \text{ cm}^3\text{s}^{-1}$  \\
151 & \ce{C2H6 + C6H5+ -> C8H5+} & $2.3 \times 10^{-10} \text{ cm}^3\text{s}^{-1}$  \\
152 & \ce{C2H3 + C2H4+ -> H2 + C4H5+} & $(5.2 \times 10^{-10} \text{ cm}^3\text{s}^{-1}) \cdot (T/300[\text{K}])^{-0.5}$  \\
153 & \ce{C2H3CN + C2H5+ -> C2H4 + C2H3CNH+} & $3.0 \times 10^{-9} \text{ cm}^3\text{s}^{-1}$  \\
154 & \ce{C2H3CN + HCNH+ -> HCN + C2H3CNH+} & $4.5 \times 10^{-9} \text{ cm}^3\text{s}^{-1}$  \\
155 & \ce{C2H4 + CH5+ -> CH4 + C2H5+} & $1.5 \times 10^{-9} \text{ cm}^3\text{s}^{-1}$  \\
156 & \ce{C2H4 + C2H5+ -> CH4 + C2H5+} & $1.5 \times 10^{-9} \text{ cm}^3\text{s}^{-1}$  \\
157 & \ce{C2H4 + C2H2+ -> H + C4H5+} & $3.04 \times 10^{-10} \text{ cm}^3\text{s}^{-1}$  \\
158 & \ce{C2H4 + C2H4+ -> H + C4H7+} & $7.47 \times 10^{-11} \text{ cm}^3\text{s}^{-1}$  \\
159 & \ce{C2H4 + C2H5+ -> CH4 + C3H5+} & $3.6 \times 10^{-10} \text{ cm}^3\text{s}^{-1}$  \\
160 & \ce{C2H4 + C2H2+ -> H + H2C4H2+} & $2.91 \times 10^{-10} \text{ cm}^3\text{s}^{-1}$  \\
161 & \ce{C2H4 + C2H+ -> H2 + C4H3+} & $1.42 \times 10^{-9} \text{ cm}^3\text{s}^{-1}$  \\
162 & \ce{C2H4 + C3H5+ -> C5H9+} & $5.10 \times 10^{-11} \text{ cm}^3\text{s}^{-1}$ \\
163 & \ce{C2H4 + C3H5+ -> H2 + C5H7+} & $1.19 \times 10^{-10} \text{ cm}^3\text{s}^{-1}$  \\
164 & \ce{C2H4 + C4H5+ -> H2 + C6H7+} & $1.2 \times 10^{-10} \text{ cm}^3\text{s}^{-1}$ \\
165 & \ce{C2H4 + C4H5+ -> H2 + C6H7+} & $6.3 \times 10^{-11} \text{ cm}^3\text{s}^{-1}$  \\
166 & \ce{C2H4 + HC3NH+ -> H + C4H5NH+} & $1.3 \times 10^{-9} \text{ cm}^3\text{s}^{-1}$  \\
167 & \ce{C2H4 + lC3H3+ -> C5H7+} & $5.5 \times 10^{-10} \text{ cm}^3\text{s}^{-1}$  \\
168 & \ce{C2H4 + lC3H3+ -> H2 + C5H5+} & $5.5 \times 10^{-10} \text{ cm}^3\text{s}^{-1}$  \\
169 & \ce{C2H5 + H -> CH3 + CH3} & $(7.95 \times 10^{-11} \text{ cm}^3\text{s}^{-1}) \cdot \exp[-127[\text{K}]/T]$  \\
170 & \ce{C2H5 + H -> C2H6} & $6 \times 10^{-11} \text{ cm}^3\text{s}^{-1}$  \\
171 & \ce{C2H5CN + C2H5+ -> C2H4 + C2H5CNH+} & $3.0 \times 10^{-9} \text{ cm}^3\text{s}^{-1}$  \\
172 & \ce{C2H5CN + HCNH+ -> HCN + C2H5CNH+} & $3.0 \times 10^{-9} \text{ cm}^3\text{s}^{-1}$  \\
173 & \ce{C2H6 + CH3 -> C2H5 + CH4} & $(2.5 \times 10^{-31}\cdot (T/1[K])^6 \text{ cm}^3\text{s}^{-1}) \cdot \exp[-3750[\text{K}]/T]$ \\
174 & \ce{C2H6 + H -> C2H5 + H2} & $(2.5 \times 10^{-15}\cdot (T/1[K])^6 \text{ cm}^3\text{s}^{-1}) \cdot \exp[-3750[\text{K}]/T]$  \\
175 & \ce{C2H6 + H2+ -> H2 + C2H6+} & $2.94 \times 10^{-10} \text{ cm}^3\text{s}^{-1}$  \\
176 & \ce{C3H4 + C3H4+ -> C2H2 + C4H6+} & $2.2 \times 10^{-11} \text{ cm}^3\text{s}^{-1}$  \\
177 & \ce{C4H3N + C2H5+ -> C2H4 + C4H3NH+} & $3.0 \times 10^{-9} \text{ cm}^3\text{s}^{-1}$  \\
178 & \ce{C4H3N + HCNH+ -> HCN + C4H3NH+} & $3.0 \times 10^{-9} \text{ cm}^3\text{s}^{-1}$  \\
179 & \ce{C4N + CH4+ -> H2 + C5H2N+} & $1 \times 10^{-10} \text{ cm}^3\text{s}^{-1}$  \\
180 & \ce{C6H6 + C2H5+ -> C2H4 + C6H7+} & $3.0 \times 10^{-9} \text{ cm}^3\text{s}^{-1}$  \\
181 & \ce{C6H6 + HCNH+ -> HCN + C6H7+} & $3.0 \times 10^{-9} \text{ cm}^3\text{s}^{-1}$ \\
182 & \ce{C7H4 + C2H5+ -> C2H4 + C7H5+} & $3.0 \times 10^{-9} \text{ cm}^3\text{s}^{-1}$  \\
183 & \ce{C7H4 + HCNH+ -> HCN + C7H5+} & $3.0 \times 10^{-9} \text{ cm}^3\text{s}^{-1}$  \\
184 & \ce{C7H8 + C2H5+ -> C2H4 + C7H9+} & $3.0 \times 10^{-9} \text{ cm}^3\text{s}^{-1}$  \\
185 & \ce{C7H8 + HCNH+ -> HCN + C7H9+} & $3.0 \times 10^{-9} \text{ cm}^3\text{s}^{-1}$  \\
186 & \ce{H + C2H2+ -> CN + HCN+} & $4.96 \times 10^{-10} \text{ cm}^3\text{s}^{-1}$ \\
187 & \ce{H + C2H5 -> H2 + C2H4} & $3 \times 10^{-12} \text{ cm}^3\text{s}^{-1}$  \\
188 & \ce{H + C2H5+ -> H2 + C2H4+} & $1 \times 10^{-11} \text{ cm}^3\text{s}^{-1}$ \\
189 & \ce{HCNH+ + CH5+ -> CH4 + HCNH+} & $3.0 \times 10^{-9} \text{ cm}^3\text{s}^{-1}$  \\
190 & \ce{HCN + CN+ -> H + C2N2+} & $3.15 \times 10^{-10} \text{ cm}^3\text{s}^{-1}$  \\
191 & \ce{HCN + C+ -> HC3N+ + H} & $(5.3 \times 10^{-12} \text{ cm}^3\text{s}^{-1}) \cdot \exp[-770[\text{K}]/T]$  \\
192 & \ce{HCN + C2H5+ -> C2H4 + HCNH+} & $2.7 \times 10^{-9} \text{ cm}^3\text{s}^{-1}$  \\
193 & \ce{HCN + C3H5+ -> C4H5NH+} & $5.0 \times 10^{-11} \text{ cm}^3\text{s}^{-1}$  \\
194 & \ce{HCN + N2(X+) -> N2(X) + HCN+} & $1.0 \times 10^{-9} \text{ cm}^3\text{s}^{-1}$  \\
195 & \ce{HCN + lC3H3+ -> C4H3NH+} & $4.8 \times 10^{-10} \text{ cm}^3\text{s}^{-1}$  \\
196 & \ce{HCNH + H -> HCN + H2} & $(2.9 \times 10^{-11} \text{ cm}^3\text{s}^{-1}) \cdot (T/1[\text{K}])^{0.5}$  \\
197 & \ce{HC3N + C2H2+ -> C5H3N+} & $(2 \times 10^{-12} \text{ cm}^3\text{s}^{-1}) \cdot (T/300[\text{K}])^{-0.25}$  \\
198 & \ce{HC3N + C2H5+ -> C2H4 + HC3NH+} & $3.6 \times 10^{-9} \text{ cm}^3\text{s}^{-1}$  \\
199 & \ce{HC3N + HCNH+ -> HCN + HC3NH+} & $3.4 \times 10^{-9} \text{ cm}^3\text{s}^{-1}$  \\
200 & \ce{H2 + C2H5 -> C2H6 + H} & $(4.1 \times 10^{-15} \text{ cm}^3\text{s}^{-1}) \cdot (T/300[\text{K}])^{3.6} \exp[-4240[\text{K}]/T]$  \\
201 & \ce{H2 + C+ -> CH2+} & $(2.0 \times 10^{-16} \text{ cm}^3\text{s}^{-1}) \cdot (T/300[\text{K}])^{-1.3} \exp[-23[\text{K}]/T]$  \\
202 & \ce{H2 + N2(X+) -> H + N2H+} & $2.0 \times 10^{-9} \text{ cm}^3\text{s}^{-1}$  \\
203 & \ce{N(S) + CH2 -> CN + 2H} & $1.6 \times 10^{-11} \text{ cm}^3\text{s}^{-1}$  \\
204 & \ce{N(S) + CH2 -> CN + H2} & $1.6 \times 10^{-11} \text{ cm}^3\text{s}^{-1}$  \\
205 & \ce{N(S) + CH2 -> HCN + H} & $(5 \times 10^{-11} \text{ cm}^3\text{s}^{-1}) \cdot \exp[-250[\text{K}]/T]$  \\
206 & \ce{N(S) + CH3 -> HCN + H2} & $1.4 \times 10^{-11} \text{ cm}^3\text{s}^{-1}$  \\
207 & \ce{N(S) + C2H5+ -> HCN + CH+} & $2.5 \times 10^{-11} \text{ cm}^3\text{s}^{-1}$  \\
208 & \ce{N(S) + C3H4+ -> H2 + HC3NH+} & $1.00 \times 10^{-11} \text{ cm}^3\text{s}^{-1}$  \\
209 & \ce{N(S) + HCNH -> HCN + NH} & $6.7 \times 10^{-11} \text{ cm}^3\text{s}^{-1}$  \\
210 & \ce{N(S) + NH -> N2(X) + H} & $(8.6 \times 10^{-12} \text{ cm}^3\text{s}^{-1}) \cdot (T/1[\text{K}])^{0.5}$  \\
211 & \ce{N(S) + N2(X+) + N2(X+) -> N2(X) + N3+} & $(0.9 \times 10^{-29} \text{ cm}^6\text{s}^{-1}) \cdot \exp[400[\text{K}]/T]$  \\
212 & \ce{N(S) + N2(X+) -> N2(X) + N+} & $(7.2 \times 10^{-13} \text{ cm}^3\text{s}^{-1}) \cdot \exp[T/300[\text{K}]]$  \\
213 & \ce{N(S) + N3+ -> N2(X) + N2(X)+} & $6.6 \times 10^{-11} \text{ cm}^3\text{s}^{-1}$ \\
214 & \ce{NH + H -> N(S) + H2} & $(1.7 \times 10^{-12} \text{ cm}^3\text{s}^{-1}) \cdot (T/1[\text{K}])^{0.68} \exp[-950[\text{K}]/T]$  \\
215 & \ce{NH3 + CH3+ -> H2 + CH2NH2+} & $1.5 \times 10^{-9} \text{ cm}^3\text{s}^{-1}$  \\
216 & \ce{NH3 + C2H5+ -> C2H4 + NH4+} & $2.1 \times 10^{-9} \text{ cm}^3\text{s}^{-1}$ \\
217 & \ce{NH3 + HCNH+ -> HCN + NH4+} & $2.3 \times 10^{-9} \text{ cm}^3\text{s}^{-1}$  \\
218 & \ce{N2(A) + CH4 -> N2(X) + CH4} & $3.2 \times 10^{-15} \text{ cm}^3\text{s}^{-1}$  \\
219 & \ce{N2(A) + H -> N2(X) + H} & $2.1 \times 10^{-10} \text{ cm}^3\text{s}^{-1}$  \\
220 & \ce{N2(A) + H2 -> N2(X) + 2H} & $2.4 \times 10^{-15} \text{ cm}^3\text{s}^{-1}$  \\
221 & \ce{N2(A) + N(S) -> N2(X) + N(S)} & $5 \times 10^{-11} \text{ cm}^3\text{s}^{-1}$  \\
222 & \ce{N2(A) + N2(X) -> 2N2(X)} & $3 \times 10^{-16} \text{ cm}^3\text{s}^{-1}$  \\
223 & \ce{N2(A) + N2(a{'}) -> N2(X) + e + N2(X)+} & $1.5 \times 10^{-11} \text{ cm}^3\text{s}^{-1}$ \\
224 & \ce{N2(A) + N2(a{'}) -> e + N4+} & $1.5 \times 10^{-11} \text{ cm}^3\text{s}^{-1}$  \\
225 & \ce{N2(B) -> N2(A)} & $6.25 \times 10^4 \text{ s}^{-1}$  \\
226 & \ce{N2(B) + CH4 -> N2(A) + CH4} & $2.85 \times 10^{-10} \text{ cm}^3\text{s}^{-1}$  \\
227 & \ce{N2(B) + CH4 -> N2(X) + H4} & $1.5 \times 10^{-11} \text{ cm}^3\text{s}^{-1}$  \\
228 & \ce{N2(B) + H2 -> N2(A) + H2} & $2.4 \times 10^{-11} \text{ cm}^3\text{s}^{-1}$  \\
229 & \ce{N2(B) + N2(X) -> 2N2(X)} & $3 \times 10^{-11} \text{ cm}^3\text{s}^{-1}$ \\
230 & \ce{N2(C) -> N2(B)} & $1.39 \times 10^6 \text{ s}^{-1}$  \\
231 & \ce{N2(C) + N2(X) -> N2(a{'}) + N2(X)} & $1.0 \times 10^{-11} \text{ cm}^3\text{s}^{-1}$  \\
232 & \ce{N2(X) + N2(A)+ -> N(S) + N3+} & $5.5 \times 10^{-11} \text{ cm}^3\text{s}^{-1}$  \\
233 & \ce{N2(X) + N2(B)+ -> N(S) + N3+} & $5.5 \times 10^{-11} \text{ cm}^3\text{s}^{-1}$  \\
234 & \ce{N2(X) + N4+ -> 2N2(X) + N2(X)+} & $(2.1 \times 10^{-16} \text{ cm}^3\text{s}^{-1}) \cdot \exp[T/121[\text{K}]]$  \\
235 & \ce{N2(a) -> N2(X)} & $1.8 \times 10^4 \text{ s}^{-1}$  \\
236 & \ce{N2(a) -> N2(a{'})} & $1.91 \times 10^2 \text{ s}^{-1}$  \\
237 & \ce{N2(a) + CH4 -> N2(X) + CH4} & $5.2 \times 10^{-10} \text{ cm}^3\text{s}^{-1}$ \\
238 & \ce{N2(a) + N2(X) -> N2(a{'}) + N2(X)} & $2.0 \times 10^{-11} \text{ cm}^3\text{s}^{-1}$  \\
239 & \ce{N2(a{'}) + CH4 -> N2(X) + CH4} & $3.0 \times 10^{-10} \text{ cm}^3\text{s}^{-1}$  \\
240 & \ce{N2(a{'}) + H -> N2(X) + H} & $2.1 \times 10^{-10} \text{ s}^{-1}$  \\
241 & \ce{N2(a{'}) + H2 -> N2(X) + 2H} & $2.6 \times 10^{-11} \text{ s}^{-1}$ \\
242 & \ce{N2(a{'}) + N2(X) -> N2(B) + N2(X)} & $1.9 \times 10^{-13} \text{ cm}^3\text{s}^{-1}$  \\
243 & \ce{N2(a{'}) + N2(a{'}) -> N2(X) + e + N2(X)+} & $1.0 \times 10^{-11} \text{ cm}^3\text{s}^{-1}$  \\
244 & \ce{N2H4 + C2H5+ -> C2H4 + N2H5+} & $3.0 \times 10^{-9} \text{ cm}^3\text{s}^{-1}$  \\
245 & \ce{N2H4 + HCNH+ -> HCN + N2H5+} & $3.0 \times 10^{-9} \text{ cm}^3\text{s}^{-1}$  \\
246 & \ce{e + N2(X)+ -> 2N(S)} & $4.8 \times 10^{-7} \text{ cm}^3\text{s}^{-1}~\cdot~ (300[K]/T_e)^{0.5}$  \\
247 & \ce{e + N4+ -> 2N2(X)} & $2.0 \times 10^{-6} \text{ cm}^3\text{s}^{-1}~\cdot~ (300[K]/T_e)^{0.5}$  \\
248 & \ce{e + C4H7+ -> CH4 + clC3H3} & $1.95 \times 10^{-7} \text{ cm}^3\text{s}^{-1}~\cdot~ (300[K]/T_e)^{0.5}$  \\
249 & \ce{e + C6H7+ -> H + C6H6} & $(5 \times 10^{-7} \text{ cm}^3\text{s}^{-1})~\cdot~ (300[K]/T_e)^{0.5}$  \\
250 & \ce{C2H2 + e \longrightarrow CH+ + CH + 2e} & collis. cross section  \\
251 & \ce{C2H4 + e \longrightarrow C2H4+ + 2e} & collis. cross section  \\
252 & \ce{C2H6 + e \longrightarrow C2H6+ + 2e} & collis. cross section  \\
253 & \ce{C2H+ + C2H2 \longrightarrow C4H2+ + H} & $1.85 \times 10^{-9}$ \\
254 & \ce{C6H2+ + C2H2 \longrightarrow C8H4+} & $4.3 \times 10^{-11}$ \\
255 & \ce{C6H5+ + C2H2 \longrightarrow C8H6+ + H} & $7.8 \times 10^{-11}$  \\
256 & \ce{C8H6+ + C2H2 \longrightarrow C10H6+ + H2} & $5.0 \times 10^{-11}$  \\
257 & \ce{C10H6+ + C2H2 \longrightarrow C12H6+ + H2} & $5.0 \times 10^{-11}$  \\
258 & \ce{C2H2 + C4H2 \longrightarrow C6H4+} & $5.0 \times 10^{-11}$  \\
259 & \ce{C2H2 + C4H3+ \longrightarrow C6H5+} & $2.9 \times 10^{-11}$  \\
260 & \ce{N + CH3 \longrightarrow HCN + H2} & $1.4 \times 10^{-11}$  \\
261 & \ce{N + CH2 \longrightarrow HCN + H} & $5.0 \times 10^{-11}$  \\
262 & \ce{N + H2CN \longrightarrow HCN + NH} & $6.7 \times 10^{-11}$  \\
263 & \ce{CN + CH4 \longrightarrow HCN + CH3} & $2.9 \times 10^{-11}$  \\
264 & \ce{CN + C2H6 \longrightarrow HCN + C2H5} & $1.8 \times 10^{-11}$ \\
265 & \ce{H2CN + H \longrightarrow HCN + H2} & $2.9 \times 10^{-11}$  \\
266 & \ce{N(^2D) + CH4 \longrightarrow CH2NH + H} & $3.84 \times 10^{-11} e^{-750/T}$  \\
267 & \ce{N(^2D) + CH4 \longrightarrow NH + CH3} & $9.6 \times 10^{-12} e^{-750/T}$  \\
268 & \ce{NH + CH3 \longrightarrow CH2NH + H} & $3.12 \times 10^{-16} T^{1.55} e^{-103/T}$  \\
269 & \ce{CN + CH2NH \longrightarrow H2CNCN + H} & $2.7 \times 10^{-11}$  \\
270 & \ce{CH5+ + CH2NH \longrightarrow CH2NH2+ + CH4} & $3.0 \times 10^{-9}$  \\
271 & \ce{H + CH2NH \longrightarrow H2CN + H2} & $4.0 \times 10^{-14}$  \\
272 & \ce{NH2 + H2CN \longrightarrow NH3 + HCN} & $5.42 \times 10^{-11} (T/300)^{-1.06} e^{-60.8/T}$  \\
273 & \ce{CH3+ + NH3 \longrightarrow CH2NH2+ + H2} & $1.0 \times 10^{-9}$  \\
274 & \ce{NH + H2 \longrightarrow NH3} & $1.0 \times 10^{-9}$  \\
275 & \ce{N(^2D) + H2 \longrightarrow NH + H} & $4.2 \times 10^{-11} e^{-880/T}$ \\
276 & \ce{NH2 + H \longrightarrow NH + H2} & $5.25 \times 10^{-12}$  \\
277 & \ce{N2+ + CH4 \longrightarrow N2H+ + CH3} & $3.42 \times 10^{-11}$  \\
278 & \ce{N2H+ + C2H2 \longrightarrow N2 + C2H3+} & $1.41 \times 10^{-9}$  \\
279 & \ce{CH4 + NH2 \longrightarrow CH3 + NH3} & $3.99 \times 10^{-14}$  \\
280 & \ce{C2H2 + NH2 \longrightarrow C2H + NH3} & $1.11 \times 10^{-13}$  \\
281 & \ce{C2H4 + NH2 \longrightarrow C2H3 + NH3} & $3.42 \times 10^{-14}$  \\
282 & \ce{C2H5 + NH2 \longrightarrow C2H4 + NH3} & $4.15 \times 10^{-11}$  \\
283 & \ce{C2H6 + NH2 \longrightarrow C2H5 + NH3} & $6.14 \times 10^{-13}$  \\
284 & \ce{C2H6 + N+ \longrightarrow NH + C2H5+} & $1.0 \times 10^{-10}$  \\
285 & \ce{C2H6 + N+ \longrightarrow NH2 + C2H4+} & $5.5 \times 10^{-10}$  \\
286 & \ce{C2H6 + N+ \longrightarrow NH3 + C2H3+} & $2.5 \times 10^{-10}$ \\
287 & \ce{C2H6 + N+ \longrightarrow CH4 + HCNH+} & $1.0 \times 10^{-10}$ \\
288 & \ce{NH + C2H2 \longrightarrow HC2N + H2} & $2.01 \times 10^{-9} T^{-1.07}$  \\
289 & \ce{NH + C2H4 \longrightarrow CH3CN + H2} & $2.3 \times 10^{-12} (T/300)^{-1.09}$  \\
290 & \ce{NH + C2H6 \longrightarrow C2H5N + H2} & $6.8 \times 10^{-12}$  \\
291 & \ce{NH + C4H2 \longrightarrow C4HN + H2} & $8.24 \times 10^{-9} T^{-1.23}$  \\
292 & \ce{CH2NH2+ + e \longrightarrow CH2NH + H} & $1.0 \times 10^{-6} (300/T_e)^{0.7}$ \\
293 & \ce{CH2NH2+ + e \longrightarrow CH2 + NH2} & $1.0 \times 10^{-6} (300/T_e)^{0.7}$ \\
294 & \ce{N2H+ + e \longrightarrow NH + N} & $5.67 \times 10^{-8} (300/T_e)^{0.51}$  \\
295 & \ce{C4H2+ \longrightarrow C4H2} & S.R. \\
296 & \ce{C5H2N+ \longrightarrow C5H2N} & S.R. \\
297 & \ce{CH2NH2+ \longrightarrow CH2NH2} & S.R. \\
298 & \ce{CH2+ \longrightarrow CH2} & S.R. \\
299 & \ce{CH3CNH+ \longrightarrow CH3CNH} & S.R. \\
300 & \ce{CH3NH3+ \longrightarrow CH3NH3} & S.R. \\
301 & \ce{CH3+ \longrightarrow CH3} & S.R. \\
302 & \ce{CH4+ \longrightarrow CH4} & S.R. \\
303 & \ce{CH5+ \longrightarrow CH5} & S.R. \\
304 & \ce{CH+ \longrightarrow CH} & S.R. \\
305 & \ce{CN+ \longrightarrow CN} & S.R. \\
306 & \ce{C2H2+ \longrightarrow C2H2} & S.R.\\
307 & \ce{C2H3CNH+ \longrightarrow C2H3CNH} & S.R. \\
308 & \ce{C2H3+ \longrightarrow C2H3} & S.R. \\
309 & \ce{C2H4+ \longrightarrow C2H4} & S.R. \\
310 & \ce{C2H5CNH+ \longrightarrow C2H5CNH} & S.R. \\
311 & \ce{C2H5+ \longrightarrow C2H5} & S.R. \\
312 & \ce{C2H6+ \longrightarrow C2H6} & S.R. \\
313 & \ce{C2H+ \longrightarrow C2H} & S.R. \\
314 & \ce{C2N2+ \longrightarrow C2N2} & S.R. \\
315 & \ce{C3H2+ \longrightarrow C3H2} & S.R. \\
316 & \ce{C3H2+ \longrightarrow C3H2} & S.R. \\
317 & \ce{C3H4+ \longrightarrow C3H4} & S.R. \\
318 & \ce{C3H5+ \longrightarrow C3H5} & S.R. \\
319 & \ce{C3+ \longrightarrow C3} & S.R. \\
320 & \ce{C4H3NH+ \longrightarrow C4H3NH} & S.R. \\
321 & \ce{C4H3+ \longrightarrow C4H3} & S.R. \\
322 & \ce{C4H5NH+ \longrightarrow C4H5NH} & S.R. \\
323 & \ce{C4H5+ \longrightarrow C4H5} & S.R. \\
324 & \ce{C4H6+ \longrightarrow C4H6} & S.R. \\
325 & \ce{C4H7+ \longrightarrow C4H7} & S.R. \\
326 & \ce{C4N2H+ \longrightarrow C4N2H} & S.R. \\
327 & \ce{C4N2+ \longrightarrow C4N2} & S.R. \\
328 & \ce{C5H3N+ \longrightarrow C5H3N} & S.R. \\
329 & \ce{C5H5+ \longrightarrow C5H5} & S.R. \\
330 & \ce{C5H7+ \longrightarrow C5H7} & S.R. \\
331 & \ce{C5H9+ \longrightarrow C5H9} & S.R. \\
332 & \ce{C6H2+ \longrightarrow C6H2} & S.R. \\
333 & \ce{C6H3+ \longrightarrow C6H3} & S.R. \\
334 & \ce{C6H4+ \longrightarrow C6H4} & S.R. \\
335 & \ce{C6H5+ \longrightarrow C6H5} & S.R. \\
336 & \ce{C6H7+ \longrightarrow C6H7} & S.R. \\
337 & \ce{C7H5+ \longrightarrow C7H5} & S.R. \\
338 & \ce{C7H9+ \longrightarrow C7H9} & S.R. \\
339 & \ce{C8H2+ \longrightarrow C8H2} & S.R. \\
340 & \ce{C8H3+ \longrightarrow C8H3} & S.R. \\
341 & \ce{C8H4+ \longrightarrow C8H4} & S.R. \\
342 & \ce{C8H5+ \longrightarrow C8H5} & S.R. \\
343 & \ce{C8H6+ \longrightarrow C8H6} & S.R. \\
344 & \ce{C10H6+ \longrightarrow C10H6} & S.R. \\
345 & \ce{C12H6+ \longrightarrow C12H6} & S.R.\\
346 & \ce{C+ \longrightarrow C} & S.R. \\
347 & \ce{HCNH+ \longrightarrow HCNH} & S.R.\\
348 & \ce{HCN+ \longrightarrow HCN} & S.R. \\
349 & \ce{HC3NH+ \longrightarrow HC3NH} & S.R. \\
350 & \ce{H2+ \longrightarrow H2} & S.R.\\
351 & \ce{H+ \longrightarrow H} & S.R.\\
352 & \ce{NH4+ \longrightarrow NH4} & S.R. \\
353 & \ce{N2(A)+ \longrightarrow N2(A)} & S.R. \\
354 & \ce{N2(B)+ \longrightarrow N2(B)} & S.R. \\
355 & \ce{N2(X)+ \longrightarrow N2(X)} & S.R. \\
356 & \ce{N2H5+ \longrightarrow N2H5} & S.R.\\
357 & \ce{N2H+ \longrightarrow N2H} & S.R. \\
358 & \ce{N3+ \longrightarrow N3} & S.R. \\
359 & \ce{N4+ \longrightarrow N4} & S.R. \\
360 & \ce{N+ \longrightarrow N(S)} & S.R.\\
361 & \ce{clC3H3+ \longrightarrow clC3H3} & S.R. \\
362 & \ce{lC3H3+ \longrightarrow lC3H3} & S.R. \\

\label{Tableappx}
\end{longtable}
}
\newpage








\bibliography{full_ref}
\bibliographystyle{aasjournalv7}



\end{document}